\documentclass[a4paper,fleqn,usenatbib]{mnras}

\usepackage{txfonts,times}
\usepackage{graphicx}			
\usepackage[T1]{fontenc}	
\usepackage{type1cm}			
\usepackage{xspace}				
\usepackage{breakurl}			

\newcommand{\fif}{4U~1538$-$522\xspace}						
\newcommand{\qvnor}{QV Nor\xspace}								
\newcommand{\nh}{$N_{\mathrm{H}}$\xspace}					
\newcommand{\msol}{$M_{\sun}$\xspace}							

\newcommand{\rxte}{\textit{RXTE}\xspace}					
\newcommand{\suz}{\textit{Suzaku}\xspace}
\newcommand{\integral}{\textit{INTEGRAL}\xspace}

\newcommand{\heaversion}{6.16\xspace}							
\newcommand{\isisversion}{1.6.2-30\xspace}				

\newcommand{\sig}{$4.6\sigma$\xspace}							
\newcommand{\sigerr}{$1.2\sigma$\xspace}					

\title[Evolution of the CRSF in \fif]{Evidence for an Evolving Cyclotron Line Energy in \fif}

\author[P. B. Hemphill et al.]{
  \parbox{\textwidth}{\raggedright Paul~B.~Hemphill,$^{1}$\thanks{E-mail: pbhemphill@physics.ucsd.edu (PBH)},
  Richard~E.~Rothschild$^{1}$,
	Felix~F\"{u}rst$^{2}$,
  Victoria~Grinberg$^{3}$,
  Dmitry~Klochkov$^{4}$,
  Peter~Kretschmar$^{5}$,
  Katja~Pottschmidt$^{6,7}$,
	R\"{u}diger~Staubert$^{4}$,
  and J\"{o}rn Wilms$^{8}$
}\vspace{0.4cm}\\
$^{1}$Center for Astrophysics and Space Sciences, University of California, San Diego, 9500 Gilman Dr., La Jolla, CA 92093-0424, USA \\
$^{2}$Cahill Center for Astronomy and Astrophysics, California Institute of Technology, MC 290-17, 1200 E. California Blvd., Pasadena, CA 91125, USA\\
$^{3}$Massachusetts Institute of Technology, Kavli Institute for Astrophysics, Cambridge, MA 02139, USA \\
$^{4}$Institut f\"{u}r Astronomie und Astrophysik, Universit\"{a}t T\"{u}bingen (IAAT), Sand 1, 72076 T\"{u}bingen, Germany \\
$^{5}$European Space Astronomy Center (ESA/ESAC), Science Operations Deparment, Villanueva de la Ca\~{n}ada, Madrid, Spain \\
$^{6}$Center for Space Science and Technology, University of Maryland Baltimore County, 1000 Hilltop Circle, Baltimore, MD 21250, USA \\
$^{7}$CRESST and NASA Goddard Space Flight Center, Astrophysics Science Division, Code 661, Greenbelt, MD 20771, USA \\
$^{8}$Dr.\ Karl Remeis-Sternwarte \& Erlangen Center for Astroparticle Physics, Sternwartstr. 7, 96049 Bamberg, Germany
}

\date{Accepted XXX. Received YYY; in original form ZZZ}

\pubyear{2015}

\begin{document}
\label{firstpage}
\pagerange{\pageref{firstpage}--\pageref{lastpage}}
\maketitle

\begin{abstract}
We have performed a full time- and luminosity-resolved spectral
analysis of the high-mass X-ray binary \fif using the available
\rxte, \integral, and \suz data, examining both phase-averaged and
pulse-phase-constrained datasets and focusing on the behavior of
the cyclotron resonance scattering feature (CRSF). No statistically
significant trend between the energy of the CRSF and luminosity is
observed in the combined dataset. However, the CRSF energy appears to
have increased by $\sim\!1.5$\,keV in the $\sim\!8.5$\,years between
the \rxte and \suz measurements, with Monte Carlo simulations finding
the \suz measurement \sig above the \rxte points. Interestingly, the
increased \suz CRSF energy is much more significant and robust in
the pulse-phase-constrained spectra from the peak of the main pulse,
suggesting a change that is limited to a single magnetic pole. The
7 years of \rxte measurements do not show any strongly-significant
evolution with time on their own. We discuss the significance of the
CRSF's behavior with respect to luminosity and time in the context of
historical observations of this source as well as recent observational
and theoretical work concerning the neutron star accretion column, and
suggest some mechanisms by which the observed change over time could
occur.
\end{abstract}


\section{Introduction}
\label{sec:intro}

Many neutron stars possess magnetic fields with dipole strengths in excess of
$10^{12}$\,G, making them some of the strongest magnets in the universe.
Material that falls onto the neutron star (NS) is channeled along the field
lines and is concentrated onto the magnetic poles, forming a hot, dense column
of accreted plasma. The conditions within this accretion column are extreme: the
infalling material comes in at relativistic ($v \sim 0.5c$) velocities and must
come to a halt by the time it reaches the NS surface. Radiation pressure in the
column can play a significant role here, shaping the dynamics of the column,
which in turn influences the observed radiation spectrum \citep[see,
e.g.,][]{becker_spectral_2012}.

Cyclotron resonance scattering features (CRSFs, also referred to as ``cyclotron
lines'') are pseudo-absorption features found in the hard X-ray spectra of
approximately two dozen accreting X-ray pulsars. The first CRSF was discovered
in Hercules~X-1 by \citet{truemper_herx1_1978}. CRSFs appear as a result of the
quantized nature of electron cyclotron motion in the characteristically strong magnetic
field of young pulsars, which creates resonances in the electron-photon
scattering cross-section at the cyclotron line energies and scatters photons out
of the line of sight. These features are notable for being the only direct means
of measuring the field strength of the NS, as their centroid energy is directly
proportional to the field strength in the scattering region.

The last several years have seen a great deal of activity around cyclotron
lines, mainly focused on the variation of the CRSF energy with luminosity. The
Be/X-ray binary V~0332+53 displays a significant negative correlation between
CRSF energy and luminosity \citep{mowlavi_v0332_2006,tsygankov_v0332_2010},
while Her~X-1 \citep{staubert_herx1_2007} and GX~304-1
\citep{yamamoto_gx301crsf_2011,klochkov_gx304_2012} show positive correlations.
This relationship can be complex: \textit{NuSTAR} observations of Vela~X-1
\citep{fuerst_vela_2014} found that the energy of the first harmonic of the CRSF
was positively correlated with luminosity, while the behavior of the fundamental
was more difficult to discern, while A~0535+26's CRSF is fairly constant at most
luminosities \citep{caballero_a0535_2007} but does display a positive
correlation between the CRSF energy and flux in certain pulse phase bins
\citep{klochkov_pulseampresolved_2011,mueller_a0535_2013}.  A~0535+26 may also
have a positive correlation in phase-averaged spectra at its highest
luminosities \citep{sartore_a0535_2015}.

A superb, well-studied example of complicated CRSF behavior can be found in
Her~X-1, whose CRSF shows a positive E$_{\mathrm{cyc}}$-luminosity correlation
\citep{staubert_herx1_2007,vasco_herx1_2011}, variability with pulse phase
\citep{vasco_herx1_2013}, and variation with the phase of Her~X-1's 35\,d
super-orbital period. Most recently, \citet{staubert_herx1_2014} showed that for
Her~X-1, on top of all these observed trends, there is additional variability in
the CRSF energy that can only be explained by a long-term decrease in the CRSF
energy. Recent \textit{NuSTAR} observations of Her~X-1 have confirmed this trend
\citep{fuerst_herx1_2013}. This result suggests that there is the possibility
for some long-term evolution within the accretion column that is not observable
either in the overall spectral shape or the luminosity of the source.

The accreting X-ray pulsar \fif was discovered by the \textit{Uhuru} satellite
\citep{giacconi_third_1974}, and the system was identified as an X-ray pulsar by
\citet{becker_1538_1977,davison_1538_1977a}, and \citet{davison_1538_1977b}.
The system consists of a $\sim\!1$\,\msol NS accreting from the stellar wind of
\qvnor, a $\sim\!16$\,\msol B0Iab star
\citep{reynolds_optical_1538_1992,rawls_mass_2011,falanga_hmxbs_2015}.
Estimates of \fif's distance have ranged from $4.5$\,kpc
\citep{clark_chandra1538_2004} to $6.4 \pm 1.0$\,kpc
\citep{reynolds_optical_1538_1992}, with older measurements by
\citet{crampton_1538_1978} and \citet{ilovaisky_1538_1979} finding $6.0 \pm
0.5$\,kpc and $5.5 \pm 1.5$\,kpc, respectively. The system's binary parameters
have similarly been difficult to constrain: while the $3.7$\,d orbital period
was established by some of the earliest observations
\citep{becker_1538_1977,davison_1538_1977b}, the orbital parameters found by
\citet{makishima_spectra_1987,clark_orbit_2000}, and
\citet{mukherjee_orbital_2006} disagree on whether this orbit is circular or
elliptical.  While we adopt an eccentricity of $0.174 \pm 0.015$ from
\citet{clark_orbit_2000} and \citet{mukherjee_orbital_2006}, we note that this
choice has only minimal effects on this analysis. Interestingly,
\citet{rawls_mass_2011} estimated the neutron star mass to be $0.874 \pm
0.073$\,\msol when using the elliptical orbital solution and $1.104 \pm
0.177$\,\msol for a circular orbit --- using either orbital solution, their
results clearly suggest that \fif contains a surprisingly low-mass neutron star.

The pulse period of \fif has an interesting history. Around the time of its
discovery, its pulse period was $528.93 \pm 0.10$\,s \citep{becker_1538_1977};
over the next decade this increased to at least $530.43 \pm 0.014$\,s
\citep{makishima_spectra_1987,corbet_orbit_1993}, but \textit{CGRO}-BATSE
observations revealed that the source underwent a torque reversal sometime in
1989 or 1990 \citep{rubin_observation_1997}. The spin-up trend continued
\citep{clark_orbit_2000,coburn_magnetic_2001,mukherjee_orbital_2006,baykal_recent_2006}
for approximately 20 years, until another torque reversal in $\sim\!2008$ put
the source on its current spin-down trend, as revealed by
\textit{Fermi}-GBM\footnote{See
\url{http://gammaray.nsstc.nasa.gov/gbm/science/pulsars}}
\citep{finger_gbm_2009}, \integral \citep{hemphill13}, and \suz
\citep{hemphill14}. The pulse period is currently $\sim\!526$\,s.

\fif's $\sim\!20$\,keV CRSF was discovered by \citet{clark_discovery_1990} in
\textit{Ginga} observations. The \textit{Ginga} spectra were
further analyzed by \citet{makishima_spectra_1987}. The feature has since been
observed in data from \rxte \citep{coburn_magnetic_2001,rodes-roca_first_2009},
\textit{BeppoSAX} \citep{robba_bepposax_2001}, \integral
\citep{rodes-roca_first_2009,hemphill13}, and \suz \citep{hemphill14}. A direct
comparison of these results is somewhat difficult, as the various authors used
different models for the spectral continuum and the CRSF, as well as different
energy bands when calculating fluxes. Discussions of the effects of model
choice on the measured CRSF energy can be found in
\citet{muller_nocorrelation_2012} and \citet{hemphill13}. However, limiting
ourselves to results using the same models, there is a noticeable change
between the early \rxte observations in 1996--1997, which found the CRSF at
$20.66^{+0.05}_{-0.06}$\,keV \citep{coburn_magnetic_2001}, and the 2012
observation by \suz, where \citet{hemphill14} found the feature at
$22.2^{+0.8}_{-0.7}$\,keV. It should be noted, however, that the early \rxte
spectral fit of \citet{coburn_magnetic_2001} had a very poor reduced $\chi^{2}$
of $\sim\,2.2$, so its small error bars should not be viewed as authoritative.

In this paper, we re-analyze the archival \rxte, \integral, and \suz
observations of \fif using consistent spectral models to better understand and
quantify this apparent trend. We also produce the best-yet characterization of
the CRSF's variability with luminosity. After a summary of the data used and the
data reduction procedure in Section \ref{sec:obs} and a brief timing analysis of
the source in Section \ref{sec:timing}, we present our spectral analysis and
results in Sections \ref{sec:spectral} and \ref{sec:results}. A discussion of
these results in the context of recent theoretical work can be found in Section
\ref{sec:discussion}. All plots display 90\% error bars, and we generally
present 1-sigma (68\%) confidence intervals on linear fits to our results,
unless otherwise indicated.

\section{Observations \& data analysis}
\label{sec:obs}

\begin{figure*}
  \centering
  \includegraphics[width=\textwidth]{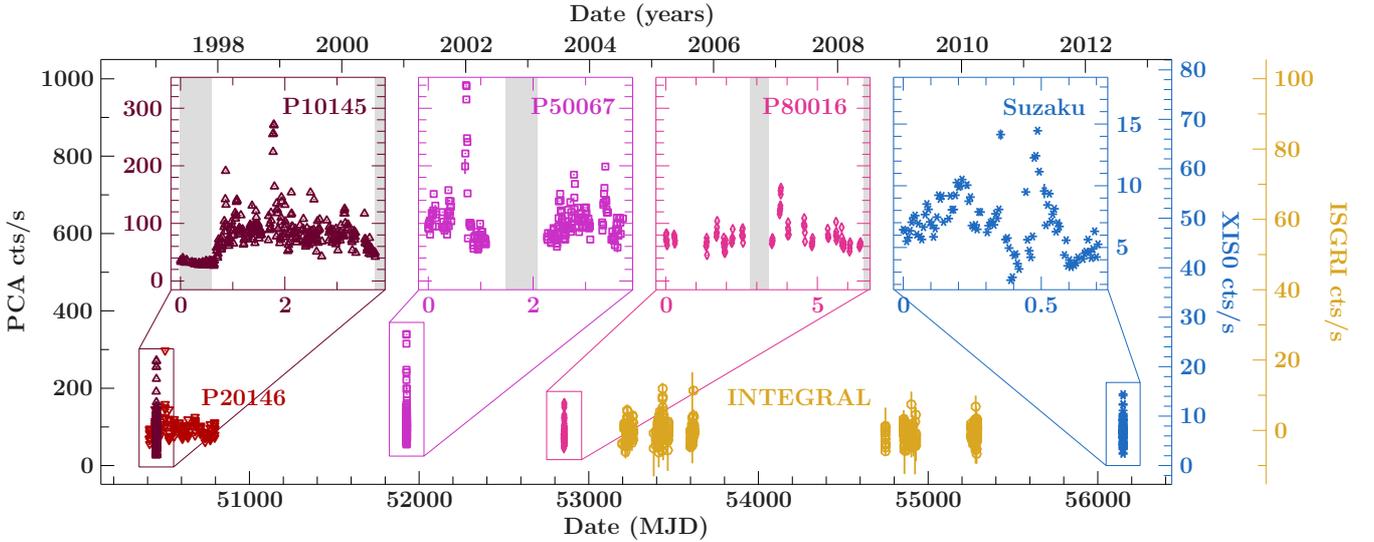}
  \caption{The 3--60\,keV \rxte-PCA (PCU2), 20--40\,keV \integral-ISGRI, and
    1--10\,keV \suz-XIS0 lightcurves.  The \rxte and \suz lightcurves are binned
    at the pulse period (see Table \ref{tab:period}), while the \integral data
    are binned at 10 times the pulse period. \rxte proposals
    P10145, P20146, P50067, and P80016 are plotted in dark red triangles, red
    inverted triangles, violet squares, and pink diamonds, respectively, while
    \integral is plotted in gold circles and \suz is plotted using blue
    stars. The inset plots zoom in on the three focused \rxte proposals and the
		\suz observation, with eclipses marked by gray shaded regions; the
		horizontal axis in the inset plots is in days since the start of the
		depicted observation. Note that the scaling between the PCA, XIS0, and ISGRI
		lightcurves is arbitrary.}
  \label{fig:lc}
\end{figure*}

A search of the NASA High Energy Astrophysics Science Archive (HEASARC) finds
70 \rxte observation IDs (obsids) containing \fif between the years 1996 and
2004. There is additionally a single \suz observation from 2012. We supplement
our results with $\sim\!700$\,ks of \integral data lying mostly between the \rxte
and \suz observations; more in-depth analyses of the available \integral data
can be found in \citet{rodes-roca_first_2009} and \citet{hemphill13}. Overall,
our dataset for \fif spans 16 years, from the earliest \rxte observations in
1996 to the 2012 \suz observation. We summarize the analyzed observations of
\fif in Table~\ref{tab:obs}. The \rxte, \integral, and \suz lightcurves are
plotted in Fig.~\ref{fig:lc}.

Unless otherwise stated, all spectral and lightcurve analysis was performed
using version \isisversion of the Interactive Spectral Interpretation System
\citep[ISIS;][]{isis_2000}. 

\begin{table*}
  \centering
  \caption{\rxte, \suz, and \integral observations of \fif}
  \label{tab:obs}
  \begin{tabular}{lrrrr}
    \hline
    Observation & Start (MJD) & End (MJD) & \multicolumn{2}{c}{Exposure (ks)} \\
    \hline
    \textit{RXTE} proposal & & & PCA & HEXTE \\
    10145     & 50450.62 & 50453.63 & 114.1 &  72.9 \\
    20146     & 50411.96 & 50795.15 &  56.4 &  36.8 \\
    50067     & 51924.88 & 51928.39 &  99.1 &  65.0 \\
    80016     & 52851.95 & 52858.35 &  53.4 &  36.3 \\
    \hline
    \textit{INTEGRAL} revolutions & & & JEM-X 1 & ISGRI \\
    0200-0299 & 53198.10 & 53439.40 &  84.6 & 234.0 \\
    0300-0399 & 53465.10 & 53620.90 &  39.6 & 107.1 \\
    0700-0799 & 54747.90 & 54928.60 &  77.2 & 240.2 \\
    0900-0999 & 55252.70 & 55288.90 &  20.5 & 127.3 \\
    \hline
    \textit{Suzaku} obsID & & & XIS 0 & HXD/PIN \\
    407068010 & 56149.02 & 56149.73 &  46.0 &  36.3 \\
    \hline
  \end{tabular}
\end{table*}

\subsection{\rxte data}
\label{ssec:rxte_data}

The Rossi X-ray Timing Explorer \citep[RXTE;][]{bradt_rxte_1993} carried two
instruments relevant to this study: the Proportional Counter Array \citep[PCA;][]{jahoda_pca_1996,jahoda_pca_2006},
a set of five proportional counter units (PCU 0--4) with a nominal energy range
of 2--60\,keV, and the High Energy X-ray Timing Experiment
\citep[HEXTE;][]{rothschild_hexte_1998}, which consists of two independent
clusters of phoswich scintillation detectors (HEXTE-A and HEXTE-B), each with
an energy range of 15--250\,keV. The HEXTE detectors rocked between on-source and
off-source positions to obtain near-real-time background data; while there were
times later in the mission where this rocking mechanism failed, all \rxte
observations of \fif were taken while both HEXTE rocking mechanisms were
functional.

Four \rxte proposals included observations of \fif: P10145, P20146, P50067, and
P80016. P20146 was a monitoring campaign: a year's worth of monthly snapshot
observations between 1996 and 1997, each with $\sim\!2$\,ks exposure, The
remaining three proposals, from 1997, 2001, and 2003, respectively, were
dedicated pointed observations with many observations within a few $3.7$\,d
orbital periods. Each proposal's data are divided into multiple
observation IDs with exposures ranging from a few to a few dozen
kiloseconds each; after excluding observations taken close to and during
the X-ray eclipse, our final \rxte dataset comprises 50 obsids. We
extracted PCA and HEXTE spectra and lightcurves from each obsid using
the standard \rxte pipeline found in version \heaversion of the HEASOFT
software distribution. We then determined the pulse period of the source
and extracted spectra in luminosity and phase bins of interest to this
analysis.

Calibration uncertainties in the background modeling for the \rxte PCA
at high energies can result in the background count rate being over-
or under-estimated by a few percent. Thus, during spectral fitting, we
correct the background in the PCA via the \texttt{corback} procedure in
\textit{ISIS}. The magnitude of this shift was typically on the order
of a few percent, on average reducing the background counting rate by
$\sim\!2$\%. This correction allowed us to take PCA spectra between 3 and
60\,keV; HEXTE spectra were used between 18 and 80\,keV.

\subsection{\suz data}
\label{ssec:suz_data}

\suz carries two sets of instruments: four X-ray Imaging Spectrometers
\citep[XIS 0-3;][]{xis_2007} and the Hard X-ray Detector
\citep[HXD;][]{hxd_2007}.  The XIS telescopes are imaging CCD detectors with
$0.2$--12\,keV energy ranges.  XIS2 was taken offline in 2006 after a
micrometeorite impact, and so we only use XIS0, XIS1, and XIS3 data. The HXD
consists of a set of silicon PIN diodes (energy range 10--70\,keV) and a GSO
scintillator (40--600\,keV); we only used HXD/PIN data in the hard X-ray band,
as the GSO signal-to-noise ratio was very low.

\suz observed \fif on 10 August 2012 for $61.9$\,ks. The reduction of the \suz
data is explained at length in \citet{hemphill14}. In this analysis, we focus
on the phase-averaged data from the first half of the observation (as the
second half contains significantly higher variability), as well as the pulse
phase-constrained spectrum of the peak of the main pulse and the
secondary pulse \citep[phase bins 1 and 4 in][]{hemphill14}. The
data were reprocessed and spectra were extracted using the standard
\suz pipeline in HEASOFT v\heaversion. We used XIS data between 1 and
12\,keV, taking the standard step of ignoring bins between $1.6$ and $2.3$\,keV 
due to calibration uncertainties in that range. Data below 1\,keV was
ignored to avoid having to model the soft excess at those energies,
which is outside the scope of this work. We rebinned the XIS spectra
according to the binning scheme used by \citet{nowak_2012}, which
attempts to best account for the spectral resolution of the XIS
detectors. The HXD/PIN data were used between 15 and 60\,keV and
rebinned to a minimum of 100 counts per bin.

\subsection{\integral data}
\label{ssec:integral_data}

We used two of the instruments aboard \integral to supplement the \rxte and
\suz analysis: ISGRI, a 15\,keV to 2\,MeV CdTe imager which forms the upper
layer of the coded-mask IBIS telescope, and JEM-X, a pair of coded mask X-ray
monitors which work in the 3--35\,keV band. There are a total of 870
$\sim\!2$\,ks exposure Science Windows (SCWs) which include \fif within the
$9\degr\times 9\degr$ fully-coded field of view (FCFOV) of ISGRI and 211
with the source inside the $4.8\degr$-diameter FCFOV of JEM-X. We extracted
ISGRI and JEM-X spectra using the standard analysis pipelines found in version
10.0 of the Offline Scientific Analysis (OSA) software package.


The available data lie mostly in a few \integral revolutions, so we produced
four spectra each for ISGRI and JEM-X 1, adding together SCWs from revolutions
0200 through 0299, 0300 through 0399, 0700 through 0799, and 0900 through 0999,
totalling 667 ISGRI and 125 JEM-X 1 science windows (there were not enough JEM-X
2 SCWs in this dataset to produce good spectra). These cover the years
2004--2006 and 2009--2010. This extraction provides long exposures (long
exposures are needed due to the low signal-to-noise inherent in coded-mask
detectors) while still maintaining some time resolution.

The background for each pixel in a coded-mask detector contains
contributions from every point on the sky in the field of view, so it
was necessary to compile a list of bright sources in the FOV for each
SCW. This list was fed back into the background subtraction routines
in the standard spectral extraction procedure. These data give us
\integral results that lie in the temporal gap between the \rxte and
\suz observations. Several \integral revolutions from after 2010 include
many SCWs with \fif in the field of view; however, these data are rather
sparse and, due to the long-term evolution of the ISGRI detector, the
recommendation for the most recent observations is to ignore data below
22\,keV\footnote{see version 10.0 of the IBIS Analysis User Manual}. For
these reasons, these data are not suitable for this analysis.

\section{Timing analysis} 
\label{sec:timing}

To determine the pulse period of the source, we extracted barycentered
lightcurves for each \rxte obsid. After applying a binary orbit
correction using orbital parameters from \citet{clark_orbit_2000}
and \citet{mukherjee_orbital_2006} (see Table~\ref{tab:orb}), we used
epoch folding \citep{leahy_epfold_1983} to determine an initial guess
for the pulse period in each obsid. By folding the lightcurve on the
pulse period, we produced a ``reference'' pulse profile for each obsid,
which was then compared via cross-correlation to each individual pulse
in the lightcurve (the source is bright enough that individual pulses
are clearly visible in the \rxte/PCA lightcurve, so no averaging was
necessary). The peak in the cross-correlation results gives the phase
shift between the reference pulse profile and the individual pulse.
By fitting the time-of-arrival and phase-shift results for each \rxte
proposal with a polynomial in pulse frequency, a more precise picture of
the pulse period can be obtained, as a linear trend in the phase shift
over time indicates a shift in the frequency from the originally assumed
value, while higher-order terms in the fit return the derivatives of the
pulse frequency.
\begin{equation}
	\delta\varphi(t) = \varphi_{0} + \delta \nu(t - t_{0}) + \frac{1}{2}\dot{\nu}(t-t_{0})^{2}
\end{equation}
Here, $t$ contains the times-of-arrival of the pulses, $\varphi_{0}$ is the
phase at $t = t_{0}$ (in our case, we define $\varphi_{0}$ and $t_{0}$ such that
the peak of the main pulse is $\varphi = 0.0$), $\delta \nu$ is the deviation
of the true pulse frequency from the originally assumed value computed by epoch
folding, and $\dot{\nu}$ is the pulse frequency derivative. With the exception
of \rxte proposal P20146, the \rxte data were over short enough timespans that
the evolution of the pulse period was not needed in the model, and we fixed
$\dot{\nu}$ to zero in these cases. This returns results which are broadly in
line with the analyses of \citet{coburn_magnetic_2001} and
\cite{baykal_recent_2006}. Our pulse period measurements, with 90\% confidence
intervals, are displayed in Table~\ref{tab:period}. We additionally include the
pulse period during the \suz observation, as reported in
\citet{hemphill14}.

\begin{figure*}
	\centering
	\includegraphics[width=\columnwidth]{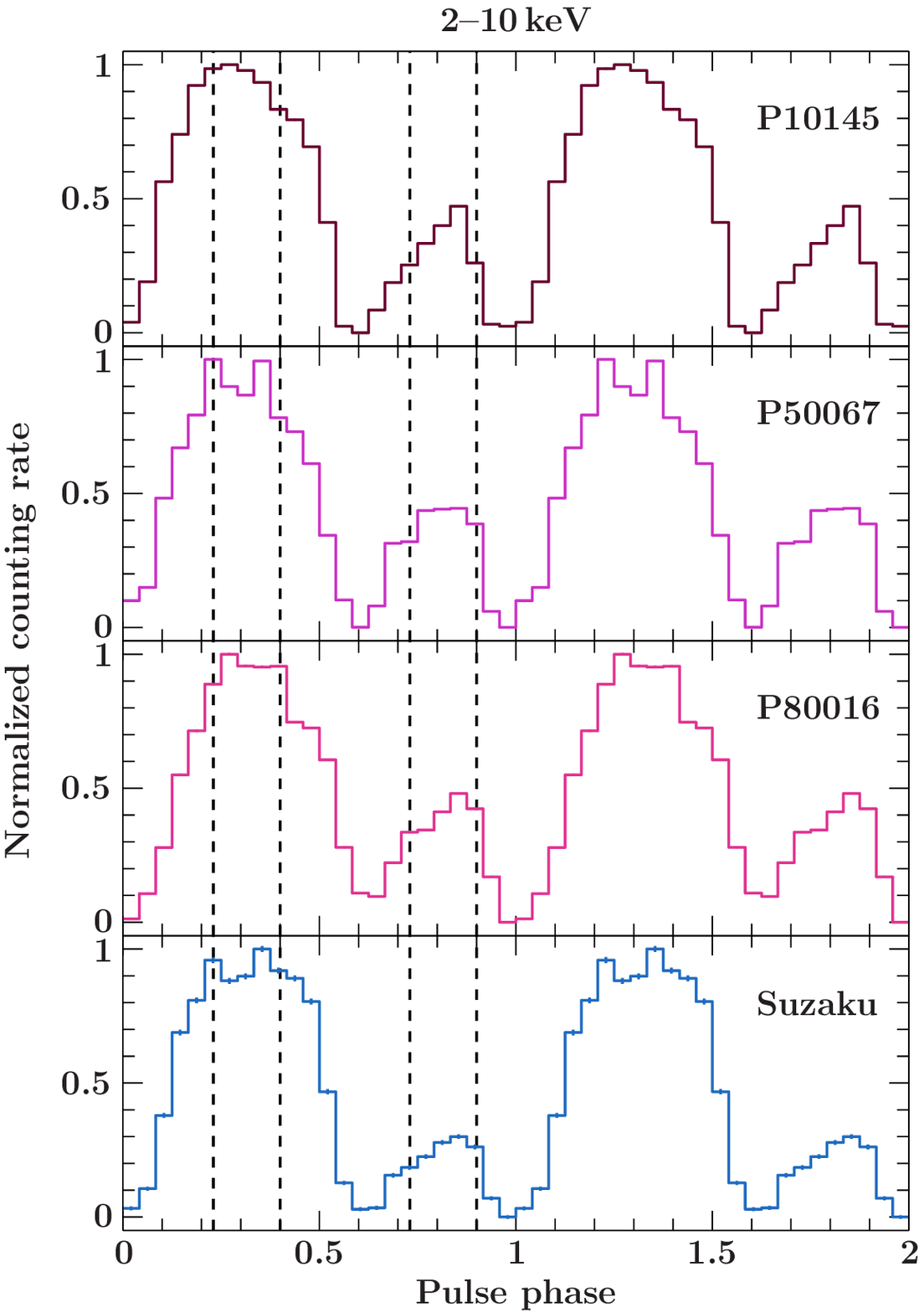}
	\includegraphics[width=\columnwidth]{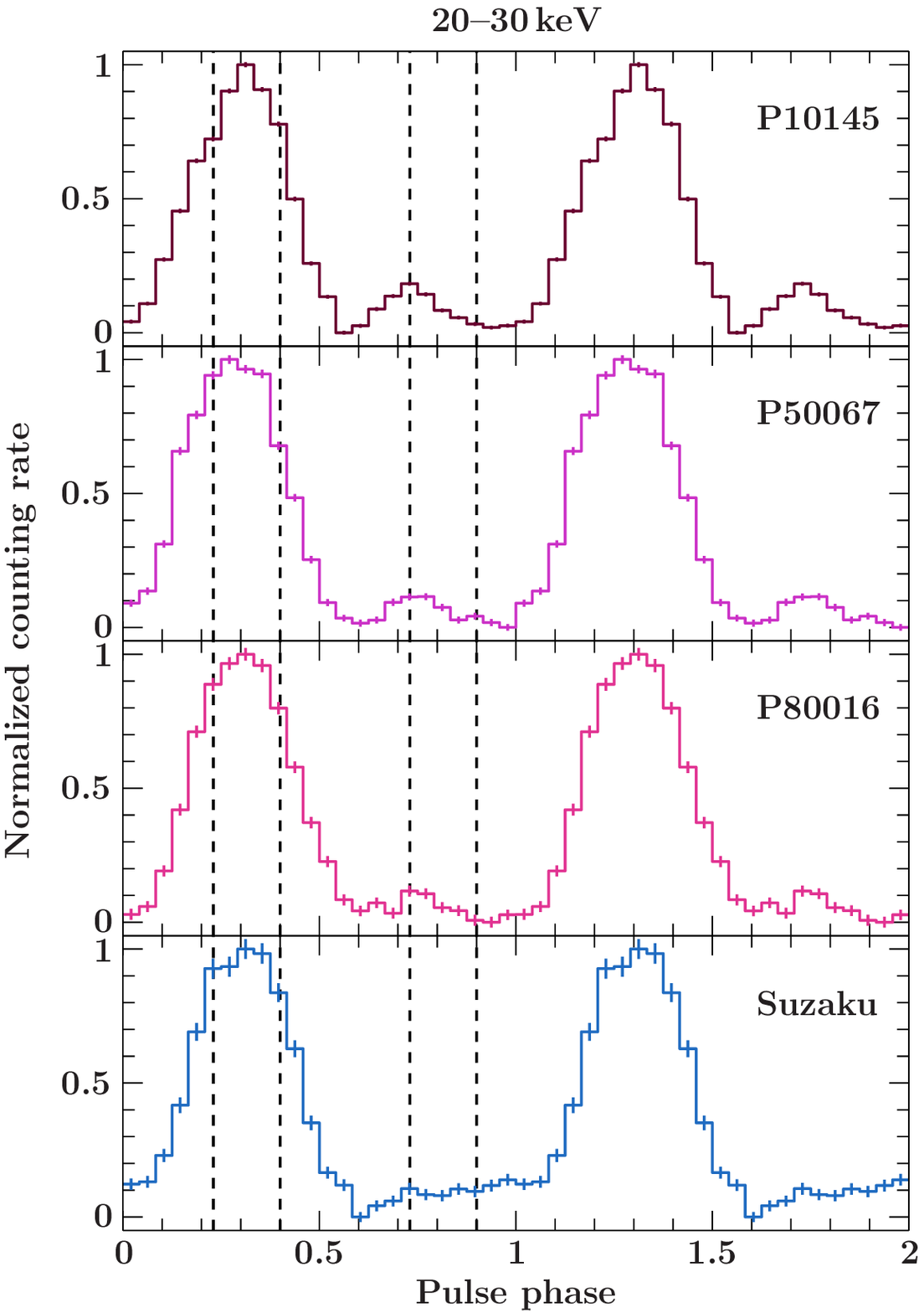}
 \caption{Left panels: the 2--10\,keV pulse profiles using
	\rxte-PCA and \suz-XIS. Right panels: 20--30\,keV pulse profiles using
	\rxte-PCA and \suz-HXD/PIN. The \rxte proposal ID for each PCA
	profile is indicated. Profiles were obtained by folding the respective
	lightcurve on the pulse periods from Table~\ref{tab:period}. The phase
	intervals for pulse-phase-constrained spectral extraction are indicated
by dashed lines.}
	\label{fig:profile}
\end{figure*}

In Fig.~\ref{fig:profile}, we plot the 2--10\,keV and
20--30\,keV pulse profiles for the three focused \rxte proposals
(P10145, P50067, and P80016) using PCA as well as the \suz observation
(using the XIS for the 2--10\,keV profile and the PIN for the
20--30\,keV profile), with the phase bins used in our spectral
extraction indicated. Due to the long timespan and relatively large
uncertainty in $\dot{P}$ for \rxte proposal P20146, we do not plot the
profile for that set of obsids. At higher energies, the primary pulse
narrows and the secondary pulse weakens considerably; the pulse profiles
for \rxte and \suz are qualitatively similar in both bands with the
exception that the secondary pulse appears to vanish completely in the
PIN profile while the PCA profile still shows a weak secondary pulse.
There are no pronounced phase shifts in the positions of the peaks
or major changes in the overall structure of the pulse profile with
energy, unlike, e.g., 4U~0115+63 \citep{ferrigno_4u0115_2009} or
Vela~X-1 \citep{kreykenbohm_velax1_2002}.


\begin{table}
  \centering
  \caption{Orbital parameters of \fif}
  \label{tab:orb}
  \begin{tabular}{llr}
    \hline
    Parameter & Units & \citet{mukherjee_orbital_2006} value \\
    \hline
    $ a\sin(i) $           & lt-s     & $ 53.1 \pm 1.5 $  \\
    $ e $                  &          & $ 0.18 \pm 0.01 $ \\
    $ P_{\mathrm{orb}} $   & d        & $ 3.728382 \pm 0.000011 $ \\
    $ T_{\pi/2} $          & MJD      & $ 52851.33 \pm 0.01 $ \\
    $ \omega_{\mathrm{d}} $&          & $ 40\degr  \pm 12\degr $  \\
    \hline
  \end{tabular}
\end{table}

\begin{table}
  \centering
  \footnotesize
	\caption{Pulse period measurements for \fif using \rxte and \suz}
	\label{tab:period}
  \begin{tabular}{lrrr}
    \hline
    Obs. & MJD range & Pulse period (s) & $\dot{P}$ ($10^{-10}$ s s$^{-1}$) \\
    \hline
    P10145 & 50450.62--50453.67 & $528.824 \pm 0.014$ & \\
    P20146 & 50411.96--50795.17 & $527.9775^{+0.0014}_{-0.0006}$ & $-5.9 \pm 0.3$ \\
    P50067 & 51924.62--51928.56 & $527.596 \pm 0.009$ & \\
    P80016 & 52851.96--52858.38 & $526.834 \pm 0.009$ & \\
		\suz$^{\mathrm{a}}$ & 56149.02--56149.73 & $525.59 \pm 0.04$ \\
    \hline
  \end{tabular} \\
  \flushleft\footnotesize{$^{a}$from \citet{hemphill14}}
\end{table}

\section{Spectral analysis}
\label{sec:spectral}

Our aim in this paper is to examine and quantify the change over time in the
CRSF energy of \fif. This necessitates controlling for other parameters which
may influence the measured line energy. Thus, we perform our analysis on four
different selections of data. The first, and simplest, selection is the spectra
from each of the \rxte obsids. This is a phase-averaged dataset, with no cuts
based on luminosity or pulse phase, although we do exclude observations taken
during the eclipse. This dataset comprises 50 \rxte spectra and the single \suz
observation, and covers a fairly wide range of fluxes. Second, we determined the peak PCU2 counting rate in each individual pulse
and produced a set of good time intervals (GTIs) for four counting rate bins,
with cuts at 103, 123, 140, and 171 counts per second (the \suz
observation's mean flux is approximately in line with the 123--140 count/s \rxte
bin). These GTIs were used to extract a set of phase-averaged and
luminosity-selected spectra. Note that this dataset is constructed on a
pulse-by-pulse basis, adding up all individual pulses in a proposal that are in
a given range of counting rates. This avoids averaging over too broad
of a range of fluxes while also ensuring that the dataset is fully
phase-averaged, with better statistics than the obsid-by-obsid dataset.
Finally, using the pulse period of the source, we produced GTIs and
extracted spectra for the peak of the main pulse and for the secondary
pulse.

For our pulse-phase-constrained analysis, in order to ensure
that we selected the same pulse phase bins across all datasets, we
computed the cross-correlation between the 2--10\,keV \rxte PCA and \suz
XIS profiles; the shift in the PCA profile which gives the largest value of the
correlation coefficient is thus the phase shift between that PCA profile
and the XIS profile, which was then used to define phase 0.0\ in all
datasets. For \rxte proposals P10145, P50067, and P80016, we used the
pulse profile from the full proposal lightcurve, as the change in the
pulse period is small over the few days that each proposal spans. \rxte
proposal P20146 was treated on an obsid-by-obsid basis due to its
extended duration and the long gaps between observations. The pulse
phase bins used for the \rxte data thus correspond to the 1st and 4th
phase bins used in \citet{hemphill14}, each of which has a width of
one-sixth of the pulse. The phase bins used are plotted in
Fig.~\ref{fig:profile}.

The main pulse spectra were restricted to
a PCA counting rate between 60 and 200 counts/s and the secondary
pulse spectra were restricted to a 50--160 counts/s range in order
to avoid major dips and flares. This phase-resolved analysis is
essential, as the CRSF energy varies over \fif's pulse by 5--10\,\%
\citep[1--2\,keV;][]{hemphill14}. Since the phase-averaged CRSF energy
is the weighted average of the observed CRSF across all phase bins, any
long-term evolution in either the pulse shape or the CRSF's variability
with pulse phase could potentially influence the measured CRSF energy in
phase-averaged spectra. Performing this pulse phase-constrained analysis
allows us to account for this effect.

For \suz, we produced a single set of XIS and HXD/PIN spectra for
the first half of the observation \citep[this is effectively the
sum of phase-averaged spectra 1 through 4\ in ][]{hemphill14} for
comparison to the phase-averaged and luminosity-resolved \rxte spectra,
while for comparison to the pulse-phase-constrained \rxte spectra
we took the corresponding bins (phase-resolved spectra 1 and 4)
from the phase-resolved spectra of \citet{hemphill14}. The quality
of the \integral data is not high enough to warrant phase-resolved
spectroscopy, so we simply use the four sets of phase-averaged spectra
as described in Section~\ref{ssec:integral_data} for comparison to the
phase-averaged \rxte and \suz datasets.

\subsection{Spectral model}
\label{ssec:models}

The choice of spectral model can influence the measured CRSF parameters
\citep[see, e.g.,][]{muller_nocorrelation_2012}. Thus, an important first step
is to ensure that every spectrum is fit with the same model. There are several
different phenomenological continuum models used for accreting X-ray pulsars, and while no model
is devoid of problems, the model that we found worked best overall for \rxte,
\suz, and \integral was a powerlaw of photon index $\Gamma$ modified by the
standard \texttt{highecut} high-energy cutoff \citep{white_highecut}:

\begin{eqnarray}
  \mathtt{plcut}(E) &=& \left\{\begin{array}{cc}
    A E^{-\Gamma} & E < E_{\mathrm{cut}} \\
    A E^{-\Gamma}\exp\left( \frac{E_{\mathrm{cut}} - E}{E_{\mathrm{fold}}} \right) & E \ge E_{\mathrm{cut}}
	\end{array}\right. 
  \label{eqn:plcut}
\end{eqnarray}.

This piecewise model can result in spurious features around the
cutoff energy $E_{\mathrm{cut}}$; we account for this by including
a narrow (width frozen to $0.01$\,keV) negative Gaussian with its
energy tied to the cutoff energy $E_{\mathrm{cut}}$ \citep[see,
e.g.,][]{coburn_magnetic_2002,fuerst_herx1_2013}. For simplicity,
we will refer to this modified powerlaw-cutoff continuum model as
\texttt{mplcut}.

The CRSF at $\sim\!21$\,keV is modeled using a Gaussian optical depth profile,
\texttt{gauabs}. This model component is identical to the XSPEC model
\texttt{gabs} with a slightly different definition of the line depth parameter,
which here represents the maximum optical depth in the line, $\tau_{0}$:

\begin{eqnarray}
  \mathtt{gauabs}(E) &=& e^{-\tau\left(E\right)} \\
  \tau(E) &=& \tau_{0}\exp\left( -\frac{\left(E-E_{0}\right)^{2}}{2\sigma^{2}}\right)
  \label{eqn:gauabs}
\end{eqnarray}.

The first harmonic of the CRSF is at $\sim\!50$\,keV in \fif
\citep{rodes-roca_first_2009,hemphill13}. We include the harmonic CRSF in the
phase-averaged, luminosity-selected spectra, using a \texttt{gauabs} feature
with its depth free to vary but with energy and width fixed to $50$ and
$5$\,keV, respectively. The feature is only detected (i.e., depth inconsistent
with zero) in the brightest spectra, but we include it in all the phase-averaged
spectra in order to ensure consistency in this dataset. The harmonic CRSF was
not detected in the individual obsids or the pulse phase-constrained data and
thus was not included in the final model for those datasets. We performed fits
for \suz with and without the harmonic CRSF, but the only noticeable effect on
the fitted parameters was a slight decrease in the source flux (due mainly to
our chosen energy range of 3--50\,keV including part of the harmonic CRSF).

Several X-ray pulsars, \fif included, show a peculiar feature in
their $\sim\!8$--$12$\,keV spectrum. This typically appears as
broad emission around $\sim\!10$--$11$\,keV (as such, it is usually
called the ``10-keV bump'' or ``10-keV feature''), although it
can be modeled as an absorption feature at somewhat lower energy
\citep{muller_sleeping_J1946_2012}. A discussion of this feature, along
with several examples, can be found in \citet{coburn_magnetic_2001},
although currently no satisfactory physical explanation exists.
In this work, most ($\sim\!80$\%) of our spectra were poorly fit
without a 10-keV feature, with reduced $\chi^{2} > 1.4$. In these
poorly-fitted cases, we include a \texttt{gauabs} feature in the model
with energy $\sim\!8.5$\,keV and width frozen to 1\,keV. However, in the
$\sim\!20$\% of spectra which were well-fit without such a feature, its
inclusion tended to severely overfit the spectrum (reduced $\chi^{2}
\la 0.5$). We thus include the feature only in the poorly-fit
cases, in order to bring the reduced $\chi^{2}$ down to $\sim\!1$ and
obtain realistic error bars on the other fitted parameters. There do
not seem to be any systematic factors which determine whether or not
a particular spectrum will need an 8-keV dip; the distributions of
fitted parameters and uncertainties for spectra with and without 8-keV
features are entirely consistent with each other. However, spectra which
did not need an 8-keV dip to obtain a good fit do tend to have lower
total counts, with an average of $7.7\times 10^{5}$ counts compared to
$1.8\times 10^{6}$ for the spectra with an 8-keV dip. Thus, the lack of
an 8-keV dip in some spectra is an issue of data quality and is not of
physical origin. We also tried fitting with an emissive feature at
$\sim\!11$\,keV as has been used before
\citep{rodes-roca_first_2009,hemphill13}; however, the broad width of
the 10-keV feature led to it interfering with the parameters of the
CRSF.

We do include an 8-keV dip in the model for \suz, in order to bring that model
as close as possible to the one we used for \rxte. Its depth and energy are
consistent with the values found by \rxte. The \integral spectra suffered from
the same issues as the lower-quality \rxte spectra in this matter, so we did not
include an 8-keV feature in those spectra.

The spectral model is further modified by an additive Gaussian emission
line modeling the iron K\,$\alpha$ at $\sim\!6.4$\,keV, as well as
photoelectric absorption, using the latest version \footnote{See \url{http://pulsar.sternwarte.uni-erlangen.de/wilms/research/tbabs/}} of
the \texttt{tbnew} absorption model \citep{wilms_tbnew_2010}. We used the
abundances from \citet{wilms_tbabs} and the cross-sections from
\citet{verner_xsect}. The iron K\,$\alpha$ energy is fixed to $6.4$\,keV in the
\rxte spectra, and the line width is fixed to $0.01$\,keV in all spectra, as it
is unresolved even at the energy resolution of \suz's XIS.

\begin{figure*}
	\centering
	\includegraphics[width=\columnwidth]{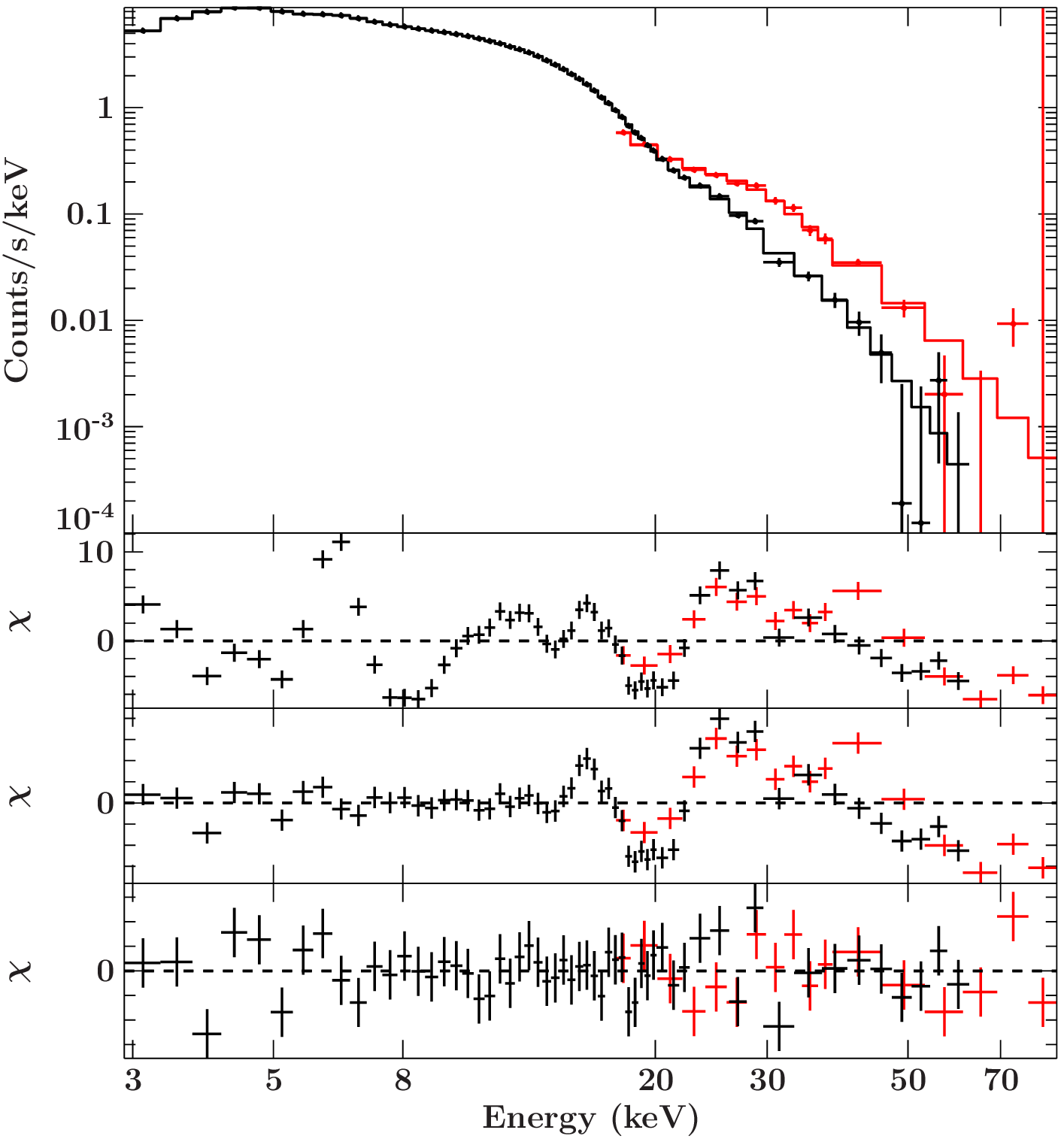}
	\includegraphics[width=\columnwidth]{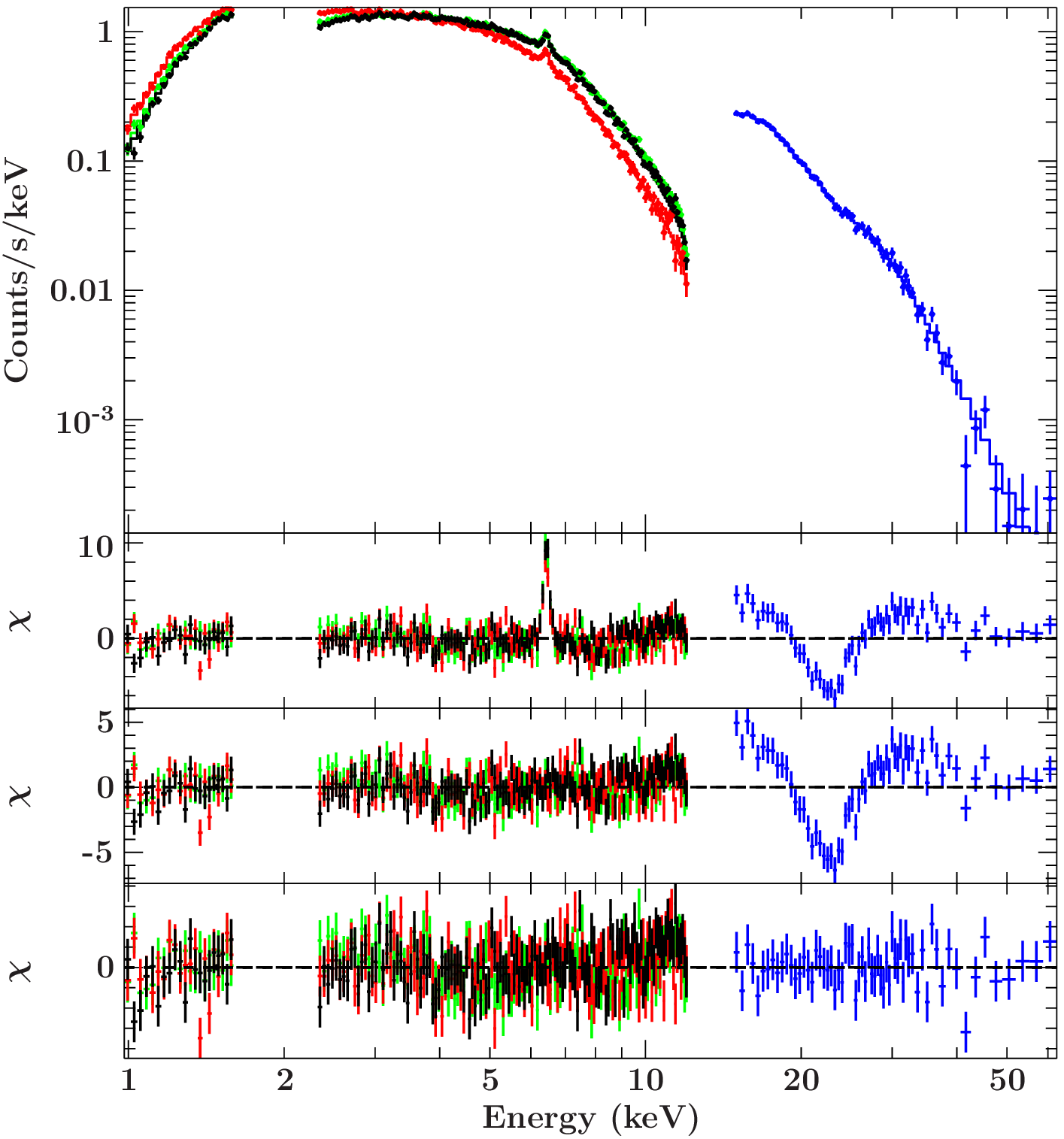}
 \caption{Representative phase-averaged \rxte (left panel) and \suz
 (right panel) spectra. The \rxte spectra are from proposal P50067 in
 the 140--171 counts/s flux bin, with PCA plotted in black and HEXTE
 plotted in red. For \suz, XIS 0, 1, and 3 are plotted in black, red, and
 green, respectively, and the HXD/PIN is plotted in blue. In each plot,
 the top panel displays the counts spectrum with the best-fit model
 overplotted, while the lower panels display, from top to bottom,
 the residuals for a fit with only a \texttt{mplcut} continuum, the
 \texttt{mplcut} continuum with a $6.4$\,keV iron K\,$\alpha$ line and an
 8-keV dip, and the best-fit residuals including the CRSF.}
	\label{fig:spectra}
\end{figure*}

To finish out the model, we apply a multiplicative constant to each
instrument to account for flux calibration differences between the PCA and
HEXTE, the XIS and HXD/PIN, and \integral's JEM-X and ISGRI. The XIS0, PCA, and JEM-X
constants were fixed to 1, while HEXTE and ISGRI's constants were allowed to
vary, with the HEXTE constant typically taking values around $\sim\!0.85$ and
ISGRI's constant being found between $0.8$ and 1. For \suz, we left the HXD/PIN
calibration constant free to vary in the phase-averaged data, finding a value of
$1.02^{+0.08}_{-0.02}$. The phase-constrained spectra were very difficult to
constrain with the calibration constant left free to vary, so we fixed it to the
recommended value of $1.16$ in that dataset. We plot a representative \rxte
spectral fit and the best-fit \suz spectrum and model in Fig.~\ref{fig:spectra}.

As a check, we also performed fits with a ``Fermi-Dirac cutoff''
continuum \citep{tanaka_fdcut}:

\begin{equation}
	\mathtt{pow*fdcut}(E) = A E^{-\Gamma} \times \left[1 + \exp\left( \frac{E - E_{\mathrm{cut}}}{E_{\mathrm{fold}}} \right)\right]^{-1}
	\label{eqn:fdcut}
\end{equation}.

This, unlike \texttt{plcut}, is not piecewise. However, \texttt{fdcut}
tended to fit the \rxte data worse than \texttt{mplcut}; while we
were eventually able to obtain comparable goodness-of-fit to our
\texttt{mplcut} results, it was necessary to include a significantly
deeper 8-keV dip, with central optical depths greater by a factor of
$\sim\!2$--$10$. In the cases with the most pronounced 8-keV dip, the
remaining spectral parameters were very poorly constrained. Fitting
with an emissive Gaussian at $\sim\!11$\,keV resulted in somewhat better
constraints on the continuum parameters, but lead to the same difficulty
disentangling the CRSF and 10-keV bump as we encountered using
\texttt{mplcut}. Thus, we will rely primarily on our \texttt{mplcut}
results, supplementing with the \texttt{fdcut} results when applicable.

In all cases we calculate the unabsorbed continuum flux in the 3--50\,keV
band by convolving our continuum model, 8-keV dip, and CRSFs with
an \texttt{enflux} spectral component in \textit{ISIS}. We fit the \rxte
and \suz spectra with this model (omitting the harmonic CRSF in the case of the
individual \rxte obsids), and calculated 90\% error bars on all parameters. We
used a considerably simpler model for the \integral spectra, omitting the
absorption, iron line, and 8-keV dip --- the low spectral sensitivity and 5\,keV
lower energy bound of JEM-X meant these features were undetected.

\section{Results}
\label{sec:results}

Our analysis finds that \fif displays few significant correlations
between spectral parameters. The spectral shape is relatively
stable over time, although the parameters are often quite broadly
scattered, especially in the obsid-by-obsid dataset. We present the
spectral parameters for the pulse-by-pulse, phase-averaged \rxte
dataset in Table~\ref{tab:avg_pars}, and the parameters for the
pulse-phase-constrained datasets in Table~\ref{tab:phase_pars}. The \suz
and \integral parameters are presented in Tables~\ref{tab:suz_pars} and
\ref{tab:integral_pars}, respectively. It was necessary in a few cases
to freeze spectral parameters to obtain a stable fit: the 3\,keV lower
energy bound of the \rxte/PCA meant we had to freeze \nh to its fitted
value in the 123--140 count\,s$^{-1}$ bin of proposal P10145 and the
140--171 count\,s$^{-1}$ bin of proposal 80016. Both spectra found \nh
at $2.9\times 10^{22}$\,cm$^{-2}$; since this low of an \nh does not
have strong effects on the spectrum above 3\,keV, it is unlikely that
fixing this parameter has large effects on the other parameters. We
also froze the energy of the 8-keV dip to $8.5$\,keV in the model for
the secondary pulse in both \rxte and \suz due to the relatively poor
quality and low flux of those spectra. Finally, it was necessary to
freeze both the photon index to $1.2$ and the CRSF width to 3\,keV in
the \integral spectra from the 900--999 revolutions to obtain a stable
fit.

\begin{table*}
	\caption{Pulse-by-pulse spectral parameters from \rxte, \textbf{with 90\% confidence intervals} (note: table continues on next page)}
  \label{tab:avg_pars}
  \begin{tabular}{llrrrr}
    \hline \hline
\multicolumn{2}{r}{PCU2 counting rate bin} & 171-575 & 140-171 & 123-140 & 103-123 \\ 
\hline 
\multicolumn{6}{l}{\textbf{RXTE proposal P10145}} \\
    3--50\,keV flux & $10^{-9}$ erg cm$^{-2}$ s$^{-1}$ & $1.884^{+0.027}_{-0.026}$ & $1.060^{+0.022}_{-0.015}$ & $0.927\pm0.019$ & $0.766^{+0.019}_{-0.011}$ \\ 
    $N_{\rm H}$ & $10^{22}$ cm$^{-2}$ & $1.65\pm0.30$ & $1.8\pm0.4$ &  2.9 (frozen) & $4.5\pm0.4$ \\ 
    $\Gamma$ &  & $1.047^{+0.016}_{-0.015}$ & $1.079^{+0.019}_{-0.011}$ & $1.1571^{+0.0073}_{-0.0020}$ & $1.176^{+0.014}_{-0.000}$ \\ 
    $E_{\rm cut}$ & keV & $14.1\pm0.4$ & $14.9^{+0.5}_{-0.4}$ & $14.7^{+0.5}_{-1.4}$ & $14.5^{+0.4}_{-0.7}$ \\ 
    $E_{\rm fold}$ & keV & $11.3^{+0.6}_{-0.5}$ & $10.1^{+0.8}_{-0.5}$ & $11.3\pm0.9$ & $11.00^{+1.19}_{-0.23}$ \\ 
    $A_{\rm smooth}$ & $10^{-3}$ ph cm$^{-2}$ s$^{-1}$ & $5.4^{+1.4}_{-1.6}$ & $4.7^{+1.8}_{-1.9}$ & $< 4.9$  & $3.2^{+1.4}_{-1.7}$ \\ 
    $E_{\rm cyc}$ & keV & $20.78^{+0.25}_{-0.24}$ & $20.8\pm0.4$ & $20.67^{+0.29}_{-0.12}$ & $20.492^{+0.307}_{-0.024}$ \\ 
    $\sigma_{\rm cyc}$ & keV & $2.79^{+0.26}_{-0.19}$ & $2.94^{+0.30}_{-0.25}$ & $2.64^{+0.17}_{-0.76}$ & $2.42^{+0.22}_{-0.42}$ \\ 
    $\tau_{\rm cyc}$ &  & $0.45\pm0.04$ & $0.55\pm0.06$ & $0.548^{+0.018}_{-0.090}$ & $0.531^{+0.019}_{-0.070}$ \\ 
    $\tau_{\rm harm}$ &  & $1.2^{+1.0}_{-0.6}$ & $< 2.8$  & $< 0.9$  & $< 1.2$  \\ 
    $A_{\rm Fe}$ & $10^{-3}$ ph cm$^{-2}$ s$^{-1}$ & $0.76\pm0.12$ & $0.28^{+0.09}_{-0.08}$ & $0.32\pm0.07$ & $0.29\pm0.07$ \\ 
    $E_{\rm dip}$ & keV &  & $8.45^{+0.29}_{-0.34}$ & $8.5\pm0.4$ & $8.83^{+0.22}_{-0.23}$ \\ 
    $\tau_{\rm dip}$ &  &  & $0.045^{+0.013}_{-0.011}$ & $0.039^{+0.011}_{-0.010}$ & $0.063\pm0.012$ \\ 
    HEXTE constant &  & $0.803\pm0.017$ & $0.768\pm0.025$ & $0.765^{+0.024}_{-0.023}$ & $0.79\pm0.04$ \\ 
    $\chi^{2}_{\rm red}$ (dof) & & 1.28 (65)& 1.22 (63)& 1.34 (64)& 0.96 (63) \\ 
\hline 
\multicolumn{6}{l}{\textbf{RXTE proposal P20146}} \\
    3--50\,keV flux & $10^{-9}$ erg cm$^{-2}$ s$^{-1}$ & $1.45\pm0.04$ & $1.108^{+0.018}_{-0.013}$ & $0.884^{+0.032}_{-0.030}$ & $0.731^{+0.030}_{-0.026}$ \\ 
    $N_{\rm H}$ & $10^{22}$ cm$^{-2}$ & $2.1\pm0.4$ & $2.49^{+0.30}_{-0.36}$ & $1.9\pm0.5$ & $3.4\pm0.5$ \\ 
    $\Gamma$ &  & $1.104\pm0.019$ & $1.115^{+0.023}_{-0.018}$ & $1.172^{+0.023}_{-0.000}$ & $1.192^{+0.023}_{-0.026}$ \\ 
    $E_{\rm cut}$ & keV & $13.55^{+0.50}_{-0.25}$ & $14.9^{+0.6}_{-1.7}$ & $13.90^{+0.59}_{-0.30}$ & $14.0^{+1.2}_{-0.9}$ \\ 
    $E_{\rm fold}$ & keV & $11.9^{+1.0}_{-0.9}$ & $10.5^{+1.4}_{-0.5}$ & $11.0^{+1.6}_{-1.4}$ & $11.4^{+1.9}_{-2.0}$ \\ 
    $A_{\rm smooth}$ & $10^{-3}$ ph cm$^{-2}$ s$^{-1}$ & $< 2.1$  & $3.7\pm1.7$ & $< 1.5$  & $< 5.8$  \\ 
    $E_{\rm cyc}$ & keV & $20.7\pm0.4$ & $20.68^{+0.34}_{-0.30}$ & $21.0\pm0.6$ & $20.3^{+0.6}_{-0.8}$ \\ 
    $\sigma_{\rm cyc}$ & keV & $2.2\pm0.4$ & $2.9^{+0.4}_{-1.0}$ & $2.3\pm0.6$ & $2.3^{+0.8}_{-0.9}$ \\ 
    $\tau_{\rm cyc}$ &  & $0.46\pm0.06$ & $0.54^{+0.07}_{-0.12}$ & $0.46\pm0.10$ & $0.43^{+0.08}_{-0.10}$ \\ 
    $\tau_{\rm harm}$ &  & $< 0.5$  & $< 1.0$  & $< 2.0$  & $< 5.0$  \\ 
    $A_{\rm Fe}$ & $10^{-3}$ ph cm$^{-2}$ s$^{-1}$ & $0.56\pm0.12$ & $0.39^{+0.08}_{-0.09}$ & $0.41\pm0.09$ & $0.27\pm0.07$ \\ 
    $E_{\rm dip}$ & keV &  & $8.45^{+0.31}_{-0.28}$ &  & $8.9^{+0.4}_{-0.5}$ \\ 
    $\tau_{\rm dip}$ &  &  & $0.042^{+0.013}_{-0.009}$ &  & $0.057^{+0.017}_{-0.015}$ \\ 
    HEXTE constant &  & $0.777^{+0.027}_{-0.026}$ & $0.792\pm0.022$ & $0.80\pm0.05$ & $0.81\pm0.05$ \\ 
    $\chi^{2}_{\rm red}$ (dof) & & 0.86 (65)& 1.07 (63)& 1.29 (65)& 1.00 (63) \\ 
\hline 
\multicolumn{6}{l}{\textbf{RXTE proposal P50067}} \\
    3--50\,keV flux & $10^{-9}$ erg cm$^{-2}$ s$^{-1}$ & $1.844^{+0.016}_{-0.013}$ & $1.219\pm0.016$ & $1.002^{+0.017}_{-0.016}$ & $0.927^{+0.021}_{-0.019}$ \\ 
    $N_{\rm H}$ & $10^{22}$ cm$^{-2}$ & $2.68^{+0.30}_{-0.25}$ & $3.00^{+0.33}_{-0.24}$ & $4.6^{+0.4}_{-0.5}$ & $6.8^{+0.6}_{-0.5}$ \\ 
    $\Gamma$ &  & $1.045^{+0.018}_{-0.016}$ & $1.119^{+0.017}_{-0.016}$ & $1.105^{+0.016}_{-0.017}$ & $1.020^{+0.015}_{-0.006}$ \\ 
    $E_{\rm cut}$ & keV & $15.12^{+0.39}_{-0.28}$ & $14.3^{+0.4}_{-0.6}$ & $13.63^{+0.25}_{-0.19}$ & $14.5^{+0.4}_{-1.2}$ \\ 
    $E_{\rm fold}$ & keV & $10.1\pm0.4$ & $11.2^{+0.6}_{-0.5}$ & $10.6\pm0.6$ & $10.3^{+0.9}_{-0.6}$ \\ 
    $A_{\rm smooth}$ & $10^{-3}$ ph cm$^{-2}$ s$^{-1}$ & $4.4^{+0.5}_{-1.4}$ & $2.7^{+1.1}_{-1.4}$ & $4.3^{+3.6}_{-2.9}$ & $5.4^{+2.3}_{-3.9}$ \\ 
    $E_{\rm cyc}$ & keV & $20.69^{+0.18}_{-0.16}$ & $20.77^{+0.22}_{-0.20}$ & $20.88^{+0.26}_{-0.25}$ & $20.92^{+0.33}_{-0.10}$ \\ 
    $\sigma_{\rm cyc}$ & keV & $3.47^{+0.24}_{-0.17}$ & $2.84^{+0.23}_{-0.35}$ & $2.28^{+0.33}_{-0.27}$ & $2.92^{+0.16}_{-0.61}$ \\ 
    $\tau_{\rm cyc}$ &  & $0.516^{+0.013}_{-0.023}$ & $0.52^{+0.04}_{-0.05}$ & $0.47\pm0.05$ & $0.62^{+0.06}_{-0.08}$ \\ 
    $\tau_{\rm harm}$ &  & $0.7^{+0.5}_{-0.4}$ & $< 0.6$  & $< 1.3$  & $< 0.8$  \\ 
    $A_{\rm Fe}$ & $10^{-3}$ ph cm$^{-2}$ s$^{-1}$ & $0.84^{+0.13}_{-0.12}$ & $0.47^{+0.10}_{-0.08}$ & $0.38\pm0.07$ & $0.32\pm0.09$ \\ 
    $E_{\rm dip}$ & keV & $8.4^{+0.4}_{-0.5}$ & $8.30^{+0.26}_{-0.33}$ & $8.31^{+0.24}_{-0.27}$ & $8.96^{+0.29}_{-0.33}$ \\ 
    $\tau_{\rm dip}$ &  & $0.026^{+0.015}_{-0.009}$ & $0.043^{+0.009}_{-0.010}$ & $0.044\pm0.008$ & $0.052\pm0.014$ \\ 
    HEXTE constant &  & $0.797^{+0.010}_{-0.008}$ & $0.805\pm0.016$ & $0.836^{+0.022}_{-0.021}$ & $0.803\pm0.027$ \\ 
    $\chi^{2}_{\rm red}$ (dof) & & 1.83 (58)& 1.29 (58)& 1.01 (57)& 0.97 (58) \\ 
		\hline \hline
	\end{tabular}
	\\ \flushleft{Note: table continues on next page.}
\end{table*}

\begin{table*}
  \centering
  \contcaption{Pulse-by-pulse spectral parameters from \rxte, with 90\% confidence intervals}
  \label{tab:avg_pars_continued}
  \begin{tabular}{llrrrr}
    \hline \hline
\multicolumn{2}{r}{PCU2 counting rate bin} & 171-575 & 140-171 & 123-140 & 103-123 \\ 
\hline 
\multicolumn{6}{l}{\textbf{RXTE proposal P80016}} \\
    3--50\,keV flux & $10^{-9}$ erg cm$^{-2}$ s$^{-1}$ & $1.818\pm0.027$ & $1.271\pm0.029$ & $1.016^{+0.029}_{-0.028}$ & $0.874\pm0.017$ \\ 
    $N_{\rm H}$ & $10^{22}$ cm$^{-2}$ & $3.3\pm0.5$ &  2.9 (frozen) & $6.3\pm0.7$ & $4.7\pm0.5$ \\ 
    $\Gamma$ &  & $1.007^{+0.022}_{-0.021}$ & $1.043^{+0.015}_{-0.027}$ & $1.05\pm0.04$ & $1.085\pm0.022$ \\ 
    $E_{\rm cut}$ & keV & $14.3^{+0.5}_{-0.4}$ & $13.2^{+7.7}_{-0.5}$ & $13.8^{+1.0}_{-0.5}$ & $13.9^{+0.6}_{-0.4}$ \\ 
    $E_{\rm fold}$ & keV & $11.6^{+0.6}_{-0.4}$ & $12.7^{+1.1}_{-3.1}$ & $11.8\pm1.1$ & $10.7\pm0.8$ \\ 
    $A_{\rm smooth}$ & $10^{-3}$ ph cm$^{-2}$ s$^{-1}$ & $6.1^{+2.2}_{-1.2}$ & $< 3.6$  & $< 4.0$  & $< 2.6$  \\ 
    $E_{\rm cyc}$ & keV & $20.76^{+0.31}_{-0.30}$ & $20.9^{+0.5}_{-0.6}$ & $21.1\pm0.5$ & $20.8\pm0.4$ \\ 
    $\sigma_{\rm cyc}$ & keV & $3.04^{+0.30}_{-0.25}$ & $2.7^{+0.9}_{-0.5}$ & $2.8\pm0.5$ & $2.6\pm0.4$ \\ 
    $\tau_{\rm cyc}$ &  & $0.451^{+0.030}_{-0.017}$ & $0.45^{+0.23}_{-0.06}$ & $0.48\pm0.08$ & $0.51\pm0.06$ \\ 
    $\tau_{\rm harm}$ &  & $1.5^{+1.1}_{-0.7}$ & $2.1^{+2.7}_{-1.3}$ & $< 1.9$  & $< 5.0$  \\ 
    $A_{\rm Fe}$ & $10^{-3}$ ph cm$^{-2}$ s$^{-1}$ & $0.59\pm0.15$ & $0.38^{+0.14}_{-0.13}$ & $0.43\pm0.12$ & $0.37\pm0.08$ \\ 
    $E_{\rm dip}$ & keV &  & $8.6^{+0.7}_{-1.1}$ &  &  \\ 
    $\tau_{\rm dip}$ &  &  & $0.029^{+0.019}_{-0.017}$ &  &  \\ 
    HEXTE constant &  & $0.778\pm0.020$ & $0.762\pm0.029$ & $0.83\pm0.04$ & $0.770\pm0.026$ \\ 
    $\chi^{2}_{\rm red}$ (dof) & & 0.87 (61)& 1.28 (60)& 1.25 (61)& 1.37 (61) \\ 
    \hline \hline
  \end{tabular}
\end{table*}

\begin{table}
  \centering
  \caption{Phase-constrained spectral parameters for \rxte, \textbf{with 90\%
	confidence intervals}}
  \label{tab:phase_pars}
  \begin{tabular}{llrr}
    \hline 
\hline 
\multicolumn{2}{l}{\textbf{RXTE proposal P10145}} & Main pulse & Secondary pulse \\ 
    3--50\,keV flux & $10^{-9}$ erg cm$^{-2}$ s$^{-1}$ & $1.459^{+0.021}_{-0.007}$ & $0.701^{+0.014}_{-0.004}$ \\ 
    $N_{\rm H}$ & $10^{22}$ cm$^{-2}$ & $2.6\pm0.4$ & $3.2\pm0.4$ \\ 
    $\Gamma$ &  & $0.881^{+0.015}_{-0.026}$ & $0.9422^{+0.0016}_{-0.0009}$ \\ 
    $E_{\rm cut}$ & keV & $14.6^{+0.7}_{-0.8}$ & $19.6^{+0.5}_{-0.4}$ \\ 
    $E_{\rm fold}$ & keV & $11.0^{+0.6}_{-0.7}$ & $4.68765^{+0.00005}_{-0.14214}$ \\ 
    $A_{\rm smooth}$ & $10^{-3}$ ph cm$^{-2}$ s$^{-1}$ & $0.17^{+0.08}_{-0.09}$ & $0.19\pm0.07$ \\ 
    $E_{\rm cyc}$ & keV & $21.32^{+0.25}_{-0.29}$ & $20.747^{+0.030}_{-0.036}$ \\ 
    $\sigma_{\rm cyc}$ & keV & $3.2\pm0.5$ & $3.08^{+0.00}_{-0.07}$ \\ 
    $\tau_{\rm cyc}$ &  & $0.48\pm0.06$ & $1.54^{+0.08}_{-0.06}$ \\ 
    $A_{\rm Fe}$ & $10^{-3}$ ph cm$^{-2}$ s$^{-1}$ & $0.27^{+0.08}_{-0.10}$ & $0.38\pm0.06$ \\ 
    $E_{\rm dip}$ & keV & $8.42^{+0.24}_{-0.23}$ &  8.5 (frozen) \\ 
    $\tau_{\rm dip}$ &  & $0.054^{+0.014}_{-0.009}$ & $0.063^{+0.013}_{-0.012}$ \\ 
    HEXTE constant &  & $0.786\pm0.015$ & $0.75\pm0.04$ \\ 
    $\chi^{2}_{\rm red}$ (dof) & & 0.85 (64) & 0.75 (65) \\ 
\hline 
\multicolumn{2}{l}{\textbf{RXTE proposal P20146}} & Main pulse & Secondary pulse \\ 
    3--50\,keV flux & $10^{-9}$ erg cm$^{-2}$ s$^{-1}$ & $0.993^{+0.018}_{-0.017}$ & $0.948^{+0.014}_{-0.008}$ \\ 
    $N_{\rm H}$ & $10^{22}$ cm$^{-2}$ & $2.31^{+0.29}_{-0.32}$ & $2.64^{+0.31}_{-0.25}$ \\ 
    $\Gamma$ &  & $1.124^{+0.015}_{-0.018}$ & $1.122^{+0.019}_{-0.015}$ \\ 
    $E_{\rm cut}$ & keV & $14.3^{+1.1}_{-0.8}$ & $15.12^{+0.27}_{-0.80}$ \\ 
    $E_{\rm fold}$ & keV & $11.0^{+0.9}_{-1.0}$ & $10.6^{+0.8}_{-0.5}$ \\ 
    $A_{\rm smooth}$ & $10^{-3}$ ph cm$^{-2}$ s$^{-1}$ & $< 0.15$  & $0.14^{+0.04}_{-0.06}$ \\ 
    $E_{\rm cyc}$ & keV & $20.73^{+0.29}_{-0.31}$ & $20.76^{+0.26}_{-0.16}$ \\ 
    $\sigma_{\rm cyc}$ & keV & $2.7^{+0.5}_{-0.6}$ & $3.11^{+0.26}_{-0.34}$ \\ 
    $\tau_{\rm cyc}$ &  & $0.48^{+0.09}_{-0.07}$ & $0.56^{+0.05}_{-0.07}$ \\ 
    $A_{\rm Fe}$ & $10^{-3}$ ph cm$^{-2}$ s$^{-1}$ & $0.36\pm0.07$ & $0.40\pm0.06$ \\ 
    $E_{\rm dip}$ & keV & $8.57^{+0.22}_{-0.25}$ &  8.5 (frozen) \\ 
    $\tau_{\rm dip}$ &  & $0.050^{+0.012}_{-0.010}$ & $0.049^{+0.009}_{-0.011}$ \\ 
    HEXTE constant &  & $0.790^{+0.018}_{-0.019}$ & $0.835^{+0.017}_{-0.014}$ \\ 
    $\chi^{2}_{\rm red}$ (dof) & & 0.81 (64) & 0.77 (65) \\ 
\hline 
\multicolumn{2}{l}{\textbf{RXTE proposal P50067}} & Main pulse & Secondary pulse \\ 
    3--50\,keV flux & $10^{-9}$ erg cm$^{-2}$ s$^{-1}$ & $1.805\pm0.018$ & $0.912^{+0.014}_{-0.013}$ \\ 
    $N_{\rm H}$ & $10^{22}$ cm$^{-2}$ & $2.7\pm0.4$ & $4.0\pm0.5$ \\ 
    $\Gamma$ &  & $0.870^{+0.016}_{-0.018}$ & $0.901^{+0.025}_{-0.026}$ \\ 
    $E_{\rm cut}$ & keV & $13.79^{+1.18}_{-0.25}$ & $17.0^{+0.5}_{-0.4}$ \\ 
    $E_{\rm fold}$ & keV & $11.2^{+0.4}_{-0.8}$ & $5.6\pm0.4$ \\ 
    $A_{\rm smooth}$ & $10^{-3}$ ph cm$^{-2}$ s$^{-1}$ & $< 0.22$  & $< 0.025$  \\ 
    $E_{\rm cyc}$ & keV & $21.15^{+0.17}_{-0.20}$ & $20.35\pm0.20$ \\ 
    $\sigma_{\rm cyc}$ & keV & $2.74^{+0.52}_{-0.23}$ & $3.19^{+0.17}_{-0.16}$ \\ 
    $\tau_{\rm cyc}$ &  & $0.429^{+0.078}_{-0.023}$ & $1.29^{+0.09}_{-0.08}$ \\ 
    $A_{\rm Fe}$ & $10^{-3}$ ph cm$^{-2}$ s$^{-1}$ & $0.36\pm0.11$ & $0.55\pm0.08$ \\ 
    $E_{\rm dip}$ & keV & $8.33^{+0.23}_{-0.26}$ &  8.5 (frozen) \\ 
    $\tau_{\rm dip}$ &  & $0.052\pm0.009$ & $0.046\pm0.013$ \\ 
    HEXTE constant &  & $0.813^{+0.012}_{-0.011}$ & $0.842\pm0.025$ \\ 
    $\chi^{2}_{\rm red}$ (dof) & & 0.96 (59) & 1.28 (60) \\ 
\hline 
\multicolumn{2}{l}{\textbf{RXTE proposal P80016}} & Main pulse & Secondary pulse \\ 
    3--50\,keV flux & $10^{-9}$ erg cm$^{-2}$ s$^{-1}$ & $1.444\pm0.019$ & $0.769^{+0.016}_{-0.015}$ \\ 
    $N_{\rm H}$ & $10^{22}$ cm$^{-2}$ & $3.3\pm0.5$ & $3.2\pm0.6$ \\ 
    $\Gamma$ &  & $0.846^{+0.028}_{-0.024}$ & $0.894^{+0.030}_{-0.031}$ \\ 
    $E_{\rm cut}$ & keV & $14.9^{+0.5}_{-1.5}$ & $22.6^{+1.3}_{-3.8}$ \\ 
    $E_{\rm fold}$ & keV & $10.5^{+1.1}_{-0.5}$ & $4.0^{+1.6}_{-0.7}$ \\ 
    $A_{\rm smooth}$ & $10^{-3}$ ph cm$^{-2}$ s$^{-1}$ & $< 0.3$  & $< 0.04$  \\ 
    $E_{\rm cyc}$ & keV & $20.68^{+0.32}_{-0.23}$ & $21.8^{+0.4}_{-1.0}$ \\ 
    $\sigma_{\rm cyc}$ & keV & $3.15^{+0.28}_{-0.82}$ & $3.46^{+0.23}_{-0.22}$ \\ 
    $\tau_{\rm cyc}$ &  & $0.49^{+0.05}_{-0.10}$ & $2.12^{+0.14}_{-0.51}$ \\ 
    $A_{\rm Fe}$ & $10^{-3}$ ph cm$^{-2}$ s$^{-1}$ & $0.38^{+0.14}_{-0.12}$ & $0.39\pm0.09$ \\ 
    $E_{\rm dip}$ & keV & $8.0^{+0.4}_{-0.5}$ &  8.5 (frozen) \\ 
    $\tau_{\rm dip}$ &  & $0.045\pm0.011$ & $0.061\pm0.016$ \\ 
    HEXTE constant &  & $0.790\pm0.016$ & $0.80\pm0.04$ \\ 
    $\chi^{2}_{\rm red}$ (dof) & & 0.72 (60) & 0.80 (61) \\ 
    \hline \hline
  \end{tabular}
\end{table}

\begin{table*}
  \centering
	\caption{\suz spectral parameters, \textbf{with 90\% confidence intervals}}
  \label{tab:suz_pars}
  \begin{tabular}{llrrr}
    \hline \hline
    Parameter & Units & Phase-averaged & Pulse peak & Secondary pulse \\ 
    \hline
    3--50\,keV flux & $10^{-9}$ erg cm$^{-2}$ s$^{-1}$ & $1.132^{+0.023}_{-0.051}$ & $1.91^{+0.04}_{-0.05}$ & $0.787\pm0.014$ \\ 
    $N_{\rm H}$ & $10^{22}$ cm$^{-2}$ & $2.185\pm0.023$ & $1.96\pm0.04$ & $2.29\pm0.06$ \\ 
    $\Gamma$ &  & $1.148\pm0.012$ & $0.839^{+0.024}_{-0.023}$ & $0.890^{+0.027}_{-0.032}$ \\ 
    $E_{\rm cut}$ & keV & $18.6\pm0.9$ & $14.7^{+6.5}_{-1.0}$ & $12.3^{+14.3}_{-0.9}$ \\ 
    $E_{\rm fold}$ & keV & $9.9^{+0.5}_{-0.4}$ & $10.1^{+0.7}_{-3.3}$ & $8.1^{+0.6}_{-0.5}$ \\ 
    $A_{\rm smooth}$ & $10^{-3}$ ph cm$^{-2}$ s$^{-1}$ & $24^{+4}_{-19}$ & $< 1.9$  & $< 3.9$  \\ 
    $E_{\rm cyc}$ & keV & $22.7^{+0.6}_{-1.0}$ & $23.0^{+0.9}_{-1.0}$ & $21.2\pm0.6$ \\ 
    $\sigma_{\rm cyc}$ & keV & $2.07^{+0.13}_{-0.72}$ & $2.5^{+1.5}_{-1.2}$ & $2.6^{+1.7}_{-0.6}$ \\ 
    $\tau_{\rm cyc}$ &  & $0.63^{+0.23}_{-0.09}$ & $0.30^{+0.28}_{-0.08}$ & $0.78^{+0.14}_{-0.15}$ \\ 
    $\tau_{\rm harm}$ &  & $< 2.7$  &  &  \\ 
    $A_{\rm Fe}$ & $10^{-3}$ ph cm$^{-2}$ s$^{-1}$ & $0.377\pm0.028$ & $0.34\pm0.06$ & $0.41\pm0.05$ \\ 
    $E_{\rm dip}$ & keV & $8.2\pm0.4$ & $7.8\pm0.4$ &  8.5 (frozen) \\ 
    $\tau_{\rm dip}$ & $10^{-3}$ ph cm$^{-2}$ s$^{-1}$ & $0.031\pm0.014$ & $0.072\pm0.022$ & $0.10\pm0.04$ \\ 
    $\chi^{2}_{\rm red}$ (dof) & & 1.25 (520) & 1.17 (509) & 1.04 (494) \\ 
    \hline \hline
  \end{tabular}
\end{table*}

\begin{table*}
  \centering
	\caption{\integral spectral parameters, \textbf{with 90\% confidence intervals}}
  \label{tab:integral_pars}
  \begin{tabular}{llrrrr}
    \hline \hline
\multicolumn{2}{r}{Revolutions} & 200-299 & 300-399 & 700-799 & 900-999 \\ 
\hline 
Parameter & Units & & & & \\ 
    3--50\,keV flux & $10^{-9}$ erg cm$^{-2}$ s$^{-1}$ & $0.84\pm0.04$ & $0.80^{+0.06}_{-0.05}$ & $0.84\pm0.05$ & $0.94^{+0.09}_{-0.08}$ \\ 
    $\Gamma$ &  & $1.25^{+0.14}_{-0.17}$ & $1.16^{+0.17}_{-0.21}$ & $0.98^{+0.20}_{-0.52}$ &  1.2 (frozen) \\ 
    $E_{\rm cut}$ & keV & $13.0^{+3.9}_{-1.9}$ & $15.0^{+6.7}_{-2.6}$ & $13^{+7}_{-4}$ & $12.4^{+1.4}_{-2.0}$ \\ 
    $E_{\rm fold}$ & keV & $13.6^{+2.2}_{-2.3}$ & $11.3^{+2.8}_{-4.3}$ & $10.6^{+2.4}_{-3.5}$ & $12.4\pm3.0$ \\ 
    $A_{\rm smooth}$ & $10^{-3}$ ph cm$^{-2}$ s$^{-1}$ & $< 11.0$  & $< 20.0$  & $< 21.0$  & $< 52.0$  \\ 
    $E_{\rm cyc}$ & keV & $21.8^{+0.8}_{-0.9}$ & $21.6^{+0.9}_{-0.7}$ & $20.7^{+1.1}_{-1.5}$ & $23.0^{+1.4}_{-1.5}$ \\ 
    $\sigma_{\rm cyc}$ & keV & $2.8^{+1.5}_{-1.2}$ & $2.8^{+1.4}_{-1.7}$ & $4.1^{+2.6}_{-1.5}$ &  3.0 (frozen) \\ 
    $\tau_{\rm cyc}$ &  & $0.44^{+0.23}_{-0.13}$ & $0.73^{+0.38}_{-0.25}$ & $0.57^{+0.45}_{-0.19}$ & $0.54^{+0.44}_{-0.27}$ \\ 
    $\tau_{\rm harm}$ &  & $1.4^{+0.7}_{-0.6}$ & $< 1.6$  & $< 1.2$  & $< 1.3$  \\ 
    ISGRI constant &  & $1.02^{+0.12}_{-0.10}$ & $0.91^{+0.19}_{-0.15}$ & $0.81^{+0.14}_{-0.12}$ & $1.01^{+0.40}_{-0.26}$ \\ 
    $\chi^{2}_{\rm red}$ (dof) & & 1.13 (15)& 0.90 (16)& 0.87 (16)& 0.72 (17) \\ 
    \hline \hline
  \end{tabular}
\end{table*}

\begin{figure*}
  \centering
  \includegraphics[width=\columnwidth]{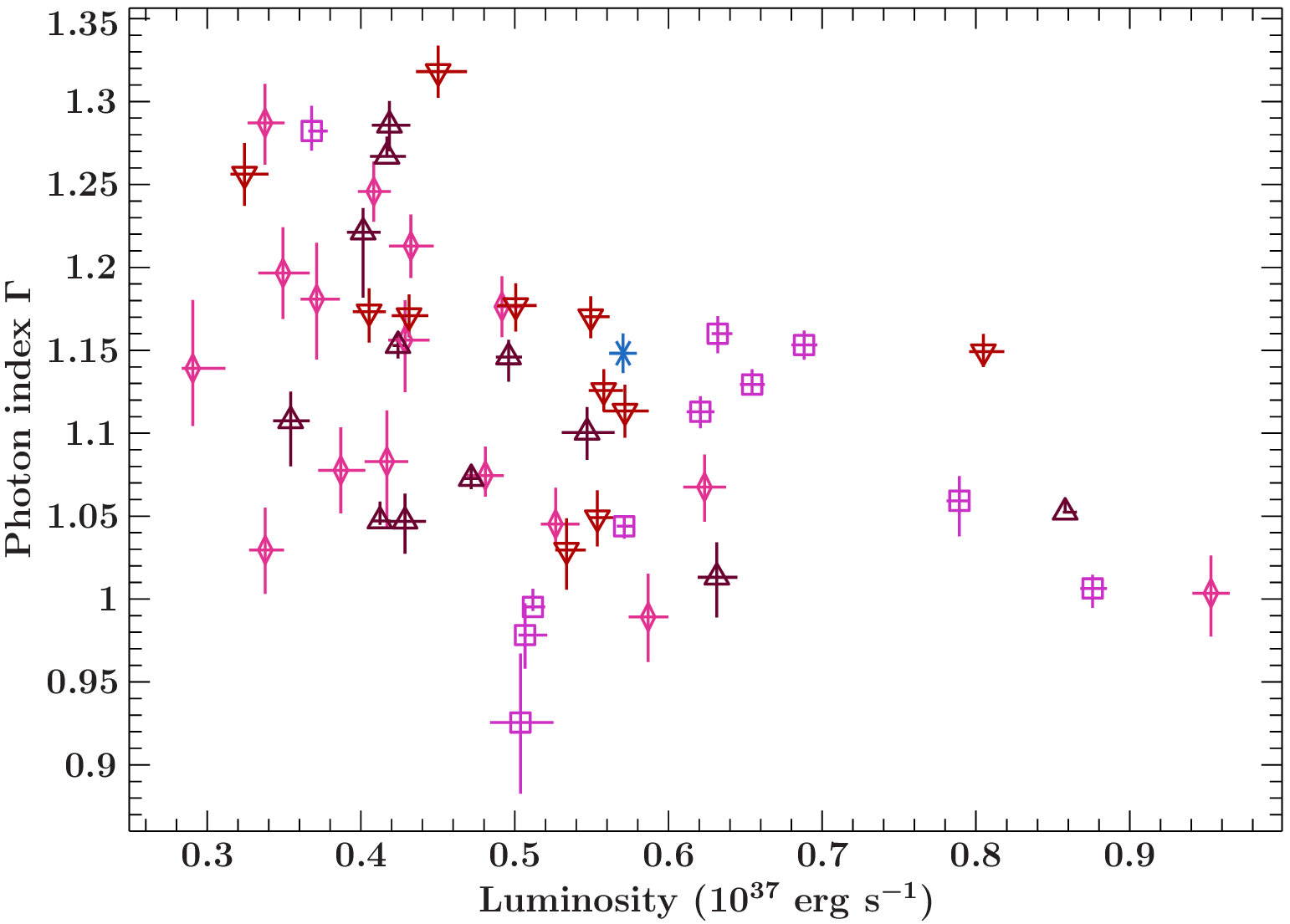}
  \includegraphics[width=\columnwidth]{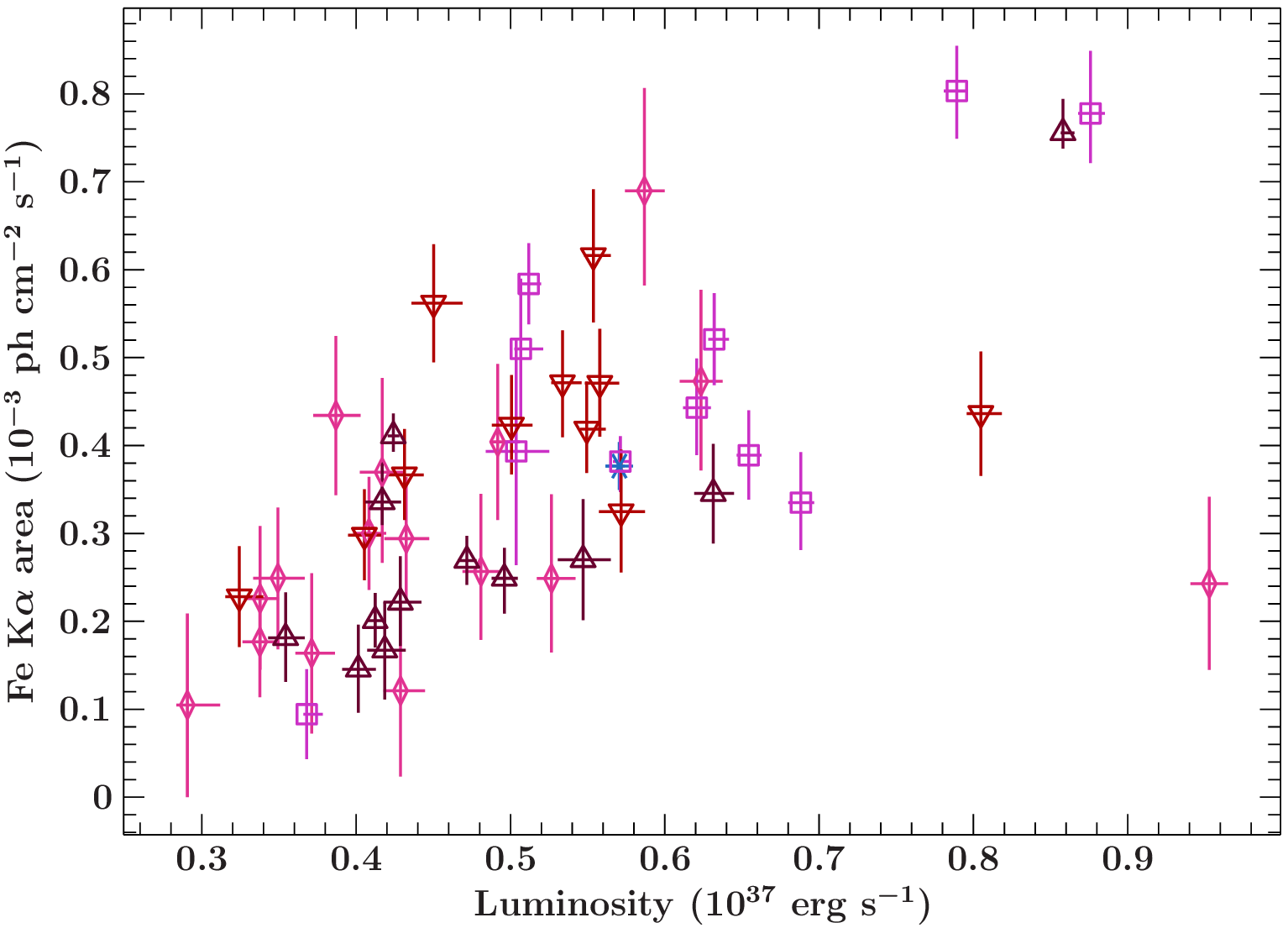} \\
  \includegraphics[width=\columnwidth]{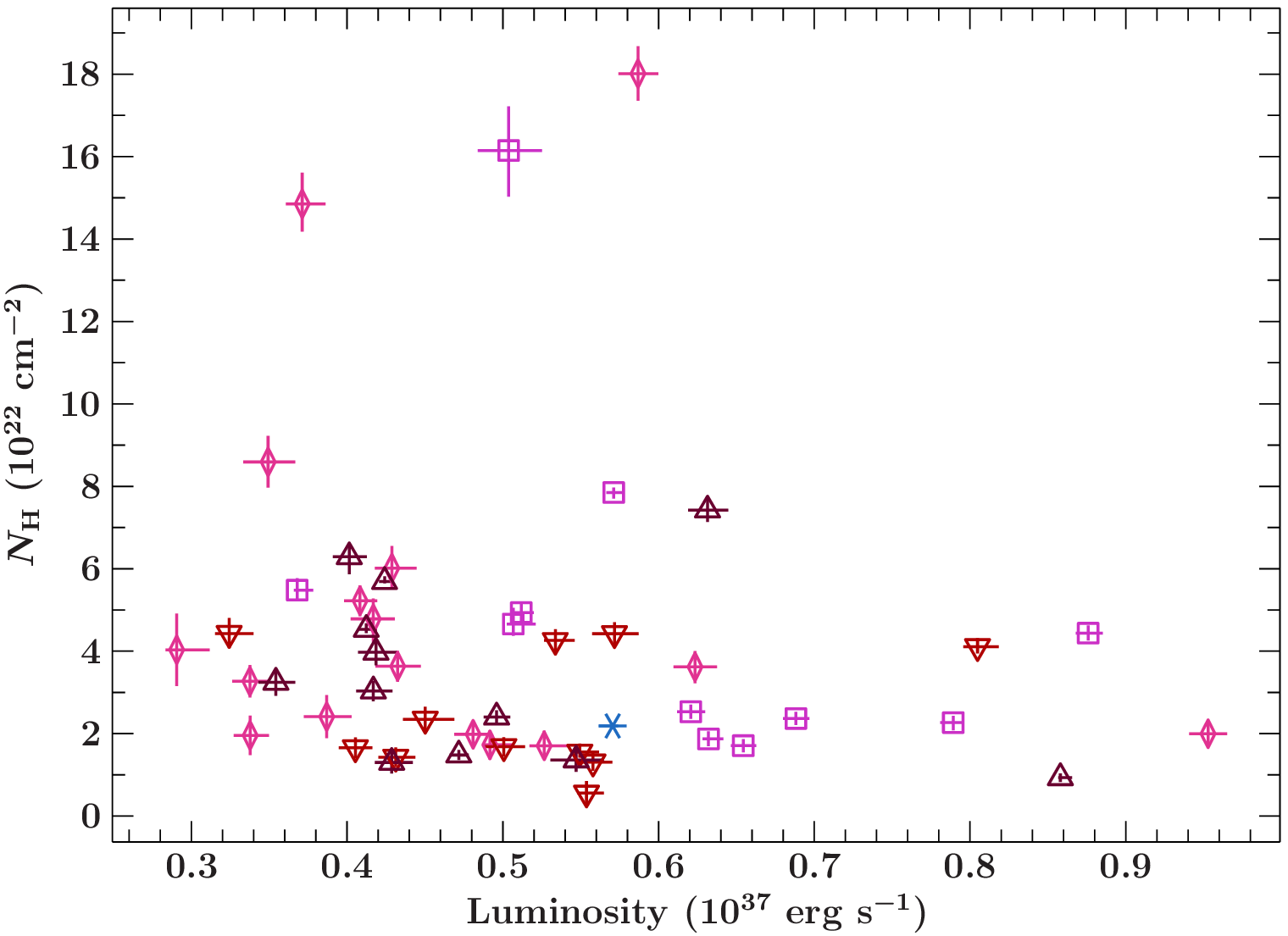}
  \includegraphics[width=\columnwidth]{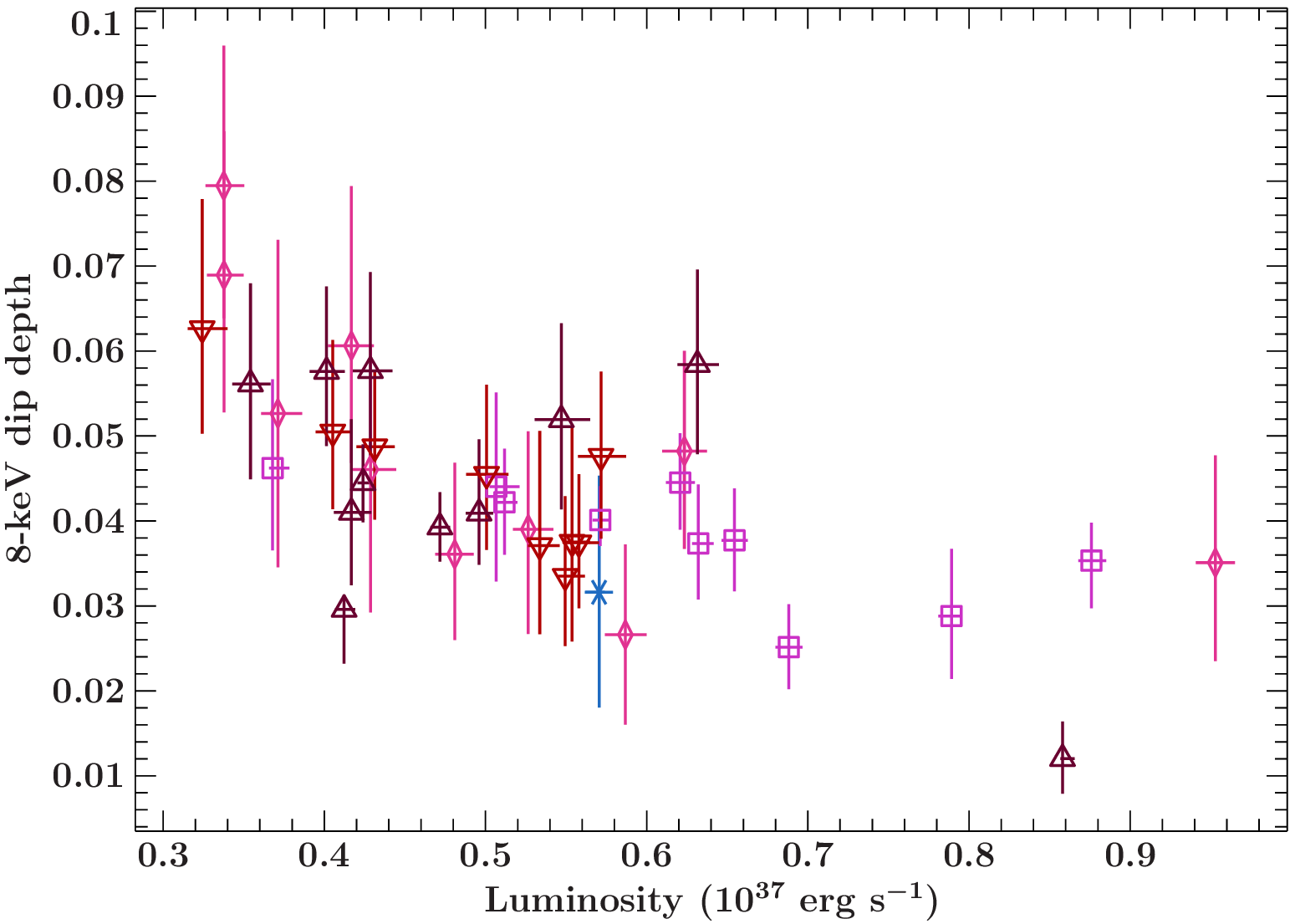}
  \caption{Clockwise from upper left: power-law index $\Gamma$,
 iron line flux, 8-keV feature optical depth, and absorbing column
 density \nh from the obsid-by-obsid dataset using the \texttt{mplcut}
 continuum, plotted against luminosity. \rxte measurements are in
 dark red triangles, red inverted triangles, violet squares, and
 pink diamonds for proposals P10145, P20146, P50067, and P80016,
 respectively, while the \suz measurement is plotted as a blue star.}
  \label{fig:plndx_fe_nh_obsid}
\end{figure*}

We plot the iron line flux, photon index, absorbing column density, and 8-keV
feature depth against luminosity (calculated from the unabsorbed 5--50\,keV
flux) in Fig.~\ref{fig:plndx_fe_nh_obsid}. The iron line flux shows a clear
positive correlation with luminosity, with a slope of $(0.93 \pm 0.03)\times
10^{-3}$\,ph\,cm$^{-2}$\,s$^{-1}$ per $10^{37}$\,erg\,s$^{-1}$ in the
obsid-by-obsid dataset, while the photon index $\Gamma$ displays a negative
correlation with luminosity, with a linear fit finding a slope of $-0.262 \pm
0.006$ in the obsid-by-obsid dataset. In both of these cases, the fitted slope
in the pulse-by-pulse dataset is highly consistent with the obsid-by-obsid
result; the positive correlation between the iron line flux and broad-band flux
is not seen in the pulse-phase-constrained results, but the iron line is near
its weakest flux in the pulse peak \citep{hemphill14} and the secondary peak
phase bin has low flux overall, making the iron line difficult to constrain in
PCA data in both phase-constrained datasets. Fits using the \texttt{fdcut}
continuum find statistically similar results for these parameters.

\begin{figure}
  \centering
  \includegraphics[width=\columnwidth]{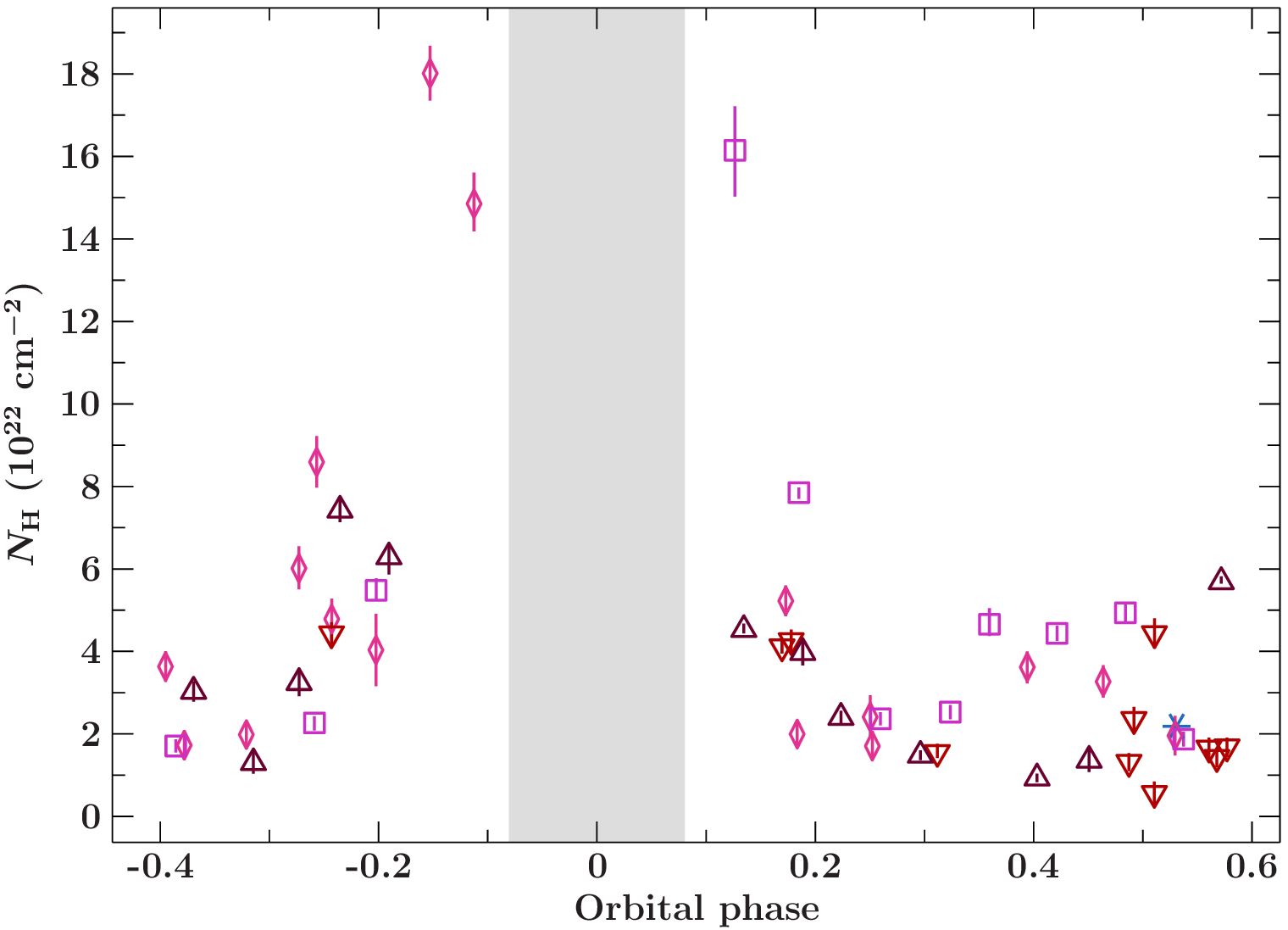}
  \caption{The absorbing column density towards \fif plotted against orbital
    phase, using the orbital parameters from Table \ref{tab:orb}. The
    approximate extent of the X-ray eclipse is shaded; phase $0.0$ is
    the eclipse center. Symbols and colors are as in
		Fig.~\ref{fig:plndx_fe_nh_obsid}.}
  \label{fig:nh}
\end{figure}

The absorbing column density \nh varies over the binary orbit, typically
sitting between $1$ and $5 \times 10^{22}$\,cm$^{-2}$ at most orbital
phases, but rising dramatically to $\sim\!10^{23}$\,cm$^{-2}$ near
eclipse. We plot \nh versus orbital phase, using results from the
obsid-by-obsid dataset, in Fig.~\ref{fig:nh}. While the increase
in \nh near eclipse is primarily a line-of-sight effect, the high
variability seen while the source is out of eclipse indicates highly
variable local absorption, consistent with a clumpy or otherwise
inhomogenous stellar wind. Our observed variability in \nh is
largely consistent with \rxte/PCA and \textit{BeppoSAX} data analyzed
by \citet{mukherjee_orbital_2006} and with \textit{MAXI} results
from \citet{rodes-roca_maxi_2015}. Qualitatively, we see somewhat
higher variability over the orbit compared to the \rxte/PCA results
of \citet{mukherjee_orbital_2006}, although this may be due to the different absorption model and abundances that we use compared to that analysis,
in addition to our considerably larger energy coverage. \nh also
appears to show some anticorrelation with the unabsorbed flux (see
Fig.~\ref{fig:plndx_fe_nh_obsid}), but this correlation is statistically
insignificant --- taking \nh measurements from obsids between orbital
phases $0.2$ and $0.7$ (i.e., ignoring eclipse ingress and egress),
Pearson's $r$ for these two parameters is $-0.12$, for a $p$-value of
$0.52$. The lowest observed \nh is $\sim\!0.6 \times 10^{22}$\,cm$^{-2}$
and most obsids find \nh above $1\times 10^{22}$\,cm$^{-2}$,
consistent with the approximate line-of-sight Galactic \nh
\citep{dickey_lockman_nh_1990,lab_nh_survey_2005,willingale_nh_2013}
\footnote{As calculated by
\url{https://heasarc.gsfc.nasa.gov/cgi-bin/Tools/w3nh/w3nh.pl}}. This
would point towards the \textit{local} absorption dropping to effectively
undetectable levels at times --- e.g., at those points we are viewing
the neutron star through a ``hole'' in the surrounding medium.

The energy of the 8-keV dip feature is uncorrelated with the source flux, with
an average energy of $7.6 \pm 1.6$\,keV in the pulse-by-pulse dataset. However,
the feature's depth is clearly inversely correlated with luminosity
(Fig.~\ref{fig:plndx_fe_nh_obsid}), trending down by $-0.057 \pm 0.015$ per
$10^{37}$\,erg\,s$^{-1}$. This decrease is in line with the fact that the
brighter spectra tend to be less likely to need the feature at all, although
with no physical explanation for the feature in the first place, the reason for
the trend is unclear.

Across all \rxte obsids and the \integral and \suz measurements, the flux
varies by a factor of $\sim\!3$, with most of the observations lying between 6
and 14 $\times 10^{-10}$\,erg\,cm$^{-2}$\,s$^{-1}$ (assuming a distance of
$6.4$\,kpc, this translates to a luminosity between 3 and $7 \times
10^{36}$\,erg\,s$^{-1}$), reaching a peak luminosity of $\sim\!1.2 \times
10^{37}$\,erg\,s$^{-1}$ during flares.

\subsection{CRSF variability with luminosity}
\label{ssec:e_vs_lum}

Previous analyses \citep{hemphill13,hemphill14} have found only tentative
evidence for correlations between \fif's CRSF energy and luminosity. With our
considerably larger dataset, we can report the best limits yet on any
correlation between these two parameters.

\begin{figure*}
	\centering
  \includegraphics[width=\columnwidth]{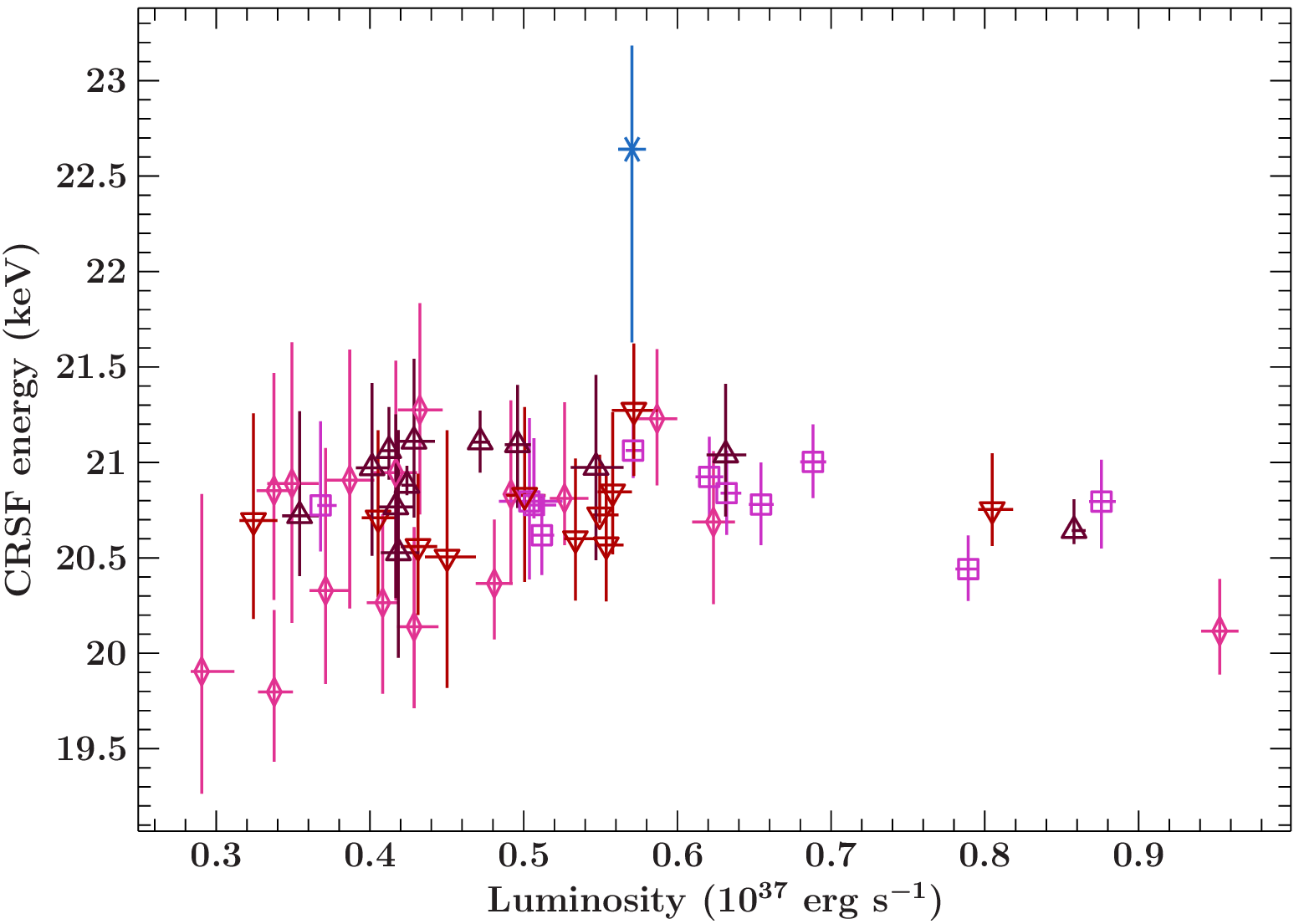}
  \includegraphics[width=\columnwidth]{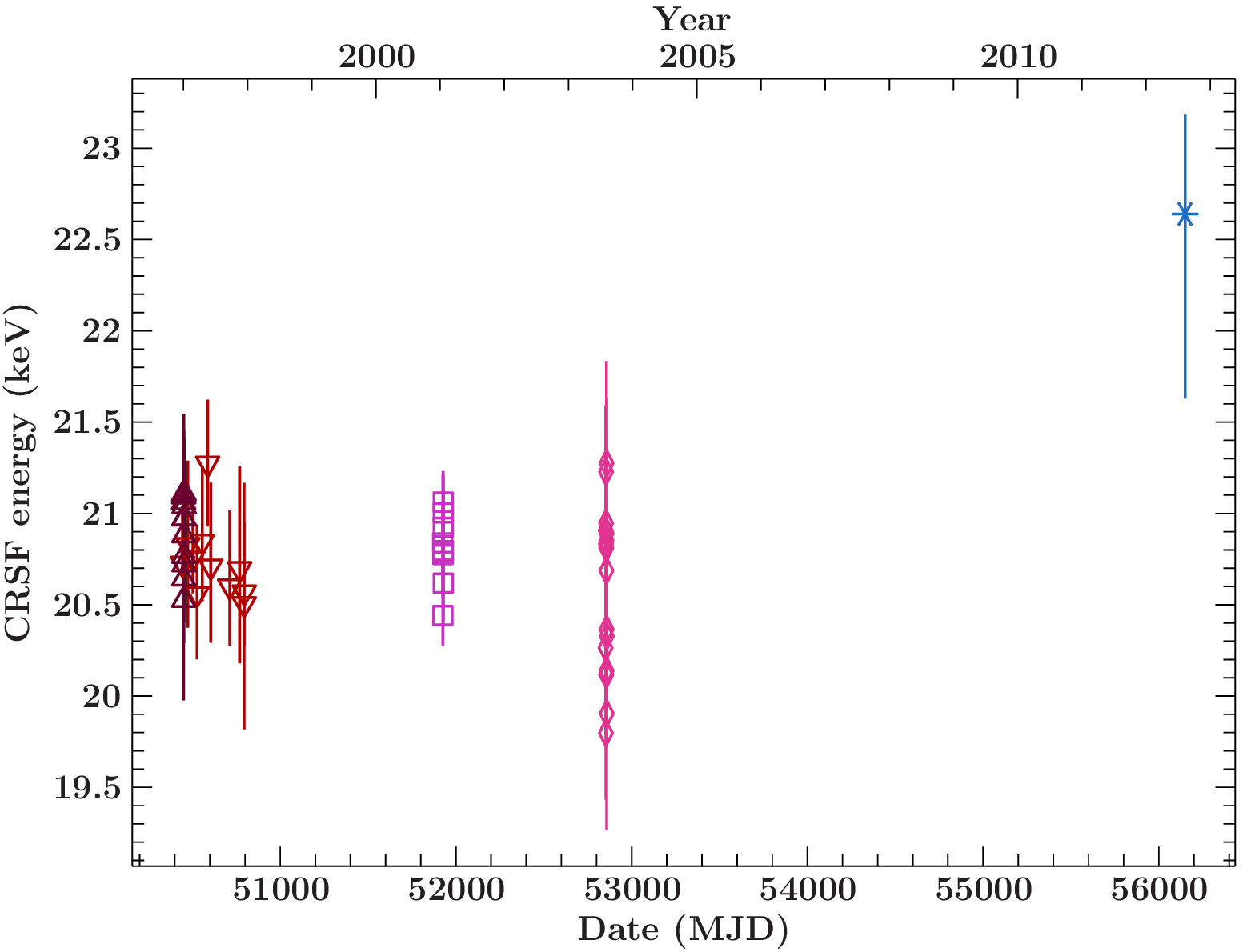} \\
  \includegraphics[width=\columnwidth]{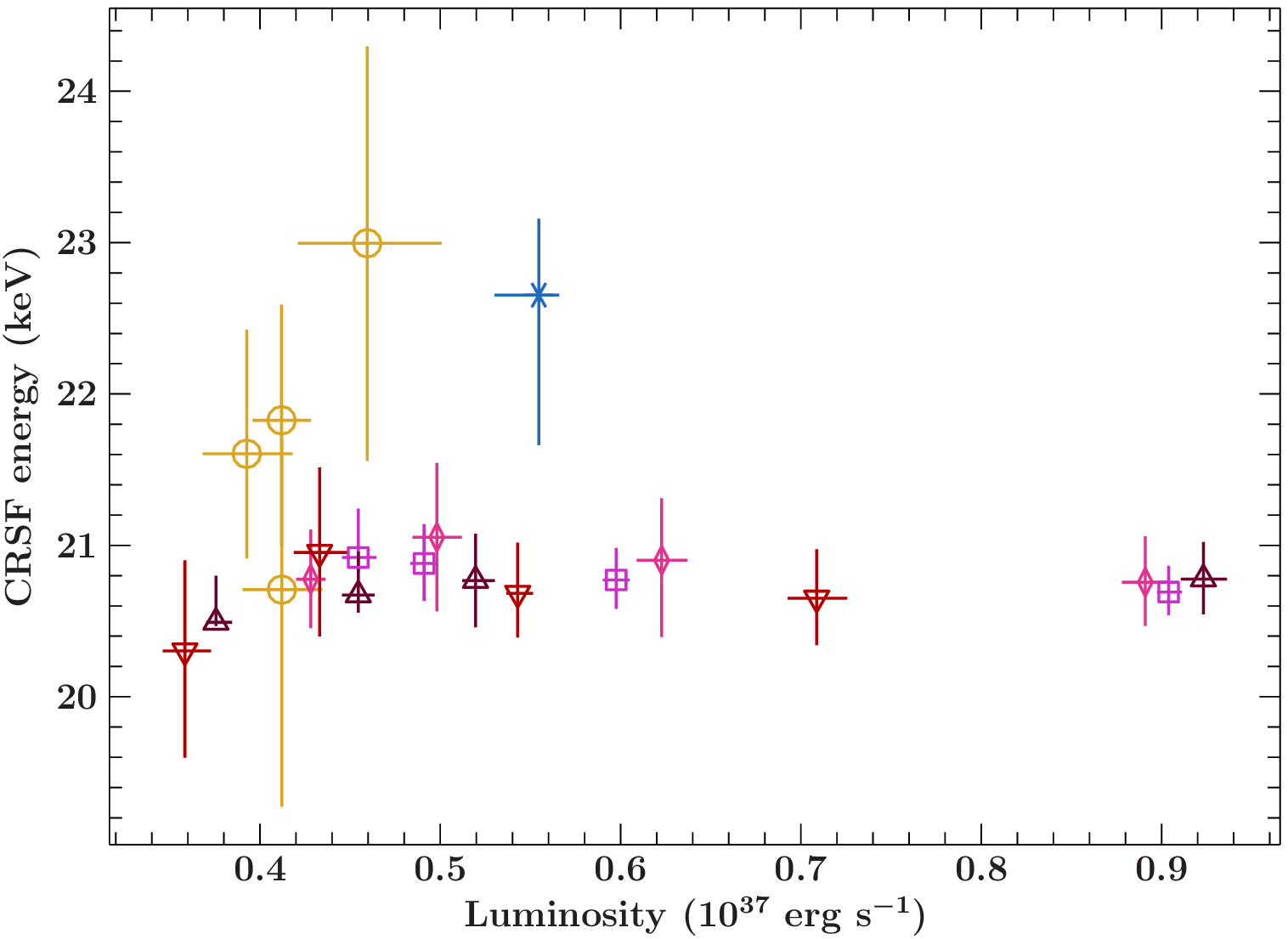}
  \includegraphics[width=\columnwidth]{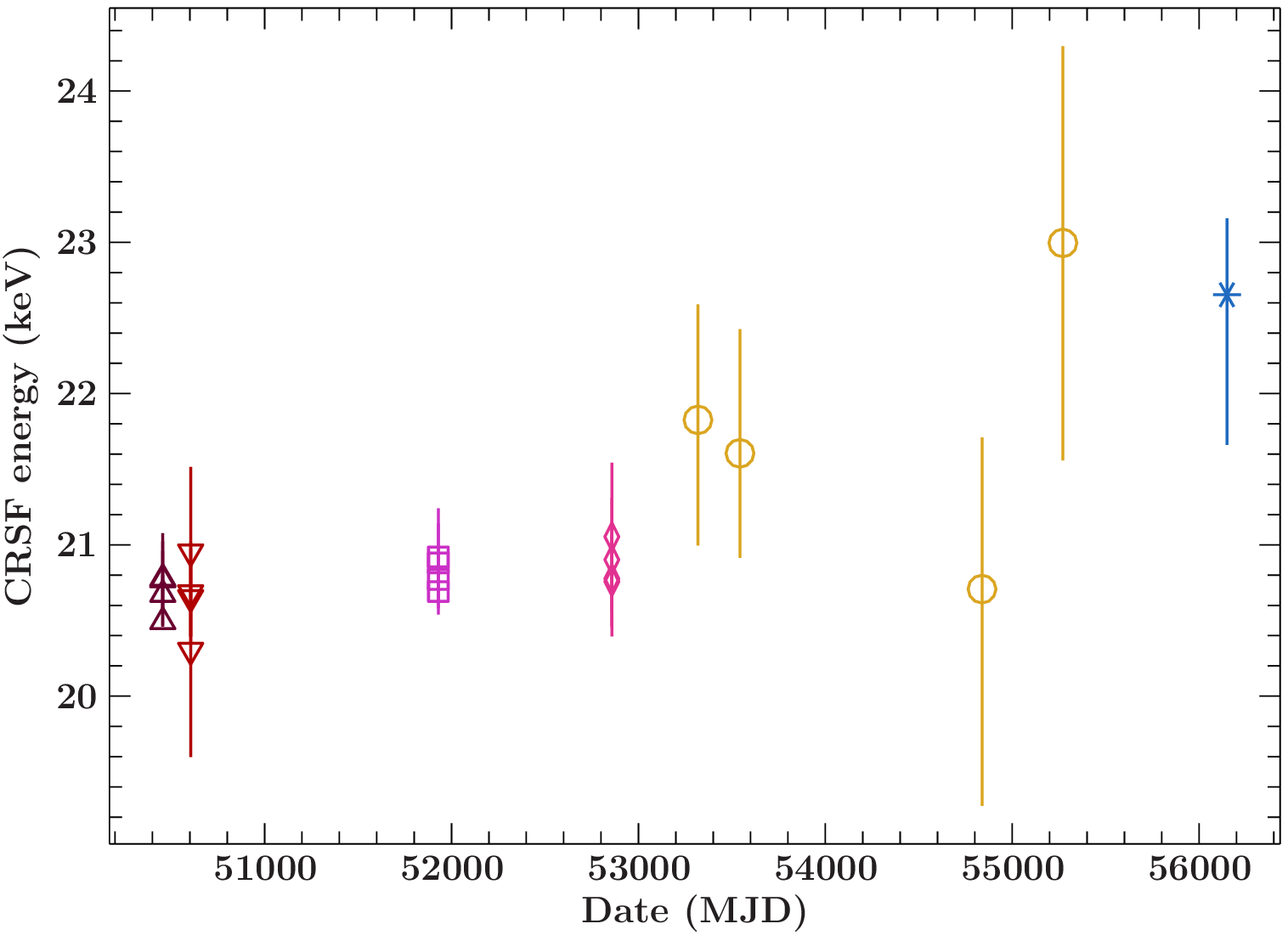} \\
 \caption{\fif's CRSF energy using the \texttt{mplcut} continuum and
	phase-averaged data. Upper plots display the results from the individual
	obsids, while the lower panels display the pulse-by-pulse results.
	Left-hand plots have the CRSF energy plotted against 3--50\,keV flux,
	and right-hand plots show the energy plotted against time. Symbols and
	colors are as in Fig.~\ref{fig:plndx_fe_nh_obsid}, with \integral
	results plotted in gold circles.}
	\label{fig:ecyc_avg}
\end{figure*}

We plot the obsid-by-obsid and pulse-by-pulse CRSF energy versus the unabsorbed
3--50\,keV flux in the left-hand panels of Fig.~\ref{fig:ecyc_avg}. Linear fits
to the obsid-by-obsid and pulse-by-pulse \rxte datasets find slopes of $-0.48
\pm 0.13$ and $0.11 \pm 0.19$\,keV per $10^{37}$\,erg\,s$^{-1}$, respectively.
No trend is visible in the \rxte peak-pulse and secondary-pulse
phase-constrained results (with the exception of the significantly higher-energy
\suz measurement), consistent with what is seen in the pulse-by-pulse results.

The obsid-by-obsid results have an intriguing hint of a change in
slope at a luminosity of $\sim\!6\times 10^{36}$\,erg\,s$^{-1}$, with
a slope of $+1.0 \pm 0.4$\,keV per $10^{37}$\,erg\,s$^{-1}$ below this
luminosity and $-1.48 \pm 0.35$\,keV per $10^{37}$\,erg\,s$^{-1}$
for higher luminosities. This is especially interesting given the
observed bimodality of $E_{\mathrm{cyc}}$-luminosity slopes across all CRSF
sources, with lower-luminosity sources showing positive slopes and
higher-luminosity sources showing negative slopes. However, the obsids
each average over a range of luminosities; the overall flatness of the
pulse-by-pulse measurements may be a more authentic representation
of the source's true behavior. Breaking the pulse-by-pulse CRSF
measurements at a luminosity of $6\times 10^{36}$\,erg\,s$^{-1}$ as
we did for the obsid-by-obsid results, the slopes are $-0.1 \pm 0.8$
and $+1.20 \pm 0.65$\,keV per $10^{-37}$\,erg\,s$^{-1}$ for high and
low luminosities, respectively. The upward trend at lower luminosities
for this dataset is only significant at the $\sim\!2\sigma$ level and
is entirely due to the two lowest-luminosity measurements; while this
suggests a break, similar to that seen in the obsid-by-obsid dataset, at
$\sim 4 \times 10^{36}$\,erg\,s$^{-1}$, the lack of measurements at very
low luminosity makes this trend highly suspect.

\subsection{CRSF variability with time}
\label{ssec:e_vs_t}

\begin{figure*}
	\centering
	\includegraphics[width=\columnwidth]{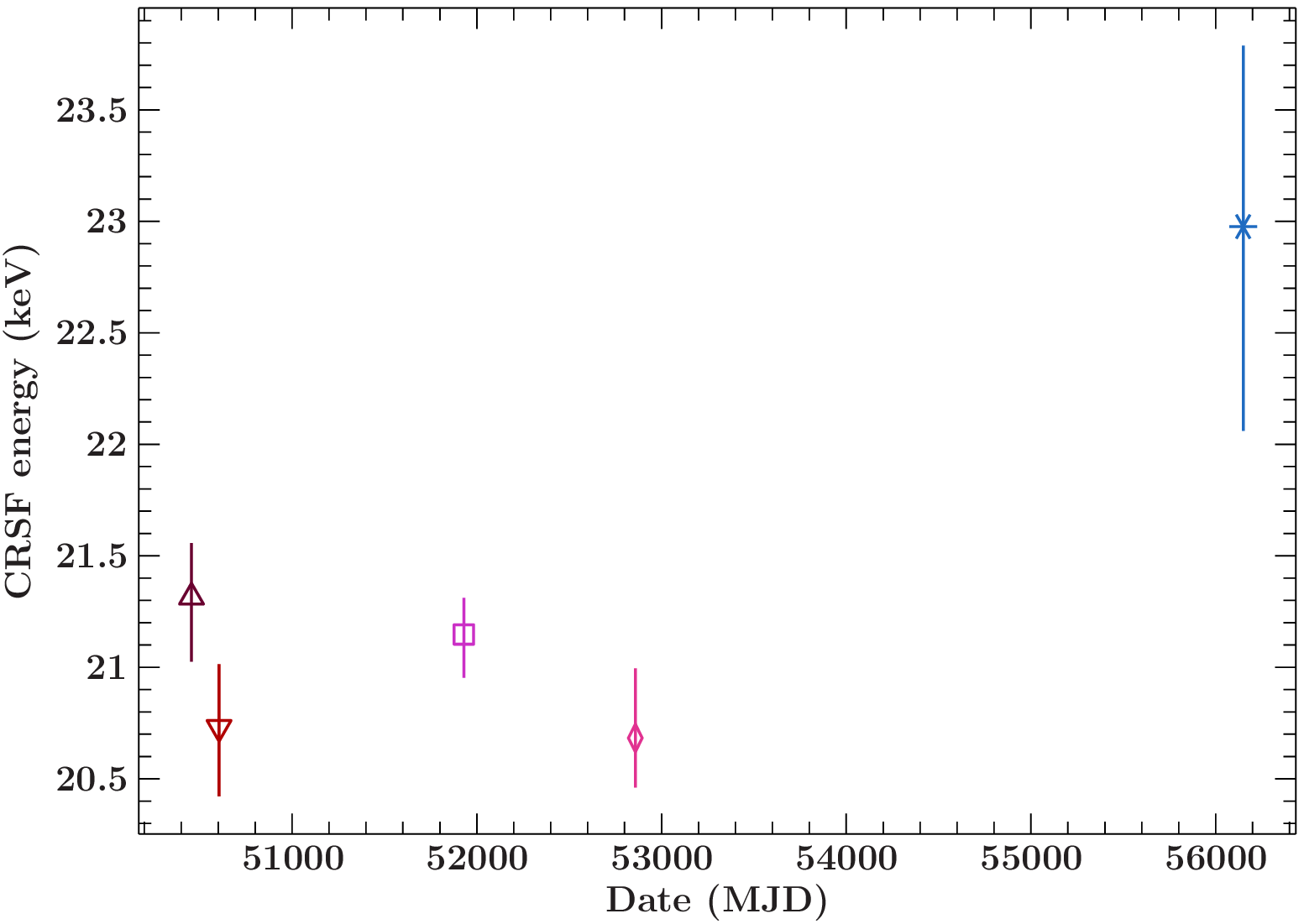}
	\includegraphics[width=\columnwidth]{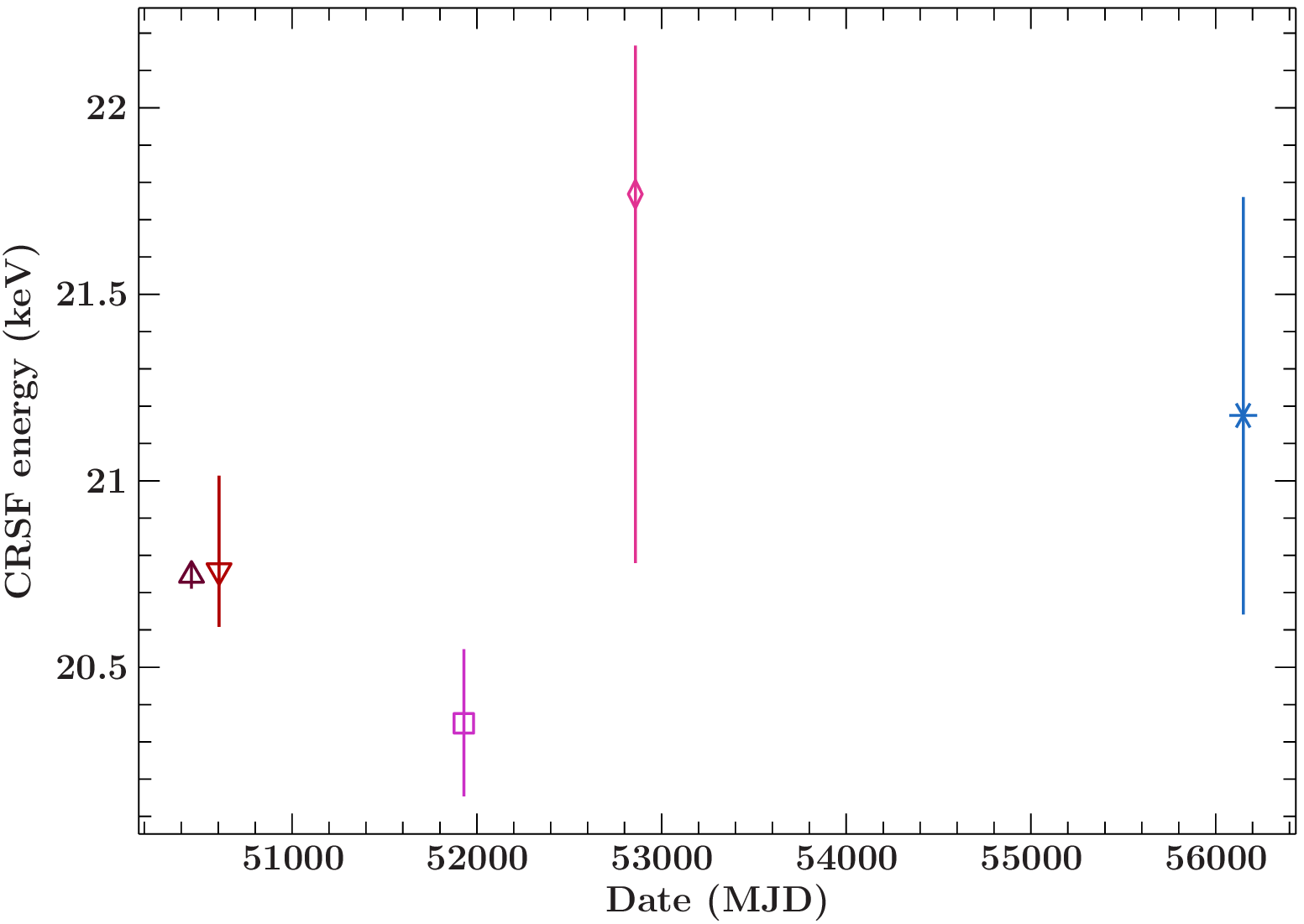}
 \caption{\fif's CRSF energy from the phase-resolved datasets, both
	plotted against time. On the left, the measurements from the peak of the
	main pulse; on the right are the measurements from the secondary pulse.
	The large increase in CRSF energy seen in the main-pulse spectrum is not
	apparent in the spectrum of the secondary pulse.}
	\label{fig:ecyc_phase}
\end{figure*}

\begin{figure}
	\centering
	\includegraphics[width=\columnwidth]{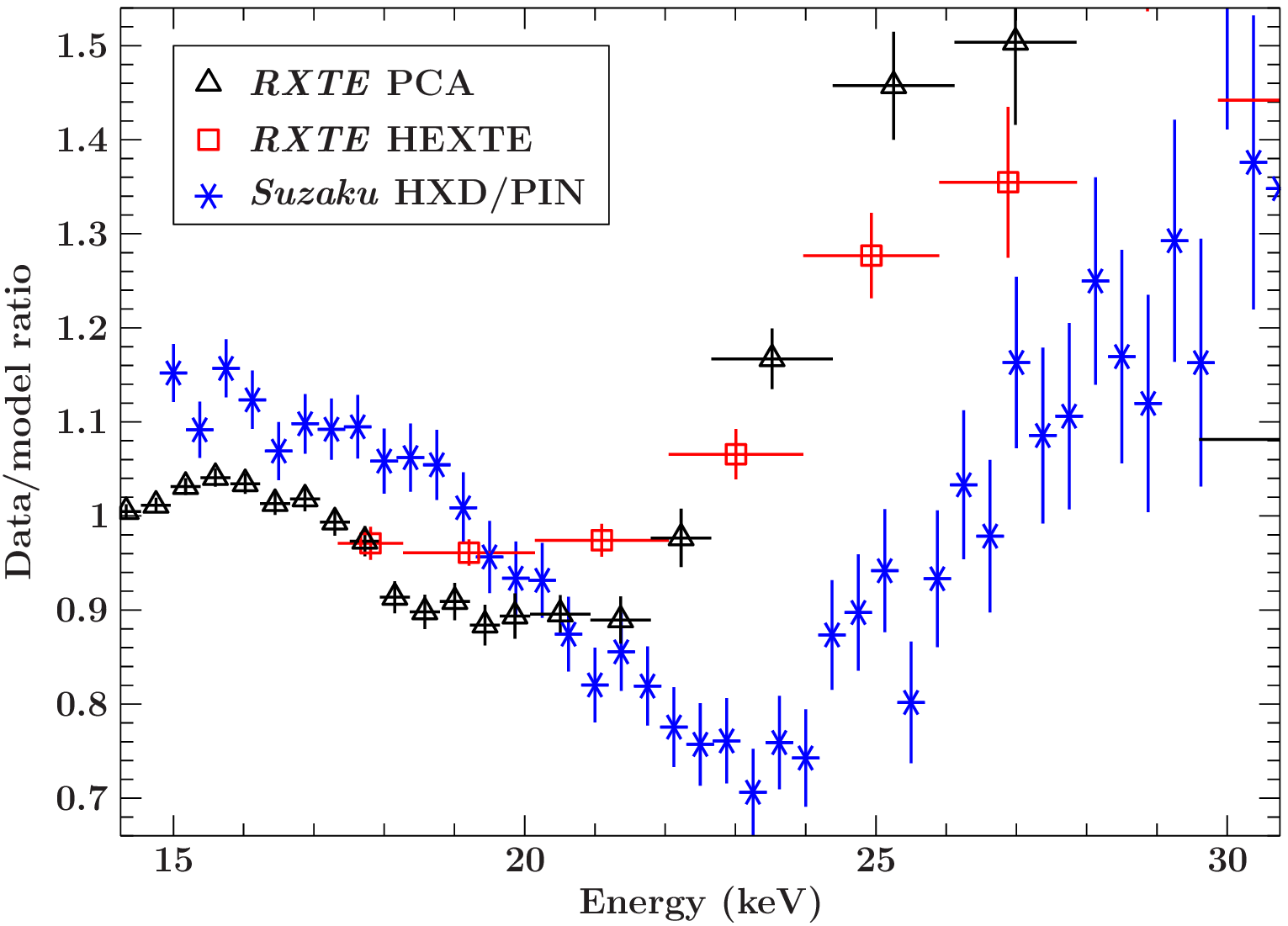}
 \caption{The ratio residuals for a fit with no CRSF for
 the 140-171 counts/s flux bin \rxte proposal P50067 and the \suz
 observation (i.e., the same data as in Fig.~\ref{fig:spectra}) in the
 15--30\,keV band. The PCA data are shown in black triangles, the HEXTE
 in red squares, and the PIN in blue stars. The CRSF energy in the \suz
 observation is clearly higher than in the \rxte data.}
	\label{fig:crsf_zoom}
\end{figure}

In the phase-averaged and main-pulse spectra, there is a jump of approximately
$1.5$\,keV in the energy of the cyclotron line between the \rxte and \suz
measurements, as can be seen in the right-hand panels of Fig.~\ref{fig:ecyc_avg}
and the left-hand panel of Fig.~\ref{fig:ecyc_phase}, where we plot the results
from the pulse-phase-constrained datasets. This shift is seen in both the
\texttt{mplcut} and \texttt{fdcut} results. However, as can be seen in the right
panel of Fig.~\ref{fig:ecyc_phase}, there is no significant increase in the CRSF
energy in the phase-constrained spectra of the secondary pulse; while the
poorly-constrained \suz results for that phase bin are at least partially to
blame here, there are some arguments that this may be a real effect.
In Fig.~\ref{fig:crsf_zoom}, we plot a closer look at the
data-to-model ratio residuals in the energy band around the CRSF when
the CRSF is excluded from the model, using the same data as plotted in
Fig.~\ref{fig:spectra}. The dip due to the CRSF in the \suz PIN data is
visibly higher-energy compared to the \rxte data.

While the \integral results help to fill in the gap between the \rxte
and \suz results, their high uncertainties limit their usefulness
for this analysis. The \rxte results taken on their own do not show
any consistent trend with time; linear fits to the obsid-by-obsid
and pulse-by-pulse results disagree, finding downward ($-0.033 \pm
0.009$\,keV\,yr$^{-1}$) and upward ($+0.034 \pm 0.015$\,keV\,yr$^{-1}$)
trends, respectively.


To better quantify the statistical significance of the \suz measurement,
and to avoid making any assumptions about the underlying distribution
of CRSF energy measurements, we took a Monte Carlo approach. We constructed
simulated obsid-by-obsid datasets by varying each measured CRSF energy
randomly according to its 1$\sigma$ uncertainty. Each trial in this
manner thus produced a set of 51 simulated CRSF energy measurements (50
\rxte points and one \suz point); we then computed the significance of
the simulated \suz point compared to the mean and standard deviation of
the simulated \rxte points. Performing $10^{6}$ trials in this manner,
the \suz point was found on average \sig above the \rxte points, with a
spread of \sigerr.

To examine the possibility of systematic, instrumental differences in
CRSF measurements between \rxte and \suz, either due to differences
in energy calibration or due to the difference in energy coverage
between the two satellites, we simulated \suz spectra for each of the
obsid-by-obsid \rxte spectral models using the \texttt{fakeit} procedure
in \textit{ISIS}. The CRSF energy as measured in these simulated spectra
was generally within the uncertainties of the \rxte measurements,
finding a mean energy of $20.6$\,keV with a standard deviation of
$0.5$\,keV across all 50 spectra, compared to $20.7 \pm 0.3$\,keV in the
real \rxte data --- i.e., \suz is not systematically likely to measure
higher CRSF energies compared to \rxte, given the same underlying
spectrum. However, this simulated approach does not account
for any additional instrumental systematics --- these results mainly
tell us that the more limited energy range covered by \suz does not
contribute to overall changes in the measured CRSF energy.

Our reported values use HXD/PIN data with a lower limit of 15\,keV and
 XIS data with an upper limit of 12\,keV. While these instruments are
usually considered to be reliable in these ranges, their calibration
is more poorly constrained as one approaches the edges of their energy
bounds. We thus checked our results by fitting the phase-averaged \suz
spectra with the PIN limited to above 18\,keV and the XIS limited to
below 10\,keV. Using these limits, we still find results consistent
with our reported values. However, the PIN normalization and cutoff
energy are considerably less well-constrained, at $1.2^{+0.3}_{-0.4}$
and $17.8^{+5.1}_{-0.7}$\,keV, respectively.

The real \suz spectrum is still fitted fairly well when the CRSF
energy is frozen to a typical \rxte value --- fixing the CRSF energy to
$20.7$\,keV results in a reduced $\chi^{2}$ value of $1.29$ for 522
degrees of freedom, compared to $1.25$ for 521 DOF when the CRSF is left
free to vary, with a reduction in $\chi^{2}$ of $19.3$. This fit also
brings the cutoff energy $E_{\mathrm{cut}}$ more into line with the measured
values found using \rxte. However, the F-test probability for this
improvement to arise by chance is $3.1 \times 10^{-5}$.

It should be noted that in the phase-averaged \suz spectra, the cutoff energy is
very high compared to \rxte and relatively close to the CRSF energy ($18.6$\,keV
for $E_{\mathrm{cut}}$ vs $22.7$\,keV for $E_{\mathrm{cyc}}$). The depth of the
``smoothing'' Gaussian added to compensate for the piecewise nature of the
\texttt{plcut} continuum is also quite high compared to the \rxte values. This
is likely in part due to the energy gap between the XIS and HXD/PIN spectra ---
if the smoothing Gaussian is omitted entirely, the cutoff energy and CRSF energy
are both found at nearly the same energy, at $\sim\!21$--$22$\,keV.

\begin{figure}
	\centering
  \includegraphics[width=\columnwidth]{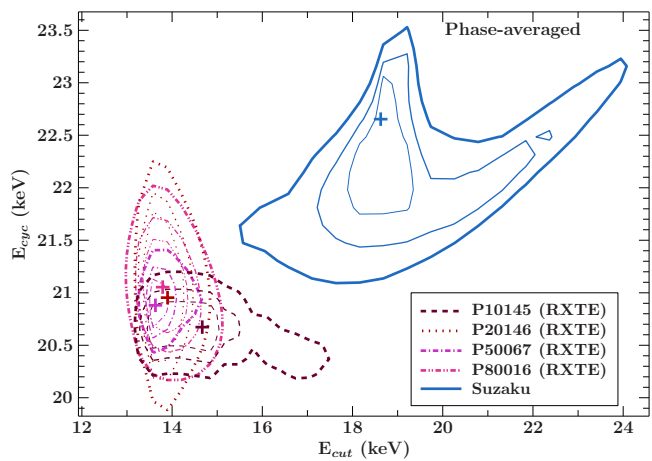}
	\caption{$E_{\mathrm{cut}}$ vs $E_{\mathrm{cyc}}$ contours for \suz and the
 123--140 counts/s \rxte results, using phase-averaged spectra. From the
 inside out, the contours represent the 68\%, 90\%, and 99\% confidence
 intervals for two parameters. See text for details regarding
	the interpretation of the \suz contours.}
 \label{fig:ecut_ecyc_contour_avg}
\end{figure}

There is no detectable correlation between $E_{\mathrm{cut}}$ and
$E_{\mathrm{cyc}}$ in the combined \rxte, \integral, and \suz datasets.
However, Fig.~\ref{fig:ecut_ecyc_contour_avg} shows confidence
contours for $E_{\mathrm{cyc}}$ vs $E_{\mathrm{cut}}$ for \suz and the
123--140\,counts\,s$^{-1}$ \rxte data, which is closest in flux to the
\suz measurements; while the \rxte results are highly inconsistent with
\suz, the \suz contours are strangely shaped, with a noticeable
correlation between $E_{\rm cyc}$ and $E_{\rm cut}$ above $E_{\rm cut}
\approx 20$\,keV. This correlation is likely artificial, due to the
piecewise continuum; the fact that the best-fit value lies in the
``spike'' above the correlated region suggests that the measured value
can be trusted.

We can get some additional insight as to the difference between the
\suz and \rxte spectra by examining different models and different
datasets. Fitting the pulse-by-pulse and obsid-by-obsid datasets
with the \texttt{fdcut} continuum still finds a significantly
higher $E_{\mathrm{cyc}}$, at $22.8^{+0.5}_{-0.3}$, compared to
$\sim\!21.2$\,keV in \rxte. However, the cutoff energy found with \suz
using \texttt{fdcut} is still significantly higher than the \rxte
measurement ($26.5 \pm 1.2$\,keV compared to $\sim\!22$\,keV in \rxte),
and the confidence contours in both \rxte and \suz are not as well
constrained as with the \texttt{mplcut} continuum.

\begin{figure}
	\centering
  \includegraphics[width=\columnwidth]{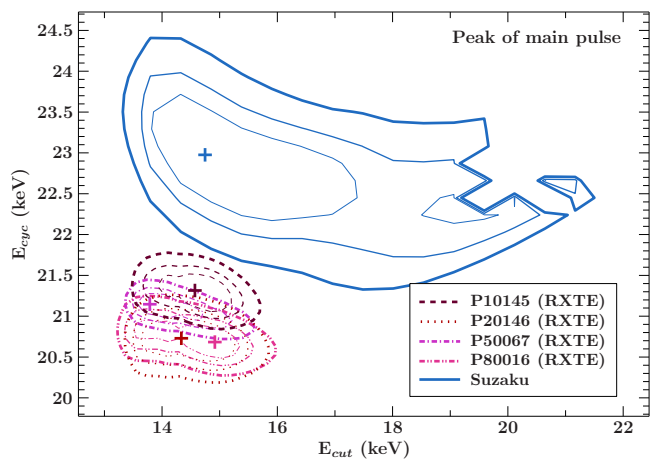}
	\caption{$E_{\mathrm{cut}}$ vs $E_{\mathrm{cyc}}$ contours for \suz and \rxte
	spectra of the peak of the main pulse. Confidence intervals are same as
	in Fig.~\ref{fig:ecut_ecyc_contour_avg}.}
 \label{fig:ecut_ecyc_contour_peak}
\end{figure}

The most interesting result is that of the phase-constrained
\suz spectrum of the peak of the main pulse, which does not suffer
from the same issues as the phase-averaged spectrum does ---
using \texttt{mplcut}, its spectral parameters including
$E_{\mathrm{cut}}$ are entirely consistent with the \rxte values, with
the exception of its CRSF energy, which at $23.0^{+0.9}_{-0.8}$
is significantly higher than the $\sim\!20.9$ found in the \rxte
spectra. This can be seen clearly in the left-hand panel of
Fig.~\ref{fig:ecyc_phase}. Fig.~\ref{fig:ecut_ecyc_contour_peak} shows
the $E_{\mathrm{cut}}$-$E_{\mathrm{cyc}}$ contours for the peak-pulse
spectra; in this case \suz clearly finds a higher-energy CRSF compared
to \rxte. There is some bimodality in the contours with a very slightly
lower-energy CRSF at a significantly higher cutoff energy, but this
is still a significantly higher-energy CRSF compared to \rxte. This
raises the possibility that the increased CRSF energy seen in the
phase-averaged data is due primarily to an increase in the CRSF energy
in this phase bin. Indeed, the spectra of the \textit{secondary} pulse
do not show any evolution of the CRSF in time (see the right-hand
panel of Fig.~\ref{fig:ecyc_phase}); this could be due to a change in
accretion geometry limited to a single pole. However, it should
be noted that the \suz spectrum of the secondary peak has a very
poorly-constrained cutoff energy.

\section{Discussion}
\label{sec:discussion}

\subsection{Relationship between $E_{\mathrm{cyc}}$ and flux}
\label{ssec:disc_ecyc_flux}

There are two generally recognized regimes for neutron star accretion,
originally laid out by \citet{basko_sunyaev_1976}, defined by the
critical luminosity $L_{\mathrm{crit}}$. When accretion is supercritical,
radiation pressure is capable of stopping the infalling material
entirely; in this case, the accreted material sinks relatively slowly
from a radiation-dominated shock to the surface, and the observed
X-rays are mainly photons that escape through the sides of the column
(``fan-beam'' emission), as the optical depth through the top of
the column is large. Under subcritical accretion, for luminosities
significantly below $L_{\mathrm{crit}}$, the infalling material impacts
on and heats the surface \citep{zeldovich_shakura_1969}, producing
an accretion mound and a hot spot at the magnetic pole; here, the
observed X-rays can escape from the top of the column (``pencil-beam''
emission) due to the lower optical depth of the weaker accretion
stream. In the transition region between these two modes, when the
accretion is still subcritical but close to the critical point, the
infalling material is likely slowed by
a combination of radiation pressure and gas pressure, and the emission
is a hybrid of the above two modes.

\citet{becker_spectral_2012} and \citet{poutanen_reflection_2013}
present two very different mechanisms for CRSF production and
variability. \citet{becker_spectral_2012} has the CRSF being produced
directly in the accretion channel, and the observed correlations
derive from the line-producing region moving upwards (when accretion
is supercritical) or downwards (for moderately-subcritical accretion)
in the column in response to increases in accretion rate. In contrast,
the CRSF in the \citet{poutanen_reflection_2013} model is produced
when light from the accretion column, which is preferentially beamed
downwards due to relativistic effects, reflects off the stellar surface,
with the CRSF energy determined by the surface magnetic field strength
of the neutron star. Under supercritical accretion, the height of the
accretion column is proportional to the accretion rate, and thus the
observed $E_{\mathrm{cyc}}$-luminosity anticorrelation results the taller
column illuminating a larger fraction of the NS surface, sampling a
lower average magnetic field (as the surface magnetic field strength
drops as one moves away from the magnetic pole). Both models make
similar qualitative predictions for the supercritical case, but the
reflection model inherently predicts smaller variations in $E_{\mathrm{cyc}}$, as the magnetic field only varies by a factor of $\sim\!2$ over
the NS surface (compared to $B \propto r^{-3}$ in the higher reaches
of the column). While \citet{poutanen_reflection_2013} do not directly
address the subcritical-accretion case in their work, highly-subcritical
sources will emit their X-rays from a hot spot on the surface, and the
reflection mechanism is not likely to produce strong variability in
these cases. Moderately-subcritical sources will still have
something approximating an accretion column; for these cases,
\citet{mushtukov_doppler_2015} point out that the velocity
distribution of infalling electrons will change as a source goes from
highly-subcritical to moderately-subcritical, with the electron velocity
at the base of the accretion channel reaching zero when the source
reaches critical luminosity. The change in the distribution of Doppler
shifts between photons and electrons as the source approaches $L_{\mathrm{crit}}$ then results in the observed positive trend between $E_{\mathrm{cyc}}$
and luminosity.

Assuming the NS mass and radius are $1$\,\msol \citep{rawls_mass_2011}
and $10$\,km, respectively, and assuming spherical accretion, the
theoretical framework laid out by \citet{becker_spectral_2012} finds
that the effective Eddington luminosity, $L_{\mathrm{crit}}$, is $1.0 \times
10^{36}$\,erg\,s$^{-1}$, somewhat lower than \fif's typical luminosity
range of $3$--$10 \times 10^{36}$\,erg\,s$^{-1}$. One can adjust the
predicted $L_{\mathrm{crit}}$ up and down depending on the parameters one
chooses; however, there is no physically-reasonable set of parameters
that results in \fif accreting subcritically for more than a small
fraction of its observed luminosity range --- pushing $L_{\mathrm{crit}}$
to above $\sim\!5\times 10^{36}$\,erg\,s$^{-1}$ would require either
assuming disk accretion or assuming that the surface magnetic field
strength is significantly higher than CRSF energy predicts.

Thus, this implies that \fif accretes supercritically, and we should
thus expect a negative $E_{\mathrm{cyc}}$-luminosity correlation. This
can only be reconciled with observations if the predicted change in
$E_{\mathrm{cyc}}$ is small enough to be hidden in the available data,
which, with some dependence on the dataset one looks at, constrains
us to changes of $\sim\!0.1$\,keV per $10^{36}$\,erg\,s$^{-1}$.
\citet{becker_spectral_2012} find a linear relationship between
luminosity and the height of the CRSF-producing region for supercritical
accretion, thus predicting a change in height by a factor of $\sim\!3$
for \fif. The exact height depends on the value of the $\xi$ parameter
from \citet{becker_spectral_2012}, which characterizes the relationship
between the flow velocity of the accreted material and the effective
``velocity'' of photon diffusion upwards through the infalling material;
taking $\xi$ to be $\sim\!10^{-2}$ results in emission heights of
$\sim\!10$--$30$\,m above the NS surface. If one assumes a dipolar
magnetic field, this range of heights corresponds to a change in the
CRSF energy on the order of $10^{-2}$--$10^{-3}$\,keV, far smaller
than any observable trend. However, the magnetic field this close to
the NS surface may deviate significantly from a dipole \citep[see,
e.g.][]{mukherjee_crsf_2012}, so this prediction should be viewed with
some care.

More recent work by \citet{mushtukov_lcrit_2015} attempts to
additionally take into account resonant scattering and photon
polarization. Their work finds that wind-accreting sources generally
have \textit{higher} $L_{\mathrm{crit}}$ compared to disk-accreting sources,
due to the larger footprint of the accretion flow. While they do
not provide calculations for \fif's exact parameters, they do find
$L_{\mathrm{crit}} \approx 2$--$4\times 10^{36}$\,erg\,s$^{-1}$
for wind-accreting sources around \fif's CRSF energy, intriguingly
close to the $\sim\!6\times 10^{36}$\,erg\,s$^{-1}$ ``break'' in the
$E_{\mathrm{cyc}}$-luminosity relationship seen in the obsid-by-obsid
dataset. \citet{mushtukov_lcrit_2015} assume a neutron star mass
of $1.4$\,\msol, while \fif's neutron star is likely closer to
$\sim\!1$\,\msol. A lower mass would result in a lower velocity for
infalling material, decreasing $L_{\mathrm{crit}}$, but would also increase
the size of the hot spot on the neutron star surface, decreasing the
temperature of the hot spot and increasing $L_{\mathrm{crit}}$. The overall
effect here is difficult to judge, given the lack of a closed-form
solution using the framework of \citet{mushtukov_lcrit_2015}.

There are a number of systematic factors that can influence the
calculated luminosity, which must be taken into account if we are to
compare our results to theoretical predictions. The uncertainty in the
distance to \fif is $\sim\!1$\,kpc, corresponding to at most a factor of
$\sim\!2$ possible change in the measured luminosity; if the distance
is closer to \citet{clark_chandra1538_2004}'s estimate of $4.5$\,kpc,
our observed luminosity range is closer to $1.5$--$4.5 \times
10^{36}$\,erg\,s$^{-1}$. Relativistic lightbending and the non-isotropic
emission of the X-ray pulsar will also push the true luminosity down,
since our luminosity calculation assumes emission over $4\pi$\,sr; when
this beaming is taken into account, the true luminosity can be a factor
of $\sim\!2$ lower compared to the computed luminosity, although this
is highly dependent on the emission geometry \citetext{M. K\"{u}hnel,
private communication, 21 July 2015; Mart\'{i}nez-N\'{u}\~{n}ez et al.
2015, in preparation}. Overall, though, the lack of any detectable
trend, the weakness of the predicted correlations, and the possible
close proximity of the scattering region to the NS surface make it
prudent to say at this stage that \fif's accretion mode is still
uncertain.

\subsection{Change in $E_{\mathrm{cyc}}$ between \rxte and \suz}
\label{ssec:disc_ecyc_time}

The CRSF energy of \fif appears to have increased between the \rxte
measurements of 1996--2004 and the 2012 \suz measurement. This is
a peculiar occurrence; while it does not seem entirely attributable to
instrumental or modeling artifacts, it may be limited to the peak of the main
pulse.

\begin{figure}
  \centering
  \includegraphics[width=\columnwidth]{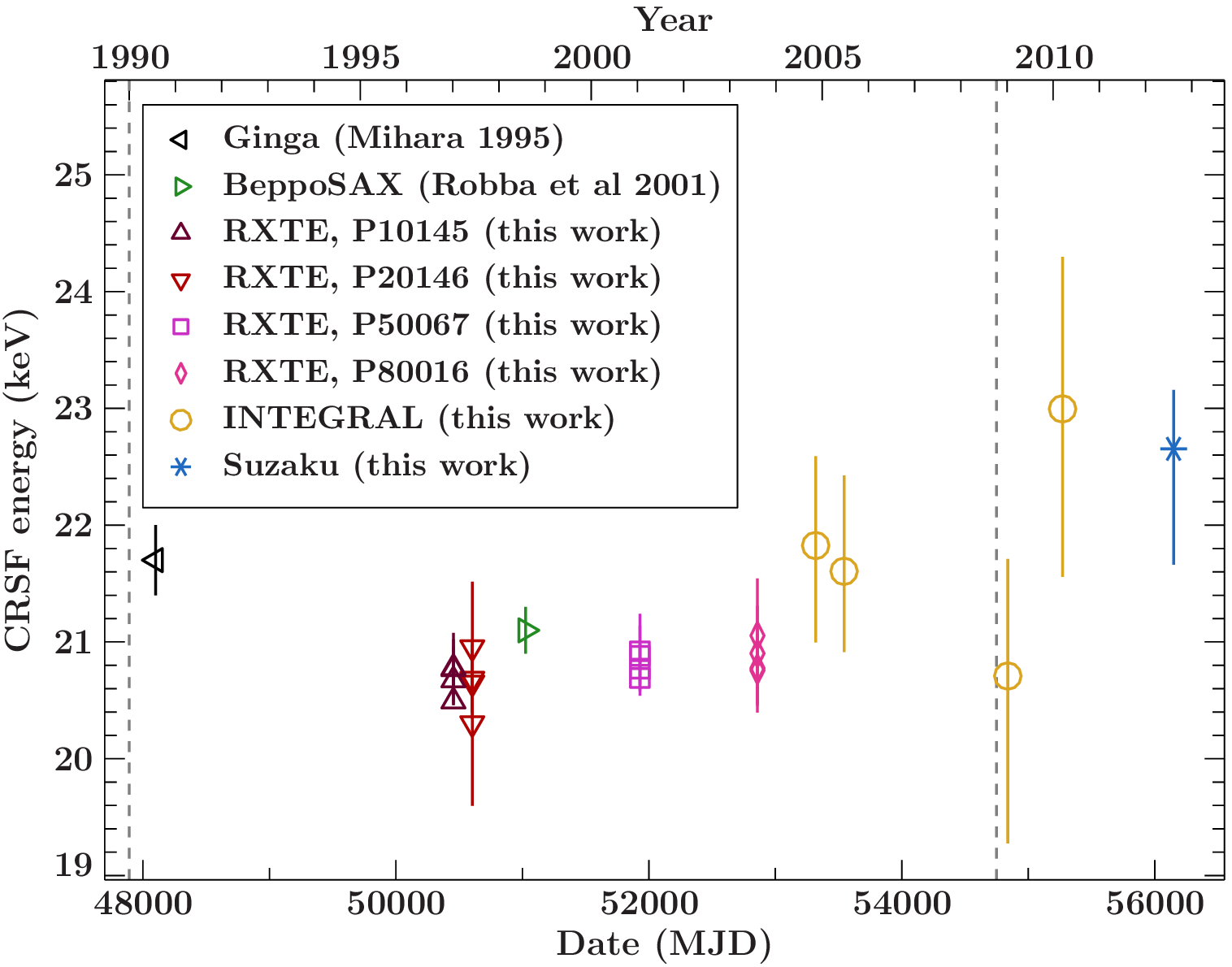}
  \includegraphics[width=\columnwidth]{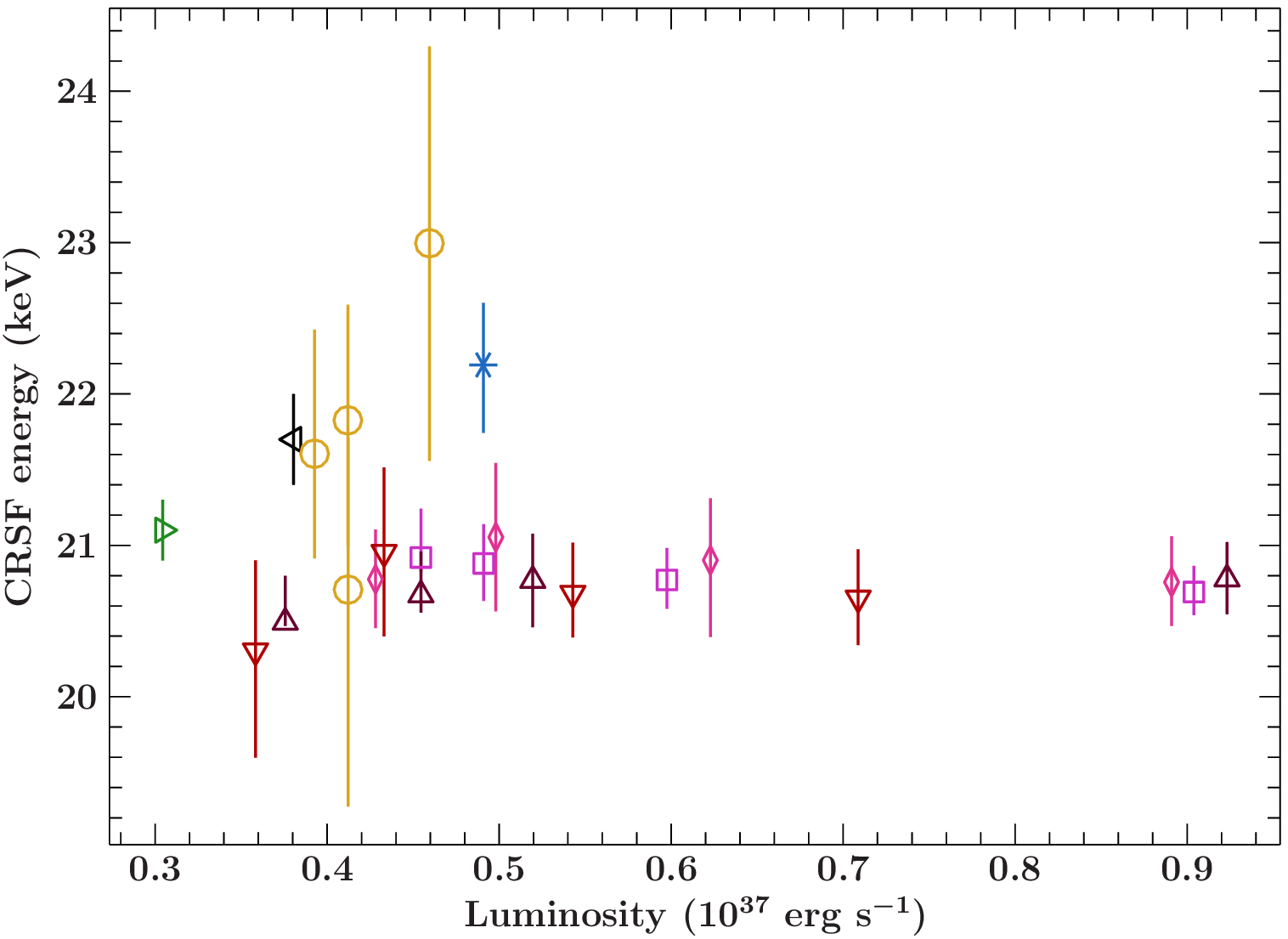}
	\caption{22 years of $E_{\mathrm{cyc}}$ measurements for \fif vs. time (left
  panel) and vs. luminosity (right panel). Luminosity is calculated from the
  3--50 keV flux, taking spectral parameters from the best-fit phase-averaged
  models used by \citet{mihara_thesis_1995} and \citet{robba_bepposax_2001}.
  All measurements use Gaussian profiles for the CRSF. The approximate
	times for \fif's torque reversals in 1990 and late 2008 are
	indicated by vertical dashed lines.}
  \label{fig:ecyc_all}
\end{figure}

We plot all $E_{\mathrm{cyc}}$ measurements for \fif in Fig.~\ref{fig:ecyc_all}.
\citet{robba_bepposax_2001}, analyzing the 1998 \textit{BeppoSAX} observation
of \fif, found a phase-averaged CRSF energy of $21.1 \pm 0.2$\,keV, entirely in
line with the \rxte measurements from around the same time. The
\textit{BeppoSAX} analysis used approximately the same model choices as this
analysis (\texttt{plcut} continuum and a \texttt{gauabs} CRSF) and found the
source at a roughly comparable luminosity to the average \rxte luminosity. It
is somewhat more difficult to compare the results of the 1988 and 1990
\textit{Ginga} observations \citep{clark_discovery_1990,mihara_thesis_1995},
when the CRSF was discovered, due to differences in model choice. The
\citeauthor{clark_discovery_1990} analysis used a model consisting of a
powerlaw modified by a \texttt{cyclabs} component, where the second harmonic in
the \texttt{cyclabs} model effectively modeled the high-energy turnover that we
model using \texttt{highecut}. \citeauthor{mihara_thesis_1995} introduced the
\texttt{npex} continuum model and compared different CRSF models \citep[see
Table F.1.1\ in ][]{mihara_thesis_1995}. Using a Gaussian optical depth
profile for the CRSF, he found an energy of $21.7 \pm 0.3$\,keV (the usually
quoted value is $20.6$\,keV, but this uses the \texttt{cyclabs} model, where
the fitted energy does not correspond to the peak energy of the CRSF).
Applying the same procedure as we used for the significance of the \suz
point, the \textit{Ginga} measurement sits above the \rxte average, but at
only $2.1\sigma$, the separation is not as great as the \rxte-\suz
split.

Currently, Her~X-1 is the only source with confirmed long-term evolution in its
CRSF independent of other observable factors
\citep{staubert_herx1_2014,klochkov_herx1_bat_2015}.  Its CRSF behavior with
respect to time displays two main features: a sharp jump
upwards by $\sim\!4$\,keV in the early 1990s, followed by a
$\sim\!0.25$\,keV\,yr$^{-1}$ decline since then. Additionally, a 2012
\integral observation found a significantly higher CRSF energy compared
to the surrounding \suz, \integral, and \textit{NuSTAR} observations,
indicating that significant changes in the CRSF energy can occur on both
short and long timescales.


In \fif's case, the \rxte and \textit{BeppoSAX} data are bracketed by
higher-energy measurements from \suz and \textit{Ginga}. The \rxte
and \textit{BeppoSAX} measurements alone cannot constrain any trend
with time; however, a linear fit to the \rxte, \textit{BeppoSAX},
\integral, and \suz measurements finds a slope of $0.058 \pm
0.014$\,keV\,yr$^{-1}$. However, the large uncertainties on the
\integral measurements make it impossible to say for certain whether the
\suz measurement represents a long-term trend or merely a short-term
increase in $E_{\mathrm{cyc}}$.

It is highly unlikely that we are observing the evolution of the
neutron star's intrinsic magnetic field; rather, any long-term change
in the CRSF energy is likely due to a change in the properties of
the scattering region where the CRSF is produced. We have discussed
some of the properties of the scattering region above in Section
\ref{ssec:disc_ecyc_flux}; the question now is how one can produce a
$\sim\!1$\,keV shift in the CRSF energy alongside minimal long-term
changes in the source flux given the properties of the accretion flow.

\fif is a young system, as indicated by its high-mass B0Iab companion
and its strong magnetic field, so it is unlikely that accretion has
been ongoing long enough to significantly ``bury'' the magnetic
field \citep[in the sense outlined by][]{payne_melatos_burial_2004};
this burial process proceeds on far too long of a timescale to
produce a change of a few percent in only a few years. However,
\citet{mukherjee_crsf_2012} find that accretion mound masses of
$\sim\!10^{-12}$\,\msol are likely sufficient to distort the magnetic
field significantly. Based on the observed CRSF energy-luminosity
relationship as discussed in \ref{ssec:disc_ecyc_flux}, the CRSF
parameters are probably a good probe of the environment around
the polar cap, and a change in the CRSF parameters could reflect
some changes in the accretion mound which might not be visible in
other observables, such as luminosity. For example, a slow growth
in the mound's height would probably not affect the broad-band
spectral parameters or luminosity of \fif very much \citep[the mound
mainly contributes to the black-body component of the spectrum,
which is a small fraction of the source's overall luminosity; see
e.g.][]{becker_continuum_2007}. However, relatively small changes in
the mound's height can affect the magnetic field in the mound quite
drastically --- \citet{mukherjee_crsf_2012} found that the magnetic
field strength could deviate from the dipole strength by upwards of a
factor of four in the sides of sufficiently large accretion mounds. A
$\sim\!5$\% increase in \fif's CRSF energy could be simply due to a
reconfiguring of the mound geometry resulting in a slightly stronger
average field strength. However, it is unclear as to whether a 5--10
year timescale is realistic for this type of process. Also, as pointed
out above, it is unclear if the shift observed by \suz is representative
of a long-term change in the CRSF energy or if it is more of a temporary
effect.

There is the additional question of whether the change in the CRSF
energy is related to a change in only one of the accretion columns,
or if it is due to changes in both columns. The fact that the changed
CRSF energy is most prominent in the phase-constrained data from the
peak of the main pulse and not detected in the secondary pulse suggests
that this effect may be limited to a single magnetic pole. Given the
possible mechanisms laid out above, this is not an unreasonable thing to
suggest --- there is no fundamental reason that the two poles' accretion
structures should move in lock-step with each other; a difference of a
few percent in the two poles is conceivable.

Finally, it is interesting to note here that the source underwent
torque reversals in $\sim\!1990$ \citep{rubin_observation_1997} and
$\sim\!2008$ \citep{hemphill13,finger_gbm_2009}, intriguingly close
to bracketing the \rxte and \textit{BeppoSAX} observations and
separating them somewhat from the higher-energy \textit{Ginga} and \suz
measurements. For an accreting neutron star, the evolution of the spin
period is driven by the torque exerted on the neutron star's magnetic
field by the accreted material; thus it is reasonable to say that a
shift in the properties of the magnetic field could be associated
with changes in the pulse period evolution. Unfortunately, there is
no spectrally-sensitive coverage of either torque reversal; the best
we can do is look at the \integral results of \citep{hemphill13},
which found no significant changes on either side of the 2008 torque
reversal. However, \integral's relative spectral insensitivity means
there is ample room for smaller changes. The \rxte All-Sky Monitor
\citep{levine_asm_1996} shows no detectable changes in flux in this
time, although \fif is very dim in the ASM --- rebinning the ASM
lightcurve to the $3.74$\,d orbital period, the counting rate is $0.7
\pm 0.8$\,cts\,s$^{-1}$ --- and as such we cannot place any strong
limits on source variability across the torque reversal based on the ASM
lightcurve.

\section{Conclusion}
\label{sec:conclusion}

We have performed a comprehensive analysis of $\sim\!15$\,years of
X-ray observations of the high-mass X-ray binary \fif, using data from
the \rxte, \suz, and \integral satellites. Spectrally, the source is
relatively stable, with the continuum parameters remaining mostly flat
with respect to changes in luminosity. The main results are the lack of
a significant correlation between the centroid energy of the fundamental
CRSF and the increase by $\sim\!1$\,keV in the CRSF energy between
the \rxte and \suz observations. The lack of a detectable correlation
between the CRSF energy and luminosity is supported by theoretical work
by \citet{becker_spectral_2012}, \citet{poutanen_reflection_2013},
and \citet{mushtukov_lcrit_2015}, although there is some uncertainty
as to exactly what theoretical scenario is being played out. The
time-dependence of the CRSF is a less easily understood issue and
requires additional work to, first, confirm or deny its reality and
second, produce a reliable explanation for the phenomenon. An upcoming
\integral campaign will help shed some light on the first point, but
what are truly needed are observations by more spectrally-sensitive
instruments, e.g. those aboard \textit{NuSTAR} or \textit{Astro-H},
which will be able to make a precise measurement of the CRSF energy.

Our analysis makes heavy use of a collection of ISIS scripts provided by
ECAP/Remeis Observatory and MIT, which can be found at
\url{http://www.sternwarte.uni-erlangen.de/isis/}. Support for VG was
provided by NASA through the Smithsonian Astrophysical Observatory (SAO)
contract SV3-73016 to MIT for Support of the Chandra X-Ray Center (CXC)
and Science Instruments; CXC is operated by SAO on behalf of NASA under
contract NAS8-03060.

\footnotesize{
\bibliographystyle{mnras}
\bibliography{refs}

\begin{thebibliography}{}
\makeatletter
\relax
\def\mn@urlcharsother{\let\do\@makeother \do\$\do\&\do\#\do\^\do\_\do\%\do\~}
\def\mn@doi{\begingroup\mn@urlcharsother \@ifnextchar [ {\mn@doi@}
  {\mn@doi@[]}}
\def\mn@doi@[#1]#2{\def\@tempa{#1}\ifx\@tempa\@empty \href
  {http://dx.doi.org/#2} {doi:#2}\else \href {http://dx.doi.org/#2} {#1}\fi
  \endgroup}
\def\mn@eprint#1#2{\mn@eprint@#1:#2::\@nil}
\def\mn@eprint@arXiv#1{\href {http://arxiv.org/abs/#1} {{\tt arXiv:#1}}}
\def\mn@eprint@dblp#1{\href {http://dblp.uni-trier.de/rec/bibtex/#1.xml}
  {dblp:#1}}
\def\mn@eprint@#1:#2:#3:#4\@nil{\def\@tempa {#1}\def\@tempb {#2}\def\@tempc
  {#3}\ifx \@tempc \@empty \let \@tempc \@tempb \let \@tempb \@tempa \fi \ifx
  \@tempb \@empty \def\@tempb {arXiv}\fi \@ifundefined
  {mn@eprint@\@tempb}{\@tempb:\@tempc}{\expandafter \expandafter \csname
  mn@eprint@\@tempb\endcsname \expandafter{\@tempc}}}

\bibitem[\protect\citeauthoryear{{Basko} \& {Sunyaev}}{{Basko} \&
  {Sunyaev}}{1976}]{basko_sunyaev_1976}
{Basko} M.~M.,  {Sunyaev} R.~A.,  1976, \mnras, \href
  {http://adsabs.harvard.edu/abs/1976MNRAS.175..395B} {175, 395}

\bibitem[\protect\citeauthoryear{Baykal, \.{I}nam  \& Beklen}{Baykal
  et~al.}{2006}]{baykal_recent_2006}
Baykal A.,  \.{I}nam S.~c.,   Beklen E.,  2006, \mn@doi [\aap]
  {10.1051/0004-6361:20054616}, 453, 1037

\bibitem[\protect\citeauthoryear{{Becker} \& {Wolff}}{{Becker} \&
  {Wolff}}{2007}]{becker_continuum_2007}
{Becker} P.~A.,  {Wolff} M.~T.,  2007, \mn@doi [\apj] {10.1086/509108}, \href
  {http://adsabs.harvard.edu/abs/2007ApJ...654..435B} {654, 435}

\bibitem[\protect\citeauthoryear{{Becker}, {Swank}, {Boldt}, {Holt},
  {Serlemitsos}, {Pravdo}  \& {Saba}}{{Becker} et~al.}{1977}]{becker_1538_1977}
{Becker} R.~H.,  {Swank} J.~H.,  {Boldt} E.~A.,  {Holt} S.~S.,  {Serlemitsos}
  P.~J.,  {Pravdo} S.~H.,   {Saba} J.~R.,  1977, \mn@doi [\apjl]
  {10.1086/182498}, \href {http://adsabs.harvard.edu/abs/1977ApJ...216L..11B}
  {216, L11}

\bibitem[\protect\citeauthoryear{{Becker} et~al.,}{{Becker}
  et~al.}{2012}]{becker_spectral_2012}
{Becker} P.~A.,  et~al., 2012, \mn@doi [\aap] {10.1051/0004-6361/201219065},
  \href {http://adsabs.harvard.edu/abs/2012A%26A...544A.123B} {544, A123}

\bibitem[\protect\citeauthoryear{{Bradt}, {Rothschild}  \& {Swank}}{{Bradt}
  et~al.}{1993}]{bradt_rxte_1993}
{Bradt} H.~V.,  {Rothschild} R.~E.,   {Swank} J.~H.,  1993, \aaps, \href
  {http://adsabs.harvard.edu/abs/1993A%26AS...97..355B} {97, 355}

\bibitem[\protect\citeauthoryear{{Caballero} et~al.,}{{Caballero}
  et~al.}{2007}]{caballero_a0535_2007}
{Caballero} I.,  et~al., 2007, \mn@doi [\aap] {10.1051/0004-6361:20067032},
  \href {http://adsabs.harvard.edu/abs/2007A%26A...465L..21C} {465, L21}

\bibitem[\protect\citeauthoryear{Clark}{Clark}{2000}]{clark_orbit_2000}
Clark G.~W.,  2000, \mn@doi [\apj] {10.1086/312926}, 542, L131

\bibitem[\protect\citeauthoryear{{Clark}}{{Clark}}{2004}]{clark_chandra1538_2004}
{Clark} G.~W.,  2004, \mn@doi [\apj] {10.1086/421764}, \href
  {http://adsabs.harvard.edu/abs/2004ApJ...610..956C} {610, 956}

\bibitem[\protect\citeauthoryear{Clark, Woo, Nagase, Makishima  \& Sakao}{Clark
  et~al.}{1990}]{clark_discovery_1990}
Clark G.~W.,  Woo J.~W.,  Nagase F.,  Makishima K.,   Sakao T.,  1990, \apj,
  353, 274

\bibitem[\protect\citeauthoryear{Coburn}{Coburn}{2001}]{coburn_magnetic_2001}
Coburn W.,  2001, Phd, University of California, San Diego

\bibitem[\protect\citeauthoryear{Coburn, Heindl, Rothschild, Gruber,
  Kreykenbohm  et~al.}{Coburn et~al.}{2002}]{coburn_magnetic_2002}
Coburn W.,  Heindl W.,  Rothschild R.,  Gruber D.,  Kreykenbohm I.,   et~al.,
  2002, \mn@doi [\apj] {10.1086/343033}, 580, 394

\bibitem[\protect\citeauthoryear{{Corbet}, {Woo}  \& {Nagase}}{{Corbet}
  et~al.}{1993}]{corbet_orbit_1993}
{Corbet} R.~H.~D.,  {Woo} J.~W.,   {Nagase} F.,  1993, \aap, \href
  {http://adsabs.harvard.edu/abs/1993A%26A...276...52C} {276, 52}

\bibitem[\protect\citeauthoryear{{Crampton}, {Hutchings}  \&
  {Cowley}}{{Crampton} et~al.}{1978}]{crampton_1538_1978}
{Crampton} D.,  {Hutchings} J.~B.,   {Cowley} A.~P.,  1978, \mn@doi [\apjl]
  {10.1086/182794}, \href {http://adsabs.harvard.edu/abs/1978ApJ...225L..63C}
  {225, L63}

\bibitem[\protect\citeauthoryear{{Davison}}{{Davison}}{1977}]{davison_1538_1977a}
{Davison} P.~J.~N.,  1977, \mnras, \href
  {http://adsabs.harvard.edu/abs/1977MNRAS.179P..35D} {179, 35P}

\bibitem[\protect\citeauthoryear{{Davison}, {Watson}  \& {Pye}}{{Davison}
  et~al.}{1977}]{davison_1538_1977b}
{Davison} P.~J.~N.,  {Watson} M.~G.,   {Pye} J.~P.,  1977, \mnras, \href
  {http://adsabs.harvard.edu/abs/1977MNRAS.181P..73D} {181, 73P}

\bibitem[\protect\citeauthoryear{{Dickey} \& {Lockman}}{{Dickey} \&
  {Lockman}}{1990}]{dickey_lockman_nh_1990}
{Dickey} J.~M.,  {Lockman} F.~J.,  1990, \mn@doi [\araa]
  {10.1146/annurev.aa.28.090190.001243}, \href
  {http://adsabs.harvard.edu/abs/1990ARA%26A..28..215D} {28, 215}

\bibitem[\protect\citeauthoryear{{Falanga}, {Bozzo}, {Lutovinov},
  {Bonnet-Bidaud}, {Fetisova}  \& {Puls}}{{Falanga}
  et~al.}{2015}]{falanga_hmxbs_2015}
{Falanga} M.,  {Bozzo} E.,  {Lutovinov} A.,  {Bonnet-Bidaud} J.~M.,  {Fetisova}
  Y.,   {Puls} J.,  2015, preprint, \href
  {http://adsabs.harvard.edu/abs/2015arXiv150207126F} {} (\mn@eprint {arXiv}
  {1502.07126})

\bibitem[\protect\citeauthoryear{{Ferrigno}, {Becker}, {Segreto}, {Mineo}  \&
  {Santangelo}}{{Ferrigno} et~al.}{2009}]{ferrigno_4u0115_2009}
{Ferrigno} C.,  {Becker} P.~A.,  {Segreto} A.,  {Mineo} T.,   {Santangelo} A.,
  2009, \mn@doi [\aap] {10.1051/0004-6361/200809373}, \href
  {http://adsabs.harvard.edu/abs/2009A%26A...498..825F} {498, 825}

\bibitem[\protect\citeauthoryear{{Finger} et~al.,}{{Finger}
  et~al.}{2009}]{finger_gbm_2009}
{Finger} M.~H.,  et~al., 2009, in Johnson W.~N.,  Thompson D.~J.,  eds, 2009
  Fermi Symposium. eConf proceedings C0911022

\bibitem[\protect\citeauthoryear{{F{\"u}rst} et~al.,}{{F{\"u}rst}
  et~al.}{2013}]{fuerst_herx1_2013}
{F{\"u}rst} F.,  et~al., 2013, \mn@doi [\apj] {10.1088/0004-637X/779/1/69},
  \href {http://adsabs.harvard.edu/abs/2013ApJ...779...69F} {779, 69}

\bibitem[\protect\citeauthoryear{{F{\"u}rst} et~al.,}{{F{\"u}rst}
  et~al.}{2014}]{fuerst_vela_2014}
{F{\"u}rst} F.,  et~al., 2014, \mn@doi [\apj] {10.1088/0004-637X/780/2/133},
  \href {http://adsabs.harvard.edu/abs/2014ApJ...780..133F} {780, 133}

\bibitem[\protect\citeauthoryear{Giacconi, Murray, Gursky, Kellogg, Schreier,
  Matilsky, Koch  \& Tananbaum}{Giacconi et~al.}{1974}]{giacconi_third_1974}
Giacconi R.,  Murray S.,  Gursky H.,  Kellogg E.,  Schreier E.,  Matilsky T.,
  Koch D.,   Tananbaum H.,  1974, \mn@doi [\apjs] {10.1086/190288}, 27, 37

\bibitem[\protect\citeauthoryear{{Hemphill}, {Rothschild}, {Caballero},
  {Pottschmidt}, {K{\"u}hnel}, {F{\"u}rst}  \& {Wilms}}{{Hemphill}
  et~al.}{2013}]{hemphill13}
{Hemphill} P.~B.,  {Rothschild} R.~E.,  {Caballero} I.,  {Pottschmidt} K.,
  {K{\"u}hnel} M.,  {F{\"u}rst} F.,   {Wilms} J.,  2013, \mn@doi [\apj]
  {10.1088/0004-637X/777/1/61}, \href
  {http://adsabs.harvard.edu/abs/2013ApJ...777...61H} {777, 61}

\bibitem[\protect\citeauthoryear{{Hemphill}, {Rothschild}, {Markowitz},
  {F{\"u}rst}, {Pottschmidt}  \& {Wilms}}{{Hemphill} et~al.}{2014}]{hemphill14}
{Hemphill} P.~B.,  {Rothschild} R.~E.,  {Markowitz} A.,  {F{\"u}rst} F.,
  {Pottschmidt} K.,   {Wilms} J.,  2014, \mn@doi [\apj]
  {10.1088/0004-637X/792/1/14}, \href
  {http://adsabs.harvard.edu/abs/2014ApJ...792...14H} {792, 14}

\bibitem[\protect\citeauthoryear{{Houck} \& {Denicola}}{{Houck} \&
  {Denicola}}{2000}]{isis_2000}
{Houck} J.~C.,  {Denicola} L.~A.,  2000, in {Manset} N.,  {Veillet} C.,
  {Crabtree} D.,  eds,  Astronomical Society of the Pacific Conference Series
  Vol. 216, Astronomical Data Analysis Software and Systems IX. p.~591

\bibitem[\protect\citeauthoryear{{Ilovaisky}, {Chevalier}  \&
  {Motch}}{{Ilovaisky} et~al.}{1979}]{ilovaisky_1538_1979}
{Ilovaisky} S.~A.,  {Chevalier} C.,   {Motch} C.,  1979, \aap, \href
  {http://adsabs.harvard.edu/abs/1979A%26A....71L..17I} {71, L17}

\bibitem[\protect\citeauthoryear{{Jahoda}, {Swank}, {Giles}, {Stark},
  {Strohmayer}, {Zhang}  \& {Morgan}}{{Jahoda} et~al.}{1996}]{jahoda_pca_1996}
{Jahoda} K.,  {Swank} J.~H.,  {Giles} A.~B.,  {Stark} M.~J.,  {Strohmayer} T.,
  {Zhang} W.,   {Morgan} E.~H.,  1996, in {Siegmund} O.~H.,  {Gummin} M.~A.,
  eds,  Society of Photo-Optical Instrumentation Engineers (SPIE) Conference
  Series Vol. 2808, EUV, X-Ray, and Gamma-Ray Instrumentation for Astronomy
  VII. pp 59--70

\bibitem[\protect\citeauthoryear{{Jahoda}, {Markwardt}, {Radeva}, {Rots},
  {Stark}, {Swank}, {Strohmayer}  \& {Zhang}}{{Jahoda}
  et~al.}{2006}]{jahoda_pca_2006}
{Jahoda} K.,  {Markwardt} C.~B.,  {Radeva} Y.,  {Rots} A.~H.,  {Stark} M.~J.,
  {Swank} J.~H.,  {Strohmayer} T.~E.,   {Zhang} W.,  2006, \mn@doi [\apjs]
  {10.1086/500659}, \href {http://adsabs.harvard.edu/abs/2006ApJS..163..401J}
  {163, 401}

\bibitem[\protect\citeauthoryear{{Kalberla}, {Burton}, {Hartmann}, {Arnal},
  {Bajaja}, {Morras}  \& {P{\"o}ppel}}{{Kalberla}
  et~al.}{2005}]{lab_nh_survey_2005}
{Kalberla} P.~M.~W.,  {Burton} W.~B.,  {Hartmann} D.,  {Arnal} E.~M.,  {Bajaja}
  E.,  {Morras} R.,   {P{\"o}ppel} W.~G.~L.,  2005, \mn@doi [\aap]
  {10.1051/0004-6361:20041864}, \href
  {http://adsabs.harvard.edu/abs/2005A%26A...440..775K} {440, 775}

\bibitem[\protect\citeauthoryear{{Klochkov}, {Staubert}, {Santangelo},
  {Rothschild}  \& {Ferrigno}}{{Klochkov}
  et~al.}{2011}]{klochkov_pulseampresolved_2011}
{Klochkov} D.,  {Staubert} R.,  {Santangelo} A.,  {Rothschild} R.~E.,
  {Ferrigno} C.,  2011, \mn@doi [\aap] {10.1051/0004-6361/201116800}, \href
  {http://adsabs.harvard.edu/abs/2011A%26A...532A.126K} {532, A126}

\bibitem[\protect\citeauthoryear{{Klochkov} et~al.,}{{Klochkov}
  et~al.}{2012}]{klochkov_gx304_2012}
{Klochkov} D.,  et~al., 2012, \mn@doi [\aap] {10.1051/0004-6361/201219385},
  \href {http://adsabs.harvard.edu/abs/2012A%26A...542L..28K} {542, L28}

\bibitem[\protect\citeauthoryear{{Klochkov}, {Staubert}, {Postnov}, {Wilms},
  {Rothschild}  \& {Santangelo}}{{Klochkov}
  et~al.}{2015}]{klochkov_herx1_bat_2015}
{Klochkov} D.,  {Staubert} R.,  {Postnov} K.,  {Wilms} J.,  {Rothschild} R.~E.,
    {Santangelo} A.,  2015, \mn@doi [\aap] {10.1051/0004-6361/201526188}, \href
  {http://adsabs.harvard.edu/abs/2015A%26A...578A..88K} {578, A88}

\bibitem[\protect\citeauthoryear{{Koyama} et~al.,}{{Koyama}
  et~al.}{2007}]{xis_2007}
{Koyama} K.,  et~al., 2007, \pasj, \href
  {http://adsabs.harvard.edu/abs/2007PASJ...59S..23K} {59, 23}

\bibitem[\protect\citeauthoryear{{Kreykenbohm}, {Coburn}, {Wilms},
  {Kretschmar}, {Staubert}, {Heindl}  \& {Rothschild}}{{Kreykenbohm}
  et~al.}{2002}]{kreykenbohm_velax1_2002}
{Kreykenbohm} I.,  {Coburn} W.,  {Wilms} J.,  {Kretschmar} P.,  {Staubert} R.,
  {Heindl} W.~A.,   {Rothschild} R.~E.,  2002, \mn@doi [\aap]
  {10.1051/0004-6361:20021181}, \href
  {http://cdsads.u-strasbg.fr/abs/2002A%26A...395..129K} {395, 129}

\bibitem[\protect\citeauthoryear{{Leahy}, {Darbro}, {Elsner}, {Weisskopf},
  {Kahn}, {Sutherland}  \& {Grindlay}}{{Leahy}
  et~al.}{1983}]{leahy_epfold_1983}
{Leahy} D.~A.,  {Darbro} W.,  {Elsner} R.~F.,  {Weisskopf} M.~C.,  {Kahn} S.,
  {Sutherland} P.~G.,   {Grindlay} J.~E.,  1983, \mn@doi [\apj]
  {10.1086/160766}, \href {http://adsabs.harvard.edu/abs/1983ApJ...266..160L}
  {266, 160}

\bibitem[\protect\citeauthoryear{{Levine}, {Bradt}, {Cui}, {Jernigan},
  {Morgan}, {Remillard}, {Shirey}  \& {Smith}}{{Levine}
  et~al.}{1996}]{levine_asm_1996}
{Levine} A.~M.,  {Bradt} H.,  {Cui} W.,  {Jernigan} J.~G.,  {Morgan} E.~H.,
  {Remillard} R.,  {Shirey} R.~E.,   {Smith} D.~A.,  1996, \mn@doi [\apjl]
  {10.1086/310260}, \href {http://adsabs.harvard.edu/abs/1996ApJ...469L..33L}
  {469, L33}

\bibitem[\protect\citeauthoryear{{Makishima}, {Koyama}, {Hayakawa}  \&
  {Nagase}}{{Makishima} et~al.}{1987}]{makishima_spectra_1987}
{Makishima} K.,  {Koyama} K.,  {Hayakawa} S.,   {Nagase} F.,  1987, \mn@doi
  [\apj] {10.1086/165091}, \href
  {http://adsabs.harvard.edu/abs/1987ApJ...314..619M} {314, 619}

\bibitem[\protect\citeauthoryear{{Mihara}}{{Mihara}}{1995}]{mihara_thesis_1995}
{Mihara} T.,  1995, PhD thesis, Dept.~of Physics, Univ.~of Tokyo

\bibitem[\protect\citeauthoryear{{Mowlavi} et~al.,}{{Mowlavi}
  et~al.}{2006}]{mowlavi_v0332_2006}
{Mowlavi} N.,  et~al., 2006, \mn@doi [\aap] {10.1051/0004-6361:20054235}, \href
  {http://adsabs.harvard.edu/abs/2006A%26A...451..187M} {451, 187}

\bibitem[\protect\citeauthoryear{{Mukherjee} \& {Bhattacharya}}{{Mukherjee} \&
  {Bhattacharya}}{2012}]{mukherjee_crsf_2012}
{Mukherjee} D.,  {Bhattacharya} D.,  2012, \mn@doi [\mnras]
  {10.1111/j.1365-2966.2011.20085.x}, \href
  {http://adsabs.harvard.edu/abs/2012MNRAS.420..720M} {420, 720}

\bibitem[\protect\citeauthoryear{Mukherjee, Raichur, Paul, Naik  \&
  Bhatt}{Mukherjee et~al.}{2006}]{mukherjee_orbital_2006}
Mukherjee U.,  Raichur H.,  Paul B.,  Naik S.,   Bhatt N.,  2006, Journal of
  Astrophysics and Astronomy, 27, 411

\bibitem[\protect\citeauthoryear{{M{\"u}ller} et~al.,}{{M{\"u}ller}
  et~al.}{2012}]{muller_sleeping_J1946_2012}
{M{\"u}ller} S.,  et~al., 2012, \mn@doi [\aap] {10.1051/0004-6361/201219580},
  \href {http://adsabs.harvard.edu/abs/2012A%26A...546A.125M} {546, A125}

\bibitem[\protect\citeauthoryear{{M{\"u}ller} et~al.,}{{M{\"u}ller}
  et~al.}{2013a}]{muller_nocorrelation_2012}
{M{\"u}ller} S.,  et~al., 2013a, \mn@doi [\aap] {10.1051/0004-6361/201220359},
  \href {http://adsabs.harvard.edu/abs/2013A%26A...551A...6M} {551, A6}

\bibitem[\protect\citeauthoryear{{M{\"u}ller}, {Klochkov}, {Caballero}  \&
  {Santangelo}}{{M{\"u}ller} et~al.}{2013b}]{mueller_a0535_2013}
{M{\"u}ller} D.,  {Klochkov} D.,  {Caballero} I.,   {Santangelo} A.,  2013b,
  \mn@doi [\aap] {10.1051/0004-6361/201220347}, \href
  {http://adsabs.harvard.edu/abs/2013A%26A...552A..81M} {552, A81}

\bibitem[\protect\citeauthoryear{{Mushtukov}, {Suleimanov}, {Tsygankov}  \&
  {Poutanen}}{{Mushtukov} et~al.}{2015a}]{mushtukov_lcrit_2015}
{Mushtukov} A.~A.,  {Suleimanov} V.~F.,  {Tsygankov} S.~S.,   {Poutanen} J.,
  2015a, \mn@doi [\mnras] {10.1093/mnras/stu2484}, \href
  {http://adsabs.harvard.edu/abs/2015MNRAS.447.1847M} {447, 1847}

\bibitem[\protect\citeauthoryear{{Mushtukov}, {Tsygankov}, {Serber},
  {Suleimanov}  \& {Poutanen}}{{Mushtukov}
  et~al.}{2015b}]{mushtukov_doppler_2015}
{Mushtukov} A.~A.,  {Tsygankov} S.~S.,  {Serber} A.~V.,  {Suleimanov} V.~F.,
  {Poutanen} J.,  2015b, \mn@doi [\mnras] {10.1093/mnras/stv2182}, \href
  {http://adsabs.harvard.edu/abs/2015MNRAS.454.2714M} {454, 2714}

\bibitem[\protect\citeauthoryear{{Nowak}, {Wilms}, {Pottschmidt}, {Schulz},
  {Maitra}  \& {Miller}}{{Nowak} et~al.}{2012}]{nowak_2012}
{Nowak} M.~A.,  {Wilms} J.,  {Pottschmidt} K.,  {Schulz} N.,  {Maitra} D.,
  {Miller} J.,  2012, \mn@doi [\apj] {10.1088/0004-637X/744/2/107}, \href
  {http://adsabs.harvard.edu/abs/2012ApJ...744..107N} {744, 107}

\bibitem[\protect\citeauthoryear{{Payne} \& {Melatos}}{{Payne} \&
  {Melatos}}{2004}]{payne_melatos_burial_2004}
{Payne} D.~J.~B.,  {Melatos} A.,  2004, \mn@doi [\mnras]
  {10.1111/j.1365-2966.2004.07798.x}, \href
  {http://adsabs.harvard.edu/abs/2004MNRAS.351..569P} {351, 569}

\bibitem[\protect\citeauthoryear{{Poutanen}, {Mushtukov}, {Suleimanov},
  {Tsygankov}, {Nagirner}, {Doroshenko}  \& {Lutovinov}}{{Poutanen}
  et~al.}{2013}]{poutanen_reflection_2013}
{Poutanen} J.,  {Mushtukov} A.~A.,  {Suleimanov} V.~F.,  {Tsygankov} S.~S.,
  {Nagirner} D.~I.,  {Doroshenko} V.,   {Lutovinov} A.~A.,  2013, \mn@doi
  [\apj] {10.1088/0004-637X/777/2/115}, \href
  {http://adsabs.harvard.edu/abs/2013ApJ...777..115P} {777, 115}

\bibitem[\protect\citeauthoryear{{Rawls}, {Orosz}, {McClintock}, {Torres},
  {Bailyn}  \& {Buxton}}{{Rawls} et~al.}{2011}]{rawls_mass_2011}
{Rawls} M.~L.,  {Orosz} J.~A.,  {McClintock} J.~E.,  {Torres} M.~A.~P.,
  {Bailyn} C.~D.,   {Buxton} M.~M.,  2011, \mn@doi [\apj]
  {10.1088/0004-637X/730/1/25}, \href
  {http://adsabs.harvard.edu/abs/2011ApJ...730...25R} {730, 25}

\bibitem[\protect\citeauthoryear{{Reynolds}, {Bell}  \& {Hilditch}}{{Reynolds}
  et~al.}{1992}]{reynolds_optical_1538_1992}
{Reynolds} A.~P.,  {Bell} S.~A.,   {Hilditch} R.~W.,  1992, \mnras, \href
  {http://adsabs.harvard.edu/abs/1992MNRAS.256..631R} {256, 631}

\bibitem[\protect\citeauthoryear{{Robba}, {Burderi}, {Di Salvo}, {Iaria}  \&
  {Cusumano}}{{Robba} et~al.}{2001}]{robba_bepposax_2001}
{Robba} N.~R.,  {Burderi} L.,  {Di Salvo} T.,  {Iaria} R.,   {Cusumano} G.,
  2001, \mn@doi [\apj] {10.1086/323841}, \href
  {http://adsabs.harvard.edu/abs/2001ApJ...562..950R} {562, 950}

\bibitem[\protect\citeauthoryear{{Rodes-Roca}, Torrej\'{o}n, Kreykenbohm,
  Mart\'{i}nez N\'{u}\~{n}ez, {Camero-Arranz}  \& Bernab\'{e}u}{{Rodes-Roca}
  et~al.}{2009}]{rodes-roca_first_2009}
{Rodes-Roca} J.~J.,  Torrej\'{o}n J.~M.,  Kreykenbohm I.,  Mart\'{i}nez
  N\'{u}\~{n}ez S.,  {Camero-Arranz} A.,   Bernab\'{e}u G.,  2009, \mn@doi
  [\aap] {10.1051/0004-6361/200912815}, 508, 395

\bibitem[\protect\citeauthoryear{{Rodes-Roca}, {Mihara}, {Nakahira},
  {Torrej{\'o}n}, {Gim{\'e}nez-Garc{\'{\i}}a}  \& {Bernab{\'e}u}}{{Rodes-Roca}
  et~al.}{2015}]{rodes-roca_maxi_2015}
{Rodes-Roca} J.~J.,  {Mihara} T.,  {Nakahira} S.,  {Torrej{\'o}n} J.~M.,
  {Gim{\'e}nez-Garc{\'{\i}}a} {\'A}.,   {Bernab{\'e}u} G.,  2015, \mn@doi
  [\aap] {10.1051/0004-6361/201425323}, \href
  {http://adsabs.harvard.edu/abs/2015A%26A...580A.140R} {580, A140}

\bibitem[\protect\citeauthoryear{{Rothschild} et~al.,}{{Rothschild}
  et~al.}{1998}]{rothschild_hexte_1998}
{Rothschild} R.~E.,  et~al., 1998, \mn@doi [\apj] {10.1086/305377}, \href
  {http://adsabs.harvard.edu/abs/1998ApJ...496..538R} {496, 538}

\bibitem[\protect\citeauthoryear{Rubin, Finger, Scott  \& Wilson}{Rubin
  et~al.}{1997}]{rubin_observation_1997}
Rubin B.~C.,  Finger M.~H.,  Scott D.~M.,   Wilson R.~B.,  1997, \mn@doi [\apj]
  {10.1086/304679}, 488, 413

\bibitem[\protect\citeauthoryear{{Sartore}, {Jourdain}  \& {Roques}}{{Sartore}
  et~al.}{2015}]{sartore_a0535_2015}
{Sartore} N.,  {Jourdain} E.,   {Roques} J.~P.,  2015, \mn@doi [\apj]
  {10.1088/0004-637X/806/2/193}, \href
  {http://adsabs.harvard.edu/abs/2015ApJ...806..193S} {806, 193}

\bibitem[\protect\citeauthoryear{{Staubert}, {Shakura}, {Postnov}, {Wilms},
  {Rothschild}, {Coburn}, {Rodina}  \& {Klochkov}}{{Staubert}
  et~al.}{2007}]{staubert_herx1_2007}
{Staubert} R.,  {Shakura} N.~I.,  {Postnov} K.,  {Wilms} J.,  {Rothschild}
  R.~E.,  {Coburn} W.,  {Rodina} L.,   {Klochkov} D.,  2007, \mn@doi [\aap]
  {10.1051/0004-6361:20077098}, \href
  {http://adsabs.harvard.edu/abs/2007A%26A...465L..25S} {465, L25}

\bibitem[\protect\citeauthoryear{{Staubert}, {Klochkov}, {Wilms}, {Postnov},
  {Shakura}, {Rothschild}, {F{\"u}rst}  \& {Harrison}}{{Staubert}
  et~al.}{2014}]{staubert_herx1_2014}
{Staubert} R.,  {Klochkov} D.,  {Wilms} J.,  {Postnov} K.,  {Shakura} N.~I.,
  {Rothschild} R.~E.,  {F{\"u}rst} F.,   {Harrison} F.~A.,  2014, \mn@doi
  [\aap] {10.1051/0004-6361/201424203}, \href
  {http://adsabs.harvard.edu/abs/2014A%26A...572A.119S} {572, A119}

\bibitem[\protect\citeauthoryear{{Takahashi} et~al.,}{{Takahashi}
  et~al.}{2007}]{hxd_2007}
{Takahashi} T.,  et~al., 2007, \pasj, \href
  {http://adsabs.harvard.edu/abs/2007PASJ...59S..35T} {59, 35}

\bibitem[\protect\citeauthoryear{{Tanaka}}{{Tanaka}}{1986}]{tanaka_fdcut}
{Tanaka} Y.,  1986, in {D.~Mihalas \& K.-H.~A.~Winkler} ed.,  Lecture Notes in
  Physics, Berlin Springer Verlag Vol. 255, IAU Colloq. 89: Radiation
  Hydrodynamics in Stars and Compact Objects. p.~198,
  \mn@doi{10.1007/3-540-16764-1_12}

\bibitem[\protect\citeauthoryear{{Tr\"{u}mper}, {Pietsch}, {Reppin}, {Voges},
  {Staubert}  \& {Kendziorra}}{{Tr\"{u}mper}
  et~al.}{1978}]{truemper_herx1_1978}
{Tr\"{u}mper} J.,  {Pietsch} W.,  {Reppin} C.,  {Voges} W.,  {Staubert} R.,
  {Kendziorra} E.,  1978, \mn@doi [\apjl] {10.1086/182617}, \href
  {http://adsabs.harvard.edu/abs/1978ApJ...219L.105T} {219, L105}

\bibitem[\protect\citeauthoryear{{Tsygankov}, {Lutovinov}  \&
  {Serber}}{{Tsygankov} et~al.}{2010}]{tsygankov_v0332_2010}
{Tsygankov} S.~S.,  {Lutovinov} A.~A.,   {Serber} A.~V.,  2010, \mn@doi
  [\mnras] {10.1111/j.1365-2966.2009.15791.x}, \href
  {http://adsabs.harvard.edu/abs/2010MNRAS.401.1628T} {401, 1628}

\bibitem[\protect\citeauthoryear{{Vasco}, {Klochkov}  \& {Staubert}}{{Vasco}
  et~al.}{2011}]{vasco_herx1_2011}
{Vasco} D.,  {Klochkov} D.,   {Staubert} R.,  2011, \mn@doi [\aap]
  {10.1051/0004-6361/201116863}, \href
  {http://adsabs.harvard.edu/abs/2011A%26A...532A..99V} {532, A99}

\bibitem[\protect\citeauthoryear{{Vasco}, {Staubert}, {Klochkov}, {Santangelo},
  {Shakura}  \& {Postnov}}{{Vasco} et~al.}{2013}]{vasco_herx1_2013}
{Vasco} D.,  {Staubert} R.,  {Klochkov} D.,  {Santangelo} A.,  {Shakura} N.,
  {Postnov} K.,  2013, \mn@doi [\aap] {10.1051/0004-6361/201220181}, \href
  {http://adsabs.harvard.edu/abs/2013A%26A...550A.111V} {550, A111}

\bibitem[\protect\citeauthoryear{{Verner}, {Ferland}, {Korista}  \&
  {Yakovlev}}{{Verner} et~al.}{1996}]{verner_xsect}
{Verner} D.~A.,  {Ferland} G.~J.,  {Korista} K.~T.,   {Yakovlev} D.~G.,  1996,
  \mn@doi [\apj] {10.1086/177435}, \href
  {http://adsabs.harvard.edu/abs/1996ApJ...465..487V} {465, 487}

\bibitem[\protect\citeauthoryear{{White}, {Swank}  \& {Holt}}{{White}
  et~al.}{1983}]{white_highecut}
{White} N.~E.,  {Swank} J.~H.,   {Holt} S.~S.,  1983, \mn@doi [\apj]
  {10.1086/161162}, \href {http://adsabs.harvard.edu/abs/1983ApJ...270..711W}
  {270, 711}

\bibitem[\protect\citeauthoryear{{Willingale}, {Starling}, {Beardmore},
  {Tanvir}  \& {O'Brien}}{{Willingale} et~al.}{2013}]{willingale_nh_2013}
{Willingale} R.,  {Starling} R.~L.~C.,  {Beardmore} A.~P.,  {Tanvir} N.~R.,
  {O'Brien} P.~T.,  2013, \mn@doi [\mnras] {10.1093/mnras/stt175}, \href
  {http://adsabs.harvard.edu/abs/2013MNRAS.431..394W} {431, 394}

\bibitem[\protect\citeauthoryear{{Wilms}, {Allen}  \& {McCray}}{{Wilms}
  et~al.}{2000}]{wilms_tbabs}
{Wilms} J.,  {Allen} A.,   {McCray} R.,  2000, \mn@doi [\apj] {10.1086/317016},
  \href {http://adsabs.harvard.edu/abs/2000ApJ...542..914W} {542, 914}

\bibitem[\protect\citeauthoryear{{Wilms}, {Lee}, {Nowak}, {Schulz}, {Xiang}  \&
  {Juett}}{{Wilms} et~al.}{2010}]{wilms_tbnew_2010}
{Wilms} J.,  {Lee} J.~C.,  {Nowak} M.~A.,  {Schulz} N.~S.,  {Xiang} J.,
  {Juett} A.,  2010, in AAS/High Energy Astrophysics Division \#11. p.~674

\bibitem[\protect\citeauthoryear{{Yamamoto}, {Sugizaki}, {Mihara}, {Nakajima},
  {Yamaoka}, {Matsuoka}, {Morii}  \& {Makishima}}{{Yamamoto}
  et~al.}{2011}]{yamamoto_gx301crsf_2011}
{Yamamoto} T.,  {Sugizaki} M.,  {Mihara} T.,  {Nakajima} M.,  {Yamaoka} K.,
  {Matsuoka} M.,  {Morii} M.,   {Makishima} K.,  2011, \pasj, \href
  {http://adsabs.harvard.edu/abs/2011PASJ...63S.751Y} {63, 751}

\bibitem[\protect\citeauthoryear{{Zel'dovich} \& {Shakura}}{{Zel'dovich} \&
  {Shakura}}{1969}]{zeldovich_shakura_1969}
{Zel'dovich} Y.~B.,  {Shakura} N.~I.,  1969, \sovast, \href
  {http://adsabs.harvard.edu/abs/1969SvA....13..175Z} {13, 175}

\makeatother
\end{thebibliography}
}

\bsp
\label{lastpage}
\end{document}